# Glassy dynamics in active epithelia emerge from an interplay of mechanochemical feedback and crowding.

Sindhu Muthukrishnan[1], Phanindra Dewan[2], Tanishq Tejaswi[1], Michelle B Sebastian[1], Tanya Chhabra[1], Soumyadeep Mondal[2], Soumitra Kolya[3], Sumantra Sarkar[2]*, Medhavi Vishwakarma[1]*

1. Department of Bioengineering, Indian Institute of Science, Bangalore, India
2. Department of Physics, Indian Institute of Science, Bangalore, India
3. Tata Institute of Fundamental Research, Hyderabad, India

Correspondence: medhavi@iisc.ac.in and sumantra@iisc.ac.in

## Abstract

Glassy dynamics in active biological cells remain a subject of debate, as cellular activity rarely slows enough for true glassy features to emerge. In this study, we address this paradox of glassy dynamics in epithelial cells by integrating experimental observations with an active vertex model. We demonstrate that while crowding is essential, it is not sufficient for glassy dynamics to emerge. A mechanochemical feedback loop (MCFL), mediated by cell shape changes through the contractile actomyosin network, is required to drive glass transition in dense epithelial tissues, as revealed via a crosstalk between actin-based cell clustering and dynamic heterogeneity in experiments. Incorporating MCFL into the vertex model reveals contrasting results from those previously predicted by theories- we show that the MCFL can counteract cell division-induced fluidisation and enable glassy dynamics to emerge through active cell-to-cell communication. Furthermore, our analysis reveals, for the first time, the existence of novel collective mechanochemical oscillations that arise from the crosstalk of two MCFLs. Together, we demonstrate that an interplay between crowding and active mechanochemical feedback enables the emergence of glass-like traits and collective biochemical oscillations in epithelial tissues with active cell-cell contacts.

## Introduction:

Glass transition is a kinetic phenomenon in which molecular motion slows dramatically as temperature decreases or density increases [1–3]. A tell-tale feature of glassy dynamics, in addition to jamming, is the coexistence of relatively fast-moving locally fluidised regions and slow-moving arrested regions, or dynamic heterogeneity[3–5]. Similar behaviour is also observed in crowded granular matter and in dense systems of self-propelled, active particles- when density increases, or activity diminishes, a globally arrested state is reached, with the existence of locally fluidised individuals[6–8]. For instance, epithelia- a dense and active tissue- demonstrate glassy dynamics in 2D cell monolayers[9–12], in organoid cultures mimicking organ architecture[13,14], and in vivo[15,16], significant to a plethora of pathophysiological processes from development[15–17] and cancer[18–21] to wound healing[22]. In such active systems, dynamic heterogeneity is believed to be dictated by a competition between crowding and internal activity, such that crowding promotes jamming, and activity promotes fluidization[8,23,24]. In fact, theoretical studies modelling epithelial cells, such as the canonical vertex models, predict glassy behaviour only under unrealistic conditions of negligible cell activity and extremely high cell density[25–27]. These predictions, however, contrast with experiments that clearly show glass-like dynamics in crowded, yet metabolically active and proliferating, epithelial monolayers. Such discrepancies suggest that crowding alone cannot explain glassy dynamics in epithelia.

Here, we identified two problems: First, if crowding alone is insufficient, what other factors drive the glass transition in epithelial tissues? And second, how can we incorporate these factors into a model so that biologically plausible features emerge from simulations?

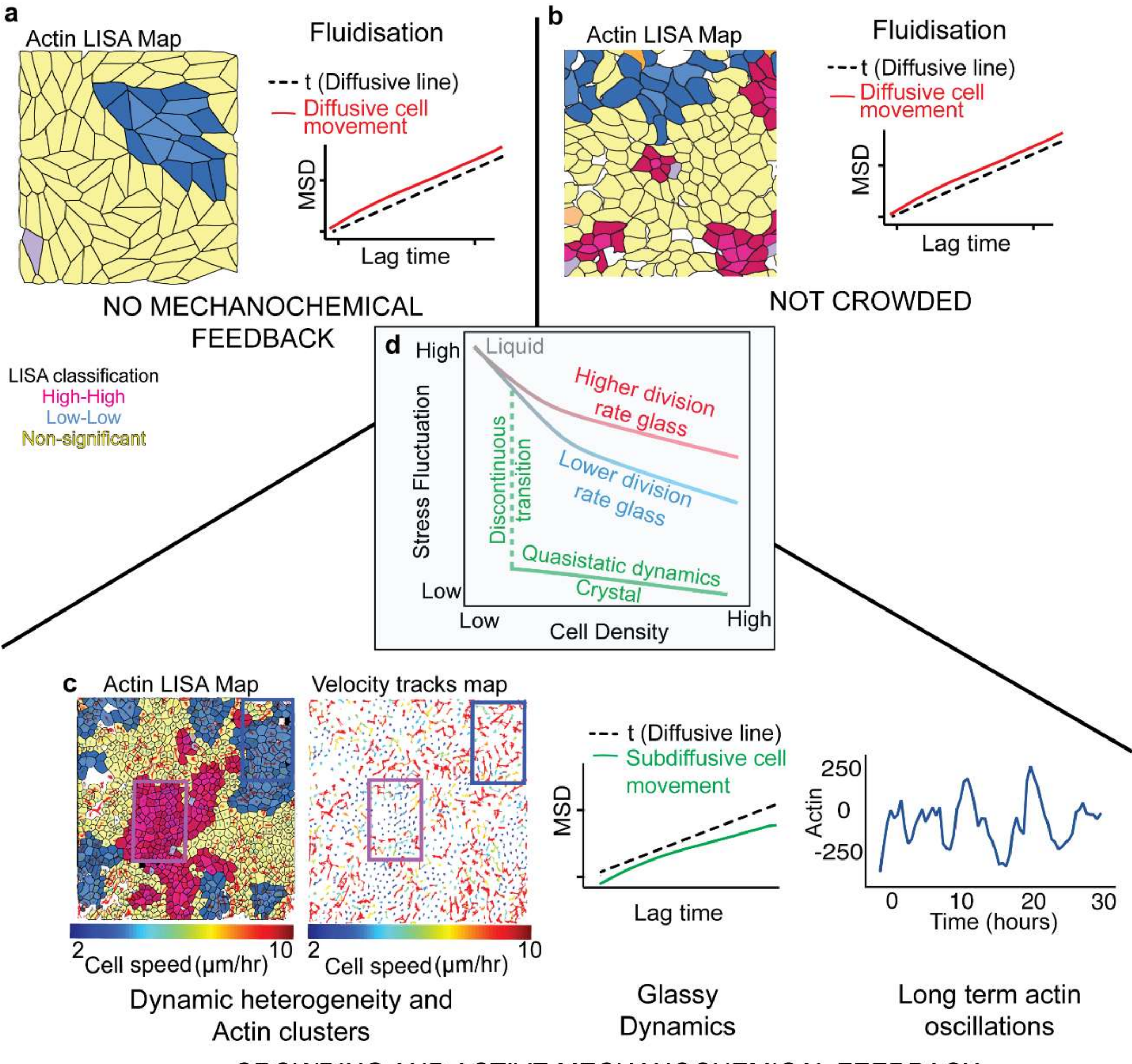

***Figure 1: Crowding and Mechanochemical Feedback are necessary for the emergence of glassy dynamics, collective oscillations and spatial clustering of actin in epithelia****: a) In the absence of mechanochemical feedback, crowding of the tissue through cell division leads to fluidisation, as revealed by the Left-highly anisotropic shapes from the 1/Area LISA (Local Indicators of Spatial Association) map from our vertex model simulation, where 1/Area is proportional to Actin levels and the Right- superdiffusive mean squared displacement (MSD) of the cells from our vertex model and previous studies17-22 . LISA clusters highlight groups of neighbouring cells with similar Actin levels (High-High: Pink and Low-Low: Blue), and cells that are not correlated with their neighbors are classified as Non-Significant (Yellow). b) Mechanochemical feedback without crowding does not lead to glassy dynamics: Right- the MSD remains superdiffusive, and Left- the cells do not form well-defined Actin clusters. c) Crowding and mechanochemical feedback together lead to glassy dynamics. We observe coexisting jammed and unjammed clusters of cells with different motilities, and the MSD becomes subdiffusive revealing dynamic heterogeneity and caging. Furthermore, the cells oscillate collectively with hour-scale time periods. d) The observed glass transition is driven by cell density. Akin to temperature-driven passive glass transition, if the cell density is changed slowly, the tissue undergoes a discontinuous fluid-solid phase transition (green). However, if the density is increased sufficiently rapidly, such as through cell division, the tissue can avoid the discontinuous transition and undergo a glass transition (red and blue) at a density that depends on the division rate: a higher division rate leads to a faster glass transition (red).*

Using actin labelled epithelial cells (LifeAct MDCK), and time-lapse imaging, we mapped actin cytoskeletal rearrangements when cells crowd and dynamic heterogeneity clusters establish, and found that mechanochemical feedback between cytoskeletal dynamics and cell mechanics drives glass transition. Incorporating this feedback into an active vertex model, linking cell geometry to actomyosin signalling, we show that dynamic heterogeneity can emerge even in highly active, dividing tissues, if the mechanochemical feedback is active (Fig. 1). Our experiments and simulations also revealed collective hour-scale actin oscillations, distinct from the minute-scale oscillations observed in isolated cells[28–30]. We show that these slower oscillations arise from mechanochemical feedback that stabilises emerging clusters and disappear when feedback strength or cell density decreases. We show that these oscillations emerge from the interaction between load-dependent actomyosin binding and a second feedback loop coupling the actin cytoskeleton to ERK signalling. Together, our results demonstrate that active mechanochemical feedback enables glass-like dynamics in epithelial tissues through the interplay of cell–cell interactions, signalling, and crowding (Fig.1).

## 2. Results:

### 2.1 Interplay of crowding and mechanochemical feedback dictates emergence of glassy dynamics

We first investigated any possible coupling between glassy dynamics and actin expression in the epithelial monolayer at homeostasis (Fig. 2a). We compared velocity fields with spatial actin distribution patterns from time-lapse imaging data of MDCK cells tagged for Actin (Supplementary video 1 and Fig. 2b). Spatial organization of F-actin was computed using local Moran's index, thus allowing us to generate Local Indicators of Spatial Association (LISA)[31] maps (Fig. 2a, c, Supplementary Fig. 1a-f), which classified cells into clusters representing regions of spatially correlated F-actin levels. Velocity tracks (Fig. 2d) revealed coexisting slow and fast-moving clusters at a densely packed state (Fig. 2c, d), indicating dynamic heterogeneity, a hallmark of glassy systems[4,5,32]. Additional features of glassy dynamics were also observed, including sub-diffusive mean squared displacement and caging (Fig. 2e), and a sub-exponential decay in the overlap function (Q(t), Fig. 2f). To capture the link between the dynamic clusters arising from glassy behavior and the biochemical organization of cells, we quantified the spatial F-actin organization using LISA and found 'High-High' & 'Low-Low' actin clusters, where neighbors correlated in actin expression. We termed those as hotspots and coldspots, respectively. Interestingly, slow-moving, jammed regions aligned with actin hotspots, whereas fast-moving, unjammed regions corresponded to actin coldspots (Fig. 2c, g), indicating that glassy dynamics and actomyosin organization are linked. From traction force microscopy experiments (Fig. 2h-k), we found that cellular tractions and actin expression are negatively correlated; specifically, unjammed coldspots displayed higher traction forces, whereas jammed hotspots displayed lower traction forces (Fig. 2i-k), as reported previously[10,33,34].

While this inverse correlation between actin intensity and both cellular movement and cellular forces may appear non-intuitive, it became clear when actin architecture was closely mapped in these regions. These experiments reveal higher basal stress fibers in unjammed regions, which correspondingly move more, exert higher tractions, and have elongated shapes (Fig. 2l), while jammed regions showed higher cortical actin, explaining higher junctional strength, but lower movement, lower tractions and more compacted shapes (Fig. 2l).

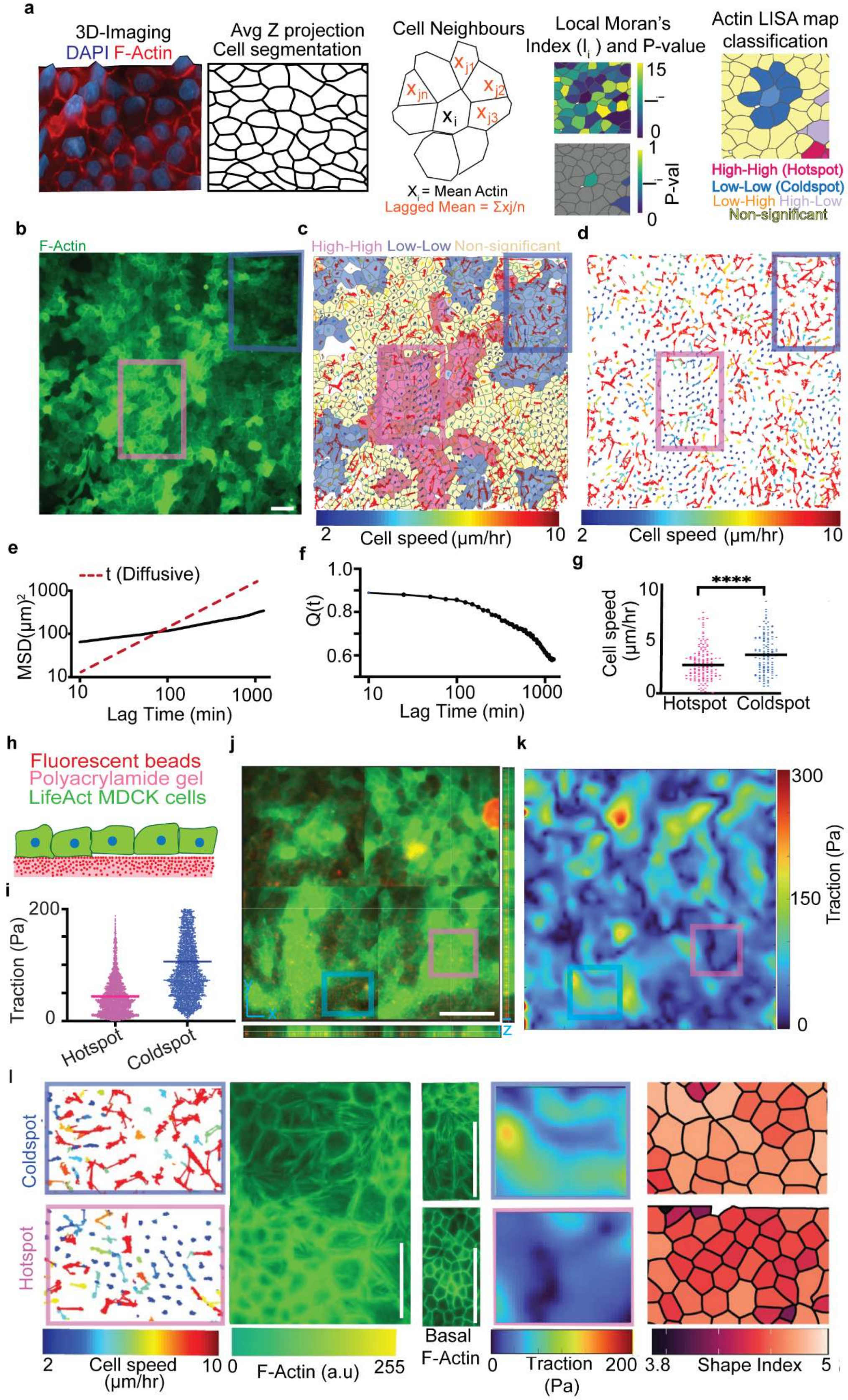

***Fig 2: Spatial distribution of Actin and forces follows the dynamic heterogeneity landscape:*** *a)Schematic of Moran's index analysis with cellpose segmentation and Moran analysis using custom R code, b) Representative*

*image from timelapse microscopy of LifeAct-MDCK cells, c) Cell tracks overlaid on F-Actin LISA map, d) Cell trajectories obtained from trackmate, e) Mean Square Displacement (MSD) plot for one representative timelapse movie of cells, f) Self-overlap function (Q(t)) plot for one representative timelapse movie of cells, g) Actin levels in groups of fast and slow cells and in non-correlated velocity regions, h) Schematic of traction force microscopy, i) Representative image of cells (green for Actin) and fluorescent beads (red), j) Traction magnitude at Actin hotspots and coldspots, k) Left- Traction heatmap (top) at low Actin (bottom) regions and right- Traction heatmap (top) at high Actin regions (bottom). Unpaired t-test: *$p < 0.05$, **$p < 0.01$, ***$p < 0.001$, ****$p < 0.0001$. Scale bars= 50μm.*

Furthermore, we observed a clear correspondence between actin LISA maps, and other mechano-transducers, such as vinculin, E-cadherin, and phospho-myosin (Supplementary Fig. 2). Together, these results suggest that dynamic heterogeneity manifests itself in biochemical signatures through mechanochemical feedback and directly affects tissue functionality by modulating cells' internal machinery. To further verify mechanochemical feedback in the system, we examined biochemistry-mediated force changes by measuring traction with and without Blebbistatin (Supplementary Fig. 3) and found that drugs that alter cellular biochemistry also dynamically alter traction (Supplementary Fig. 3e-f). Conversely, applying external force by stretching the cell substrate with a custom cell stretch device changed F-Actin levels, suggesting that external forces can dynamically change cellular biochemistry (Supplementary Fig. 3a-c).

To further ascertain that mechanochemical feedback is an important determinant of glass transition, we inhibited this feedback by contractility inhibitor Y27632, which inhibits ROCK pathway (Supplementary Fig. 4a, d). As expected, contractility inhibition led to reduced stress fibres resulting in uniform actin architectures in the monolayer (Supplementary Fig. 3h). While actin expression still showed clustering as confirmed by the presence of actin hot-spots and cold-spots (Supplementary Fig. 4b, d), the difference in movement in these clusters disappeared (Supplementary Fig. 4c, f). Furthermore, the monolayer was fluidized, with lack of jamming as confirmed by a diffusive MSD (Supplementary Fig. 4g, 4h), suggesting that the disruption of mechanochemical feedback prevents glass transition.

Since crowding is essential for dynamic heterogeneity to emerge in active and passive glass formers, we wondered whether biochemical clustering is also modulated at different levels of crowding. Indeed, LISA maps of F-actin in low and high-density monolayers (Fig. 3a and 3b) showed differences in locally correlated clusters: loosely packed monolayers had smaller clusters, and densely packed monolayers had larger clusters (Fig. 3c). Global Moran's index, a metric for spatial correlation in actin among immediate neighbours, increased with cell density, while overall spatial actin heterogeneity reduced (Fig. 3d, Supplementary Fig. 5b-e). Furthermore, in loosely packed monolayers, the mean squared displacement was not sub-diffusive (Fig. 3e) and the decay in the overlap function was faster than in a densely packed monolayer (Fig. 3f), suggesting that crowding is essential for the emergence of glassy dynamics. PIV analysis also showed diffusive dynamics in a loosely packed monolayer, but movement consistently became less diffusive with local pockets of unjamming, as density rise (Supplementary Fig. 6f-h).

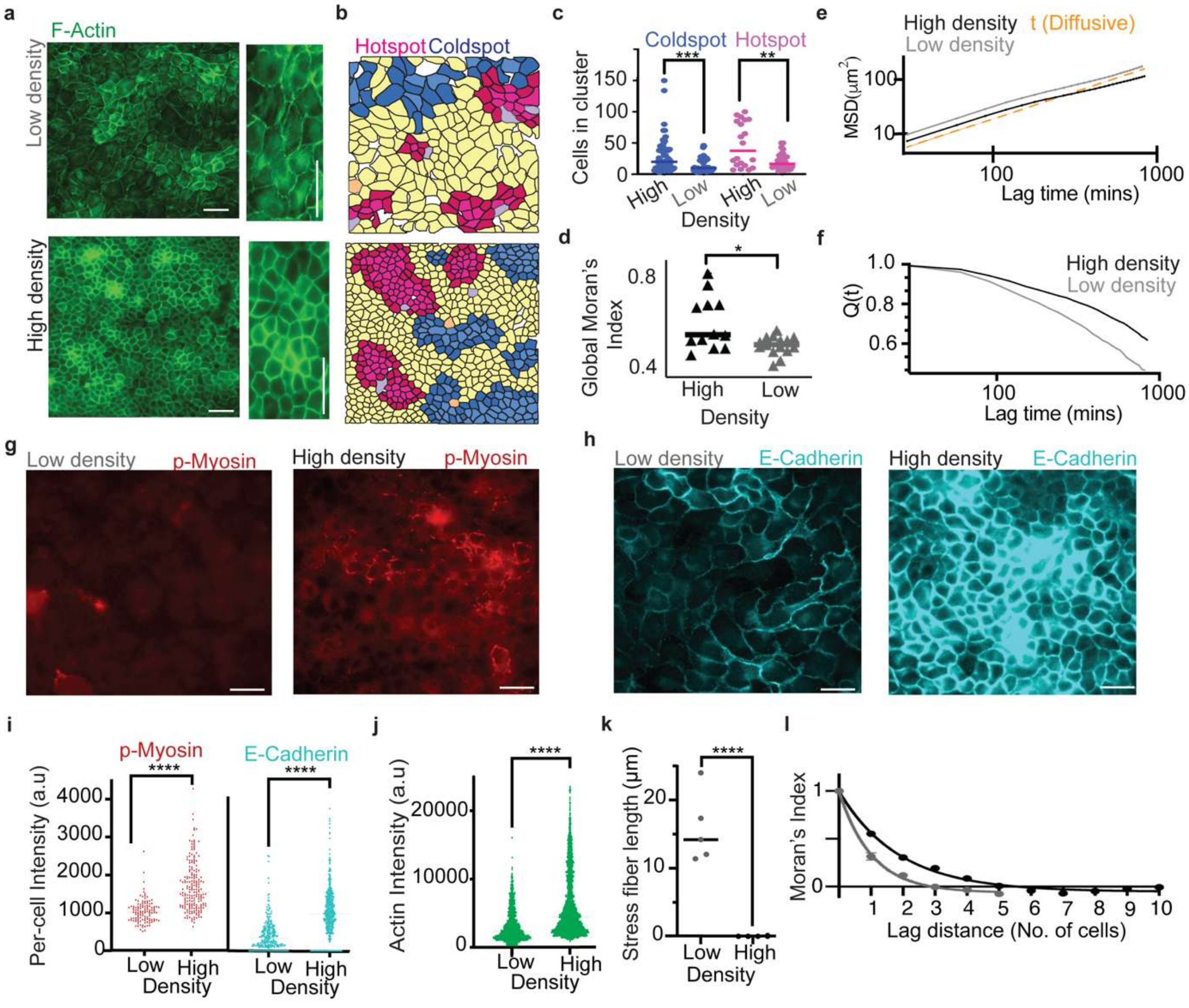

***Fig 3: Higher spatial clustering with increasing density:*** *a) Immunocytochemistry images of Actin at low (top) and high (bottom) density, b) LISA cluster maps of Actin at low (top) and high (bottom) density, c) Number of cells in hotspots and coldspots at high and low density, d) Global Moran's Index at low and high density, e) Mean Square Displacement (MSD) plots for Low density (Gray) and High density (black) monolayers, f) Self Overlap function Q(t) plots for Low density (Gray) and High density (black) monolayers, g) Immunocytochemistry average Z-projected image of Phospho-Myosin (P-Myosin) at low (left) and high (right) density, h) Immunocytochemistry average Z-projected image of E-Cadherin at low (left) and high (right) density, i) Single-cell intensity comparisons of E-Cadherin (blue) at high and low density and P-Myosin (red), j) Comparison of single-cell Actin intensity at high and low density, k) Stress fiber length at high and low density, l) Correlogram showing the decay in Moran's Index as a function of number of lags. Scale bars=50μm. *p=0.017.*

We also observed global differences in cell adhesions and contractility, as indicated by the overall lower intensity of actin, E cadherin as well as myosin in low-density monolayers (Fig. 3a, 3g-j). Longer and more numerous actin stress fibres in low-density monolayers (Fig. 3a, 3k) explains unjamming and higher movement of cells. In densely packed monolayers, cortical actin was higher, stress fibres were shorter and less numerous, and overall actin intensity was higher (Fig. 3a, 3j, 3k). Finally, inhibiting cell division by mitomycin in a homeostatic epithelium inhibits crowding or jamming (Supplementary Fig. 6a-c) and prevents glass transition, as confirmed by cell shapes and MSD plots, such that cells remain unjammed with diffusive MSD profile in the presence of mitomycin (Supplementary Fig. 6d). Together, these results indicate an active interplay between crowding and mechanochemical feedback in dictating cell clustering and glassy behavior at homeostasis.

## 2.2 An active vertex model with mechanochemical feedback and crowding undergoes a glass transition

### 2.2.1 Actomyosin contractility drives two coupled mechanochemical feedback loops

To delineate the role of crowding and mechanochemical feedback, we developed an active vertex model that integrates cell division and mechanochemical feedback loops (MCFLs). Specifically, we considered the crosstalk of two MCFLs (Fig. 4a). The first, MCFL-I, arises from load-dependent binding of myosin to actin[35–37], whereas the second, MCFL-II, arises from the feedback between ERK (extracellular signal-regulated kinase) and the actin cytoskeleton[38–40] (Fig. 4a). MCFL-I is sufficient to understand the glass transition, but the crosstalk of the two is needed to understand the collective oscillation that we describe later.

In the model, the load-dependent binding is captured by cell area ($A$) dependent variation of the vertex model parameters $A_0$ and $P_0$ (SI-theory), which we model using a key experimental observation. We observe that as the tissue matures, F-actin and myosin localise near the cell-cell junctions at the expense of the stress fibres (Fig. 3a, k), suggesting an increase in line tension, but reduced contractility as $A$ decreases upon crowding.

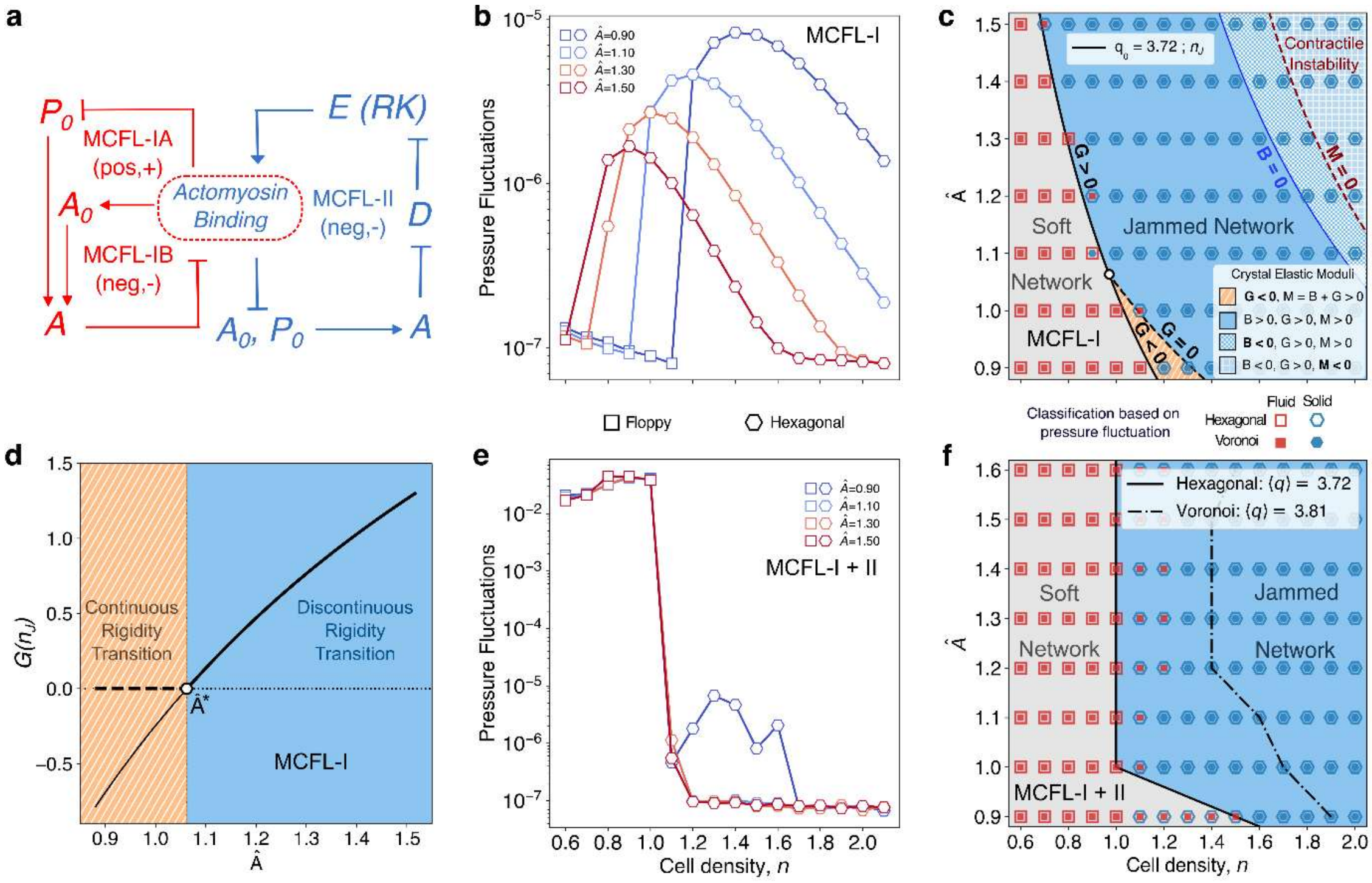

*Figure 4* ***Ground states of active vertex model with mechanochemical feedback:*** *(a) A schematic showing the crosstalk of MCFL-I and MCFL-II. (b) In constant density simulations (no cell division), pressure fluctuations change discontinuously at a density* $n_J(\hat{A})$*. This discontinuous change marks the boundary between soft fluid network to jammed network. Here results from the hexagonal initial condition are shown. (c) Using pressure fluctuation, the states can be classified into fluid (red square) and solid (blue hexagons) phases. Open and filled markers denote hexagonal and Voronoi initial conditions, respectively, which shows that the transition is agnostic to the initial condition. The solid black line corresponds to* $q_0 = P_0(A)/\sqrt{A_0(A)} = 3.72$ *and clearly demarcates the fluid and the solid phases. However, the location of this line is not density independent because of the mechanochemical feedback. Above* $n_J$*, the ground state is a hexagonal crystal for the hexagonal initial condition. The elastic moduli computed in this state can be used to classify the crystalline ground state.*

*Specifically, it shows that at sufficiently high density, the feedback generates spatially heterogeneous patterns through contractile instability. (d) Also, above a threshold feedback strength, $\hat{A}^*$ (black open circle), the shear modulus, $G(n_J)$, is positive, implying a discontinuous transition. Below this threshold, it is negative, implying that the transition is continuous. (e, f) We obtain results similar to (b, c) when both MCFL-I and II are considered.*

In the vertex model, with $\Gamma$ and $K$ as the perimeter and area elasticity coefficients, respectively, contractility is equal to $(\Gamma/K)A_0^{-1}$and tension is equal to $-(\Gamma/K)(P_0/A_0^{3/2})$ [41,42] (SI-theory). Hence, crowding, i.e. decreasing $A$, must increase $A_0$ and decrease $P_0$ (SI-theory, Fig. 4a). Crucially, this choice renders $A_0$ and $P_0$ as dynamical variables, which changes with respect to time, captured here by the following two equations:

$$\tau_A \dot{A}_0 = -\left[A_0 - \hat{A}_0(A)\right]_I$$
$$\tau_P \dot{P}_0 = -\left[P_0 - \hat{P}_0(A)\right]_I$$

Here, $\dot{X} = dX/\,dt$ and the subscript $I$ indicates that the terms originate from MCFL-I. At a constant area $A$, $A_0$ and $P_0$ relax to $\hat{A}_0 = 2a_0 p_{bound}$ and $\hat{P}_0 = 2\hat{q}_0\sqrt{a_0}(1 - p_{bound})$, with timescales $\tau_A$ and $\tau_P$, where $p_{bound} = 1/[1 + \left(A/\hat{A}\right)^2]$ is the fraction of cortical actomyosin and $1 - p_{bound}$ is the fraction of stress fibres (SI-theory). The parameter $\hat{A}$ measures the strength of MCFL-I and is inversely proportional to the binding affinity of myosin to actin. The other parameters, $a_0 = 1$ and $\hat{q}_0 = P_0(\hat{A})/\sqrt{A_0(\hat{A})}$, defines the $A_0$ and the shape index when $A = \hat{A}$, i.e., $p_{bound} = 0.5$ (SI-theory). Because the cell motility is driven by the stress fibres, we expect the motility to be directly proportional to $1 - p_{bound}$:

$$v(A) = 2v_0(1 - p_{bound}),$$

where $v_0$ is the motility when $A = \hat{A}$. Surprisingly, this simple modification (Supplementary Fig. 9) is sufficient to drive tissue solidification through cell division. In contrast, in the canonical vertex model, where $A_0$, $P_0$, and $v$ are fixed parameters of the model, cell division leads to tissue fluidisation (Supplementary video S9).

The equations governing MCFL-II are more complicated because they incorporate the ERK as a Hopf oscillator. We model ERK ($E$) as a Brusselator, but any choice of Hopf oscillators would work[40]. The equations describing $E$ are:

$$\dot{M} = a - (b+1)M + cM^2E$$
$$\dot{E} = bM - cM^2E - DE$$
$$\tau_D \dot{D} = -(D - D_0)$$

, where $M$ is a regulator of $E$ and $D$ is a degrader of $E$. $a, b, c,$ and $D_0$ are the parameters of the Brusselator (SI-theory). MCFL-II couples these equations with the cell mechanics. Specifically, deviation of $E$ from its homeostatic value $E_0$ changes $A_0$ and $P_0$ with strength $\alpha$, whereas compressing the cell below a threshold area of 1 upregulates the production of the degrader $D$ with a strength $\beta$, such that $\mu = \alpha\beta$ determines the strength of MCFL-II. Taken together, the equations governing MCFL-I and II are:

$$\dot{M} = a - (b+1)M + cM^2E$$
$$\dot{E} = bM - cM^2E - DE$$

$$\tau_D \dot{D} = -(D - D_0) - [\,\beta D(A-1)\,]_{\mathrm{II}}$$
$$\tau_A \dot{A}_0 = -\left[\, A_0 - \hat{A}_0(A) \,\right]_I - [\,\alpha(E - E_0)\,]_{\mathrm{II}}$$
$$\tau_P \dot{P}_0 = -\left[\, P_0 - \hat{P}_0(A) \,\right]_I - \left[\, \alpha(E - E_0)/\sqrt{A_0} \,\right]_{\mathrm{II}}$$

We use this final set of equations to investigate the origin of tissue solidification.

**2.2.2 In the quasistatic limit, tissue solidification is density-dependent and occurs through a discontinuous rigidity transition**

Is the observed tissue solidification a glass transition? In passive, temperature-driven glass transitions, the glass-forming liquid must be cooled through its freezing point faster than its nucleation rate. By doing so, it can transition to the metastable supercooled liquid phase, which can undergo a glass transition. Otherwise, it discontinuously transitions to the crystalline ground state[43]. It has been argued that cell density, $n$, is the driver of glass transition in tissues, such that the cell division rate is the analogue of the cooling rate[12,15]. If this analogy is exact, then we expect that if $n$ is changed quasistatically in the absence of cell division, then tissue solidification must happen through a discontinuous phase transition.

The pioneering works of Staple et.al[44]. and Bi et.al[46]. have shown that in the context of the canonical vertex model, tissue solidification is *density-independent* and is driven only by cell shapes. As the cell shape, characterized by the shape parameter $q_0 = P_0/\sqrt{A_0}$, crosses the jamming point $q_0^*$, the tissue undergoes a continuous transition from a floppy to a rigid network of cells, marked by the continuous change of the shear moduli $G$ from zero to a nonzero value. Considering these results, our hypothesis of density-driven discontinuous rigidity transition is rather surprising.

To test this hypothesis, we disabled cell division in the model, set $v = 0$, and fixed the values of $n$ and $\hat{A}$ and let the system relax to the ground state. These choices enabled us to turn off any time-dependent variation of $A_0$ and $P_0$, while preserving the density-dependent variation, such that our model can be directly compared to the earlier works. First, we turned off MCFL-II and studied the effect of just MCFL-I on the rigidity transition, using discontinuous changes in the pressure fluctuation (SI-theory), as the marker of the transition (Fig. 4b). To understand the dependence of this transition on the feedback strength $\hat{A}$ and the cell density $n$, we marked the fluid-like floppy network from the rigid network on a phase diagram (Fig. 4c). Interestingly, the rigidity transition was very similar when either disordered hexagonal lattice (hex-IC) or Voronoi tiling (vor-IC) was used as the initial condition. Consistent with the earlier works, the transition occurred when $q_0$ (for hex-IC) or, equivalently, $q = \langle P/\sqrt{A} \rangle$ (for vor-IC) was approximately 3.722, the shape index of the hexagonal lattice (Fig. 4c). However, the transition was not independent of $n$. Because $A_0$ and $P_0$ are coupled to $n$ through $p_{bound}$ in MCFL-I, $q_0 \approx 3.722$ occurs at a density $n_J$ that depends on the feedback strength $\hat{A}$. Therefore, our first important conclusion is that although tissue solidification can be understood only through cell shapes, cell shapes and cell density are intimately connected through mechanochemical feedback loops. Hence, tissue rigidity transitions are *not density independent.*

Next, we sought to understand how $G$ changes across the transition. To do so, we computed the shear modulus of a hexagonal crystal for $n \geq n_J(\hat{A})$, where it is the ground state. Because the rigidity transition was nearly identical between hex- and vor-IC, this approximation was sufficient to delineate the difference in the rigidity transition between the vertex models with and without feedback. Indeed, in our model, the transition at $n_J$ was not accompanied by the vanishing of $G$. Instead, above a threshold $\hat{A}$, referred to here as $\hat{A}^*$, $G(n_J)$ changed abruptly to a positive value, while it was negative below this threshold (Fig. 4d). The abrupt positive change in $G$ above $\hat{A}^*$ reinforces our observation that the density-driven tissue rigidity transition is a discontinuous transition. $G < 0$ below $\hat{A}^*$ indicates the fragile nature of the crystalline state for densities just above $n_J$. Specifically, these states are unstable to infinitesimal shear[47–49]. Indeed, for $\hat{A} = 0.9$, which is below $\hat{A}^*$, we observe anomalous shape and pressure fluctuations in this range when the tissue was simulated with MCFL-II (Fig. 4e) and motility (SI-theory). Hence, below $\hat{A}^*$, we expect rigid states to emerge only when $G = 0$ through a continuous transition, as in the jamming transitions[44,46,50].

We also observe a discontinuous change in pressure fluctuations when the crosstalk of MCFL-I and II is considered (Fig. 4e). The transition corresponds to $\langle q \rangle \approx 3.722$ for the hex-IC, as in the case with only MCFL-I. However, the density at which this transition happens is different from MCFL-I (Fig. 4f), as MCFL-II adds another coupling of $A_0$ to the density via the $\beta D(A-1)$ term. The discontinuous transition in the case of vor-IC also happens at densities that depend on $\hat{A}$, where $3.81 < \langle q \rangle < 3.9$. Therefore, our second important conclusion is that when the mechanochemical feedback is sufficiently strong, tissue solidification occurs discontinuously at $n_J(\hat{A})$. Hence, increasing cell density above $n_J$ with a sufficiently fast division rate should make tissues glassy. This is indeed what we observe.

### 2.2.3 Increasing cell density at a finite division rate leads to a glass transition

Increasing density through constant division rate simulations at different division rates $r$, we find that at sufficiently low densities, irrespective of the division rates, the pressure fluctuations decay at the same rate as the fluid state in the quasistatic simulations (Fig. 5a, Supplementary Fig. 10), indicating that a unique fluid state exists at these densities. As the density increases, the system exits the fluid phase at a density $n_G(r)$, that decreases with increasing division rate (triangular markers in Fig. 5a, b). Above $n_G$, the decay in pressure fluctuations or mean asphericity do not depend on $\hat{A}$, but below this density they vary with $\hat{A}$. This observation implies that cell-cell interactions are the key drivers of relaxation below $n_G$, whereas above it, crowding is the key driver[51]. This observation parallels the phenomenology of the passive glass transition, where a similar cooling-rate dependence is observed [1,3]. Furthermore, the switching of the relaxation mechanism is reminiscent of the transition between the mode-coupled and activated relaxation postulated in passive glass transition. Hence, we claim that at constant division rates, the tissue undergoes a glass transition.

Because tissue rigidity transition is driven by changes in cellular shapes, we tested whether the glass transition is also manifested in the parameters describing the cell shapes. To do so, we computed the anisotropy of the cell shape through the mean asphericity (SI-theory), $\langle \kappa \rangle$, as a function of $n$. In the quasistatic limit ($r = 0$), $\langle \kappa \rangle$ decreases monotonically (Fig. 5b), with marked difference above and below $n \approx 4$. For simulations with nonzero division rates and motility, the $\langle \kappa \rangle$ trajectory deviates significantly from the quasistatic limit (Fig. 5b, Supplementary Fig. 10). The rate-dependence of the glass transition is observed here as well; systems with lower division rates show a lower $\langle \kappa \rangle$ in the glassy phase and a higher $n_G$.

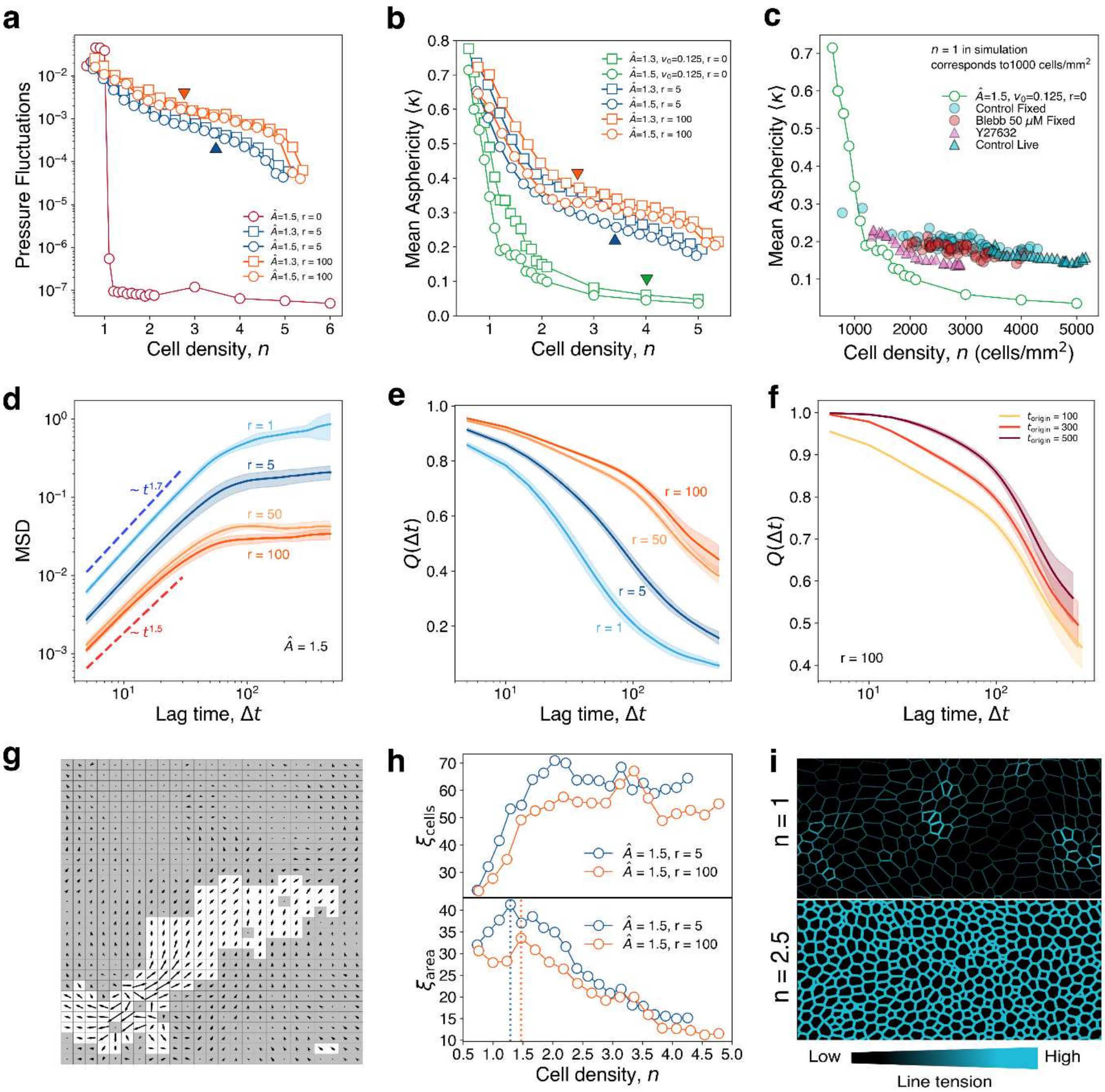

*Figure 5:* ***Glassy dynamics in the vertex model with mechanochemical feedback:*** *(a) Pressure fluctuation decay with density depends on the rate of cell division, akin to the cooling rate dependence of passive glass transition. The arrows indicate $n_G(r)$. The discontinuous constant-density simulation data is also shown for reference. (b) Mean asphericity of the cells, $\langle\kappa\rangle$, shows a similar rate dependence. $\langle\kappa\rangle$ from constant density simulations ($r = 0$) decays with $n$. The arrows indicate $n_G(r)$. (c) Mean asphericity of untreated, Blebbistatin-treated, and Y27632-treated cells shows trends similar to simulations performed at different division rates, which is consistent with the known effect of Blebbistatin and Y27632 on epithelial cells (see text). (d) Mean squared displacement (MSD) as a function of lag time ($\Delta t$) for different division rates shows faster transition to subdiffusive plateaus at higher division rates. (e) The overlap function $Q(\Delta t)$ exhibits similar behaviour, suggesting that the tissue transitions into the glassy state more rapidly at a higher division rate (see legends in d). The shaded regions indicate the standard deviation. (f) $Q(\Delta t)$ calculated with respect to different time origins (for trajectories of the cells in simulation) shows ageing. (g) The simulations show persistent dynamic heterogeneities. (h) Consistent with prior experiments, $\xi_{cells}$, the size of dynamic heterogeneity increases with cell density at low densities, peaks at an intermediate density and saturate to a slightly lower value at high densities. The normalized area of DH has been computed as $\xi_{area} = \xi_{cells}/n$. The markers show mean $\xi_h$ and $\xi_{area}$. (i) A colormap of the line tension, $\bar{\Lambda}$ at low and high densities is consistent with E-cadherin expression in the experiments. The colormaps also show contractile instabilities. For all the panels (except h and i), we simulated a system with 10×10 box size, corresponding to 1000 cells at $n = 1$. Panel h is generated from a simulation of a system with 20×20 box size. For all panels, error bars or shaded regions indicate standard deviation.*

To experimentally verify these results, we perturbed the tissue with drugs which inhibit cellular contractility: Y27632 and Blebbistatin, both of which lowered $\hat{A}$ by preventing myosin binding and reduced growth rate of cells and higher shape indices (Supplementary Fig. 12). The effect of the reduced growth rate can be captured by a lower effective growth rate $r_{eff}$ in our model. By segmenting the cells, we computed $\langle\kappa\rangle$ at different densities $n$ under various treatment conditions (Fig. 5c). Consistent with our theoretical observations, we find that the lower $r_{eff}$ at higher drug concentrations led to a delayed glass transition, as indicated by lower $\langle\kappa\rangle$ values (Fig. 5c).

To prove beyond doubt that the observed dynamics arises from an underlying glass transition, we next computed the MSD (Fig. 5d) and the overlap function ($Q(\Delta t)$, Fig. 5e), both of which showed flattening at intermediate timescales at all $r$. In particular, the relaxation of $Q(\Delta t)$ shows the classic signatures of ageing (Fig. 5f). Additionally, the flattening was more pronounced at higher $r$ (Fig. 5e), which is consistent with our claim that the glass transition occurs sooner for higher division rates. Although this observation contradicts previous theories describing glassy dynamics in biological cells, where a high division rate enhances fluidisation[26,27,53,54], it agrees with experiments describing glassy dynamics in epithelial cells[9,11,12,15,16,52,55–60]. Next, we calculated the cell velocities from the simulations, which exhibited features of dynamic heterogeneity (Fig. 5g). The size of the dynamic heterogeneity, $\xi_{cells}$ or $\xi_{area}$, changed nonmonotonically with density (Fig. 5h). These observations are consistent with prior experiments on glassy tissue dynamics[7,8,10], thereby establishing our claim.

#### 2.2.4 Mechanochemical feedback engenders dynamic heterogeneity and glassy dynamics

Because cell motility is inversely related to cell density in our model, the dynamic heterogeneity reflects the underlying spatiotemporal heterogeneity in the tissue. Indeed, experiments[12] including ours (Fig. 2l) suggest that mechanochemical feedback links heterogeneity in local cell density to heterogeneity in cell motility. Hence, the origin of dynamic heterogeneity can be traced to static density heterogeneity arising from elastic instabilities in the underlying tissues. Indeed, through the analysis of the crystalline ground state, we find that, beyond a threshold density, the P-wave modulus, $M = B + G$, becomes negative, while $G$ stays positive (Fig. 4c, SI-theory). This transition marks the onset of contractile instabilities, during which the homogeneous tissue spontaneously separates into high- and low-density phases. Because $G > 0$ and the tissue is confluent, the phase-separated tissue is stable , and the high- and low-density phases mature into high- and low-density regions (Supplementary Fig. 11) that we identify with the actin hot- and coldspots. Indeed, as in the experiments, the mechanics of the high- and low-density regions in our simulation arise from the line tension and contractility, respectively (Fig. 5i, Supplementary Fig. 10). As the high-density regions slowly grow and percolate through the system, cell motility remains high only in the spatially heterogeneous low-density regions, giving rise to the dynamic heterogeneity and slow glassy dynamics. Also, because of this reason, stresses generated by cell division mobilize only the cells in the low-density regions, ensuring tissue solidification through cell division.

### 2.3 Mechanochemical feedback stabilises activity clusters to generate long-term collective oscillations

The glass transition, driven by crowding and mechanochemical feedback, enables cells to assemble into jammed and unjammed clusters, characterized by high and low actin levels, respectively. Next, we wondered about the stability of these clusters over time and whether collective temporal patterns in biochemical signaling can emerge from dynamic heterogeneity. When we measured spatial trends in actin expression by imaging LifeAct MDCK cells over time (Fig. 6a and b), we observed collective hour-scale actin oscillations, with locally distinct time periods in hotspots and cold spots- hotspots exhibited dominant 11-hour oscillations in actin expression, while coldspots showed 4-hour oscillations (Fig. 6c, d, f), suggesting cold spots are more dynamic than hotspots. The difference in actin expression in hotspots and coldspots is large enough to not transition a hotspot into a cold spot, and vice versa, even during the oscillatory cycle (Supplementary Fig. 8a, 8b, 8e), suggesting that these regions are not interchangeable. Similar oscillatory behavior was also observed in traction force analysis, which showed global oscillations over a timescale of several hours (Supplementary Fig. 15). These results indicate that dynamic heterogeneity clusters are also synchronized in their biochemistry, with correlated and stabilized actin expression that oscillates over a timescale of several hours. To explain the biochemical origin of the observed oscillations, we consider ERK, a common biochemical oscillator.

Interestingly, oscillations in ERK and actin signal are reported in mammalian cells before, including in immune cells[28], and in embryonic stem cells[29,30]. However, in those studies, oscillations are over timescales of several minutes. Interestingly, *lifeAct* MDCK cells, when cultured as single cells, i.e. in the absence of cell-cell interactions, show oscillatory behaviour of actin with a period of approximately 20 mins (Fig. 6a). The transition from minutes timescale of actin oscillations in single cells, to hours timescale of oscillations in crowded monolayer reflects the coupling of glassy dynamics to cellular biochemistry, allowing for

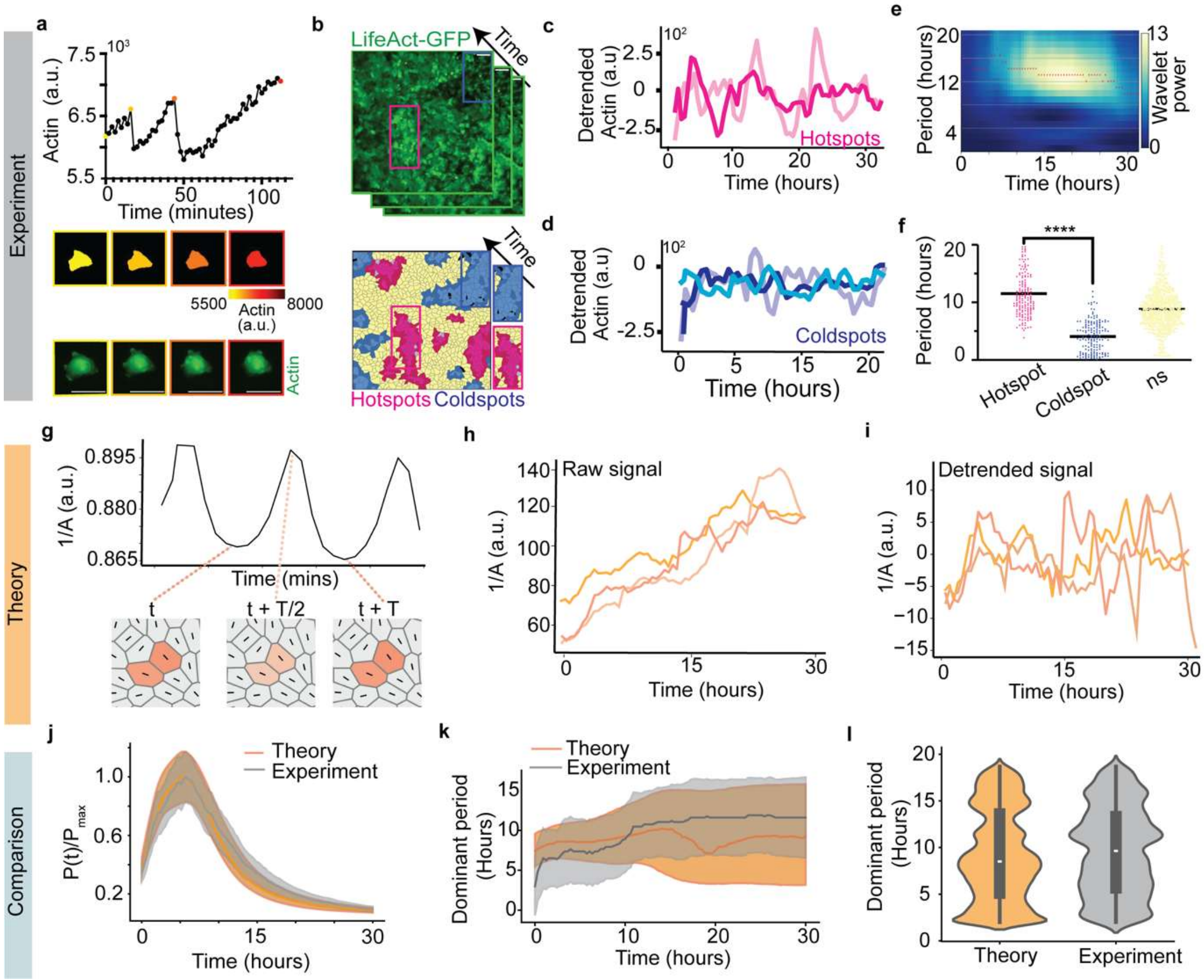

***Figure 6: Spatially collective temporal dynamics from the crosstalk of MCFL-I and MCFL-II:*** *a) Single MDCK cells show periodic actin oscillations of* $\sim 20$ *min period. Top- Plot showing Actin intensity over time in*

*single MDCK cells without neighbours. Raw images of single cells (bottom), and cell images pseudocolored for Actin (middle). The color of the box around the images (middle and bottom) correspond to the timepoint of the image corresponding to the coloured dots in the top image. b) In contrast, confluent epithelia (top) exhibit collective temporal actin oscillations that depend on spatial actin distribution (bottom), c) Actin hotspots show oscillations with ~10-hour period, d) and Actin coldspots show faster oscillations with ~4-hour period. e) Wavelet analysis of the detrended experimental signals shows decreasing wavelet power and increasing period over time, f) Comparison of dominant period at the hotspots and coldspots, g) Our model predicts single cell area oscillation of* $\sim 18$ *min period, originating from the MCFL-II. h) Because the actin level in the cell is proportional to* $1/A(1/A)$*, the latter is plotted here as a function of time, i) Detrended* $1/A$*is plotted as a function of time, j) Distributions of oscillation power over time, and k) Distributions of oscillation period over time show quantitative match between theory and experiments, l) Comparison of the dominant oscillation period from theory and experiment. Scale bars=50µm. In figures j-l, lines represent the median and the shaded regions represent* $25^{th}$ *to* $75^{th}$ *quartile.*

stabilized biochemical signalling in the local dynamic heterogeneity clusters. Furthermore, wavelet power initially increased, and then decreased with cell density, suggesting oscillations strengthen when cells crowd, but dampens with overcrowding (Fig 6e and Supplementary Fig. 7j). Finally, ERK inhibition using PD98059 led to loss in actin oscillations, as observed in temporal actin trend (Supplementary Fig. 7d, e), as well as in wavelet and Fourier power analysis (Supplementary Fig. 7f-k) suggesting a mechanochemical feedback loop, coupling ERK and actin with cell density, determines the characteristics of observed biochemical oscillations.

To probe this hypothesis, we explored the role of MCFL-II in regulating tissue dynamics. In the model, the negative feedback loop in MCFL-II couples the nonlinear chemistry of ERK with the mechanics of the actin cytoskeleton (Fig. 4a). It has been shown earlier that MCFL-II, by itself, generates collective oscillations and travelling waves in the absence of cell division[38–40]. In contrast, here, it is coupled to the glassy dynamics induced by cell division and MCFL-I, which should change the nature of the oscillations. Owing to the inverse relationship between actin and cell area in the experiments (SI-theory), we plotted $1/A$ over time to predict cellular actin dynamics. Our model predicts high-frequency cellular actin oscillations, with a period of approximately 18 minutes, for loosely packed monolayers (Fig. 6g), similar to the single-cell oscillations observed in our experiments (Fig. 6a). However, it exhibits a non-oscillating state for densely packed monolayers, punctuated by collective oscillations with a period of several hours at intermediate densities, owing to mechanochemical feedback originating from cell-cell contacts (Fig. 6h-i). Using our model, we can explain the mechanism underlying the collective oscillation as follows. Cell division generates daughter cells with smaller areas that are stabilised by MCFL-I in the crowded regions. As cell area decreases over time, ERK and actin undergo compression-induced oscillation death[40], causing MCFL-II oscillations to dampen. This manifests as a progressive loss of oscillation amplitude over time (Fig. 6j), increased median oscillation periods (Fig. 6k), and broadly distributed time periods (Fig. 6l), consistent across experiments and the model. Together, we show that in epithelia, biochemical and mechanical signalling modulate each other via mechanochemical feedback loops, resulting in glassy dynamics and unique spatiotemporal oscillations.

**Discussion:**

Epithelial homeostasis, essential for organ function, relies on regulating cell density and cell shapes. Experiments and theory reveal spatial variability in cell shapes, driven by cell activity, communication, and deformability[52,]. Experimental analysis of crowded epithelial cells also reveal glassy dynamics with typical features including caging, local cooperativity and subdiffusive movement[9,11–15,15,55]. In contrast, existing theoretical studies suggest that cellular activity should fluidise the system and prevent glass transition[7,25–27,53,62]. Due to this mismatch between theoretical predictions and experimental observations, the emergence of glassy dynamics in epithelial cells remains a matter of debate. We identified a missing link in the traditional, crowding-based explanation for glassy dynamics in epithelial tissues, and suggest that cell-to-cell communication may dictate emergent cellular dynamics. Here, we experimentally demonstrate that dynamic heterogeneity is associated with biochemical clustering and suggest that the collective physical properties of the tissue are dictated by mechanochemical feedback, which enables the effects of crowding to be propagated to cellular activity at long length scales and stabilise the locally fluidised, and locally arrested clusters. Thus, we elucidate that activity cooperates with crowding via mechanochemical feedback to slow down the system, contrasting the previous notion of competition between crowding and activity[8,23,24].

Existing models of tissue solidification focused on the emergence of rigidity only through the lens of mechanics. Yet, as our experiments show, the interplay of cellular signalling and mechanics is essential to explain glass transition. To remedy this, we developed an active vertex model that couples cell mechanics to cell signalling via mechanochemical feedback loops. Specifically, in our model, cell density controls the relative abundance of cortical actin, which increases with increasing cell density, at the expense of basal stress fibres. Because stress fibres are responsible for cell motility, mean motility decreases with increasing density, which ultimately leads to the observed glassy dynamics. A key implication of this observation is that both crowding and feedback are necessary for the glass transition. Indeed, disabling any one of them leads to tissue fluidisation in both experiments and the model.

We also theoretically established that the epithelial glass transition is density-dependent, a key difference from tissue jamming, which is density-independent and is driven entirely by shape[44,46]. Because of the mechanochemical feedback, in our model, cell shapes cannot be changed independently of cell density. As a result, although the tissue solidifies when the shape parameter crosses a threshold, the density $n_J$ at which it occurs varies with the feedback strength. Furthermore, unlike the jamming transition, the tissue solidification occurs discontinuously in the absence of motility when the feedback strength is sufficiently high. This is the fundamental difference between the tissue solidification observed here and the jamming transition, where the rigidity changes continuously across the critical shape parameter. Because a discontinuous rigidity transition is an essential requirement for the glass transition, our observation provides strong support for the existence of a glass transition in epithelial tissues. The active glass transition that we report here is driven by cell division and mechanochemical feedback and contrasts with the motility-driven active glass transition in tissues[45]. Because of the feedback, motility cannot be controlled independently of the cell shape, and a tissue is constrained to follow a particular trajectory in the shape-motility space, which changes with the feedback strength. Hence, we expect to observe distinct tissue dynamics as the feedback strength is varied in the model. Investigation in this direction may explain how cancerous tissues behave differently across tissues[63], and may provide a fundamental understanding of the emergent properties of active viscoelastic materials.

We show that mechanochemical feedback loops result not only in glassy spatiotemporal relaxation but also in unique collective biophysical oscillations as observed in cellular actin levels, area, and traction forces. We observed two interesting trends: first, time period of emergent actin oscillations is several hours, as opposed to typical minutes time scale oscillations observed previously in mammalian cell, and second, time period of oscillations are distinct for locally jammed and unjammed clusters, with the less dynamic jammed clusters oscillating over 10 hours period, and the more dynamic unjammed clusters oscillating over 4 hours period, similar to the period of cell area oscillations in MDCK monolayers reported earlier. Thus, we elucidate the mechanism of Actin oscillations, which have implications in morphogenesis and other functions[65–73]. Consistent with our findings linking actin oscillations to cell shape dynamics, coordinated actomyosin and cell shape oscillations have also been reported during several morphogenetic processes[65,66,71]. Clustering of cells and temporal oscillations are both perturbed if mechanochemical feedback is inhibited in the jammed state.

Together, we establish that an interplay between active mechanochemical feedback and cell crowding enables the emergence and stabilization of glassy dynamics in active epithelial tissues. The glass transition in biological cells has implications ranging from the development of organs to diseases[15–22,60], yet we are only now beginning to understand the biophysical crosstalk that leads to its emergence. We believe our study represnts a first step towards enhancing understanding of the spatiotemporal dynamics of epithelial cells, specifically revealing the emergence and maintenance of dynamic heterogeneity clusters through active cell-to-cell communication.

## Methods

### Cell culture:

Madin-Darby canine kidney cells MDCK cells, LifeAct-GFP MDCK cells were cultured in Dulbecco's Modified Eagle Medium (DMEM, Gibco) supplemented with 10 μg ml$^{-1}$ streptomycin (Pen Strep, Invitrogen), and 5% fetal bovine serum (FBS, Invitrogen) in a humidifier incubator maintained at 37°C and 5% $CO_2$.

**Widefield Microscopy:** Fluorescence imaging was carried out on a Zeiss Axio Observer 7 inverted microscope equipped with a scientific sCMOS camera (Iberoptics). Images were captured using both a 20× objective (air) or 40× objective (air) for live imaging and a 63× oil-immersion objective for immunofluorescence images that were used for LISA cluster analysis. For live experiments, the samples were placed in a stage-top humidified incubator and maintained at 37 °C with 5% $CO_2$ during the imaging.

### Traction force microscopy:

First, glass-bottom dishes (Cellvis) were activated by treating first with 0.1 % NaOH, followed by 4% APTMS treatment ((3-Aminopropyl)trimethoxysilane, Sigma #281778) for 15 min and 2% glutaraldehyde (Sigma #354400) treatments. Polyacrylamide gels of different stiffness were prepared by mixing different ratios of Acrylamide and Bisacrylamide with TEMED and APS to initiate polymerisation, mixed with fluorescent beads as reported previously (r). A sandwich of the gel mix between the glass bottom dish and a Sigmacote (Sigma) treated coverslip for hydrophobicity was allowed to polymerise, and the coverslip was then removed. Then, the gel surface was functionalised by coating with Sulfo-SANPAH (Invitrogen) and UV treatment, followed by coating with Fibronectin (Sigma) to allow cell adhesion to the gel. Cells were then seeded onto the gels and grown to confluence. Then images of the cells and beads were acquired, followed by drug treatment on the microscope stage followed by image acquisition of the beads and cells post addition of the drug. Then, the monolayer was detached from the gel by the addition of trypsin to obtain reference images. Then, images were aligned to correct stage drift using reference positions that were taken in cell free regions pre and post trypsin addition, followed by obtaining bead displacements. From the bead displacements, the traction field was computed the FTTC ImageJ plugin[74]. The traction field was then plotted using a custom written code on MATLAB.

For traction force microscopy experiments with Blebbistatin, pre-blebbistatin stressed bead image and Actin image was taken first. Then 20μM Blebbistatin was added on the microscope stage and 30 minutes later, post-blebbistain stressed bead and Actin images were captured. This was followed by trypsinisation and acquisition of the reference bead images.

**Immunostaining:**
Cell fixation was done with 4% formaldehyde diluted in 1× phosphate-buffered saline (PBS; pH 7.4) at room temperature (RT) for 10 min, followed by 1× PBS washes (three times). Cell permeabilization was carried out with 0.25% (v/v) Triton X-100 (Sigma) in PBS for 10 min at RT, followed by washing three times with PBS to remove the reagent. To block non-specific antibody binding, samples were incubated in 2% BSA and 5% FBS in PBS at RT for 45 min. The blocking buffer was removed after 45 min, and the primary antibody dilution (1:300) prepared in the blocking buffer was added to the samples. The samples were incubated with the primary antibody, at 4°C overnight. Then, samples were washed three times with 1× PBS. Each wash was done for 5 minutes on a gel rocker. Next, secondary antibodies tagged with a fluorophore were (1:300) prepared in 25% blocking buffer diluted in PBS and added to the sample for 60 min at RT. To counterstain cell nuclei, the samples were added with a DNA-binding dye, 4′,6-diamidino-2-phenylindole (DAPI; 1 μg/mL in PBS, Invitrogen), along with the secondary antibody solution. Then, thorough washing of the samples was done with PBS before imaging.
The following primary antibodies were used: Zo1- (1:200) (CST #5406), E-Cadherin (24E10) - (1:500) (CST #3195), P-myosin (Ser19) (1:50) (CST #3671), Beta catenin (1:500) (CST #9562). The following secondary antibodies were used: Alexa Fluor 488-conjugated anti-rabbit IgG (1:500) (CST #4412), Alexa Fluor 488-conjugated anti-mouse IgG (1:500) (Invitrogen #A32723), Alexa Fluor 594-conjugated anti-rabbit IgG (1:500) (CST #8889). F-Actin was statined with Alexa Fluor 594-conjugated phalloidin (1:1000) (CST #12877S) and 4′,6-diamidino-2-phenylindole (DAPI) (1:1000 and 1:2000) (CST #4083S) was used to stain the nucleus.

**Drug treatments:**
**Blebbistatin:**
For immunocytochemistry experiments, Blebbistatin (Sigma) was added at a final concentration of 50μM and 100μM after the monolayer reached high density, and dishes were fixed 6 hours after the addition of the drug. For live imaging experiments, the final concentration was 10μM for the duration of imaging.

**Thymidine double block:**
Cells were treated with 2mM Thymidine (Sigma) and incubated for 16 hours. After this, they were washed and cultured in regular cell culture media for 9 hours. Then, 2mM Thymidine was added again and incubated for 14 hours. Cells were then washed and used immediately for experiments.

**PD98059:**
Cells were treated with 10 μM PD98059 (CST#9900) for the duration of the live imaging. Stock solution of 20mM was prepared in DMSO, as per the instructions of the supplier. Finally, the stock solution was diluted in cell culture medium to obtain the final concentration.

**Y27632**

Cells were treated with 10 μM Y27632 (Sigma #Y0503) for the duration of the live imaging. Stock solution of 10mM was prepared in DMSO, as per the instructions of the supplier. Finally, the stock solution was diluted in cell culture medium to obtain the final concentration.

**Cell stretch:**

To perform cell stretching, a custom-made device was used, which applied mechanical strain to a PDMS substrate. PDMS was prepared by mixing the elastomer and crosslinker (SYLGARD 184, corning) at a ratio of 10:1 by weight, followed by thorough mixing and degassing the solution, which was then cast in a plexiglass mould and cured at 80 °C for 6 hours. Then, the chamber was removed from the cast and thoroughly washed with 70% Ethanol. Plasma treatment in an oxygen environment was then performed for 10 min to make the PDMS surface hydrophilic, followed by a fibronectin coating to allow cell adhesion. LifeAct-MDCK Cells were seeded on the PDMS chamber at different densities to obtain confluent and sub-confluent monolayers. Cells were then imaged in the relaxed configuration of the PDMS using 10X objective, Zeiss Axio Observer. The PDMS chamber with cells was then kept in the stretched configuration for 5 minutes using the custom cell stretcher device and immediately kept back on the microscope to obtain post-stretch images of the cells.

## Image analysis

**Cell segmentation:**

Images were segmented with cellpose[75] to get the cell masks with Actin as the primary and DAPI as the second channel in immuostained images. For segmentation in timelapse microscopy images, phase-contrast images were used and were segmented with a custom trained model on cellpose. Area, Perimeter, ellipse fit, and mean fluorescence intensity within masks were computed using LabelsToRois plugin[76] in FIJI for fixed images and using Trackmate plugin for live microscopy data. To remove wrongly segmented spots misidentified as cells by the software, cells were filtered out based on area cutoff in subsequent analysis.

**Cell tracking:**

Cells were tracked using Trackmate Plugin[78-79] from the segmented images of Cellpose using the Intersection of Union (IoU) method. Prior to tracking,wrongly segmented labels were removed using the area filter on trackmate.

**Moran's Index[77]:**

The Global Moran's Index was calculated using the following definition:

$$I = \frac{n}{\sum_{i=1}^{n}\sum_{j=1}^{n} w_{ij}} \frac{\sum_{=1}^{n}\sum_{j=1}^{n} w_{ij}(x_i - \bar{x})(x_j - \bar{x})}{\sum_{i=1}^{n}(x_i - \bar{x})^2}$$

where $w_{ij}$ represents the weights assigned to the neighbours of each cell using a custom written R code. We have assumed equal weights to all the immediate neighbours defined as the cells which are in direct contact with the cell of interest. Moran's plot was plotted with the mean intensity of cell of interest on x axis, and the average intensities of its neighbours on y axis. Correlograms were plotted by computing the global Moran's Index at different lag neighbours. The Local Moran's Index was calculated to determine the spatial autocorrelation of the proteins of interest for every cell and its neighbours which are in direct contact with it using the following definition:

$$I_i = \frac{(x_i - \bar{x})}{\sum_{k=1}^{n} \frac{(x_k - \bar{x})^2}{(n-1)}} \sum_{j=1}^{n} w_{ij}(x_j - \bar{x})$$

where $w_{ij}$ represents the weights assigned to the neighbours of each cell. We assigned equal weights to all the first nearest neighours of a cell. The "localmoran" function in the 'spdep' package in R was used for this purpose. For each cell and its neighbours, based on deviation from the global mean of the protein amount and the value of its local Moran's index, the cells were classified into four categories "High-High", "High-Low", "Low-High" and "Low-Low", while insignificant deviations were marked as such and were plotted in the form of a LISA[31] cluster map. To assess statistical significance, we used 999 Monte carlo simulations and used a p-value cutoff of 0.05. To capture the spatial clusters to its complete extent of the LISA Actin clusters, cells that are immediate neighbours to the High-High and Low-Low clusters have been represented with darker shades , since LISA analysis only captures the core of spatial clusters[80].

**Dynamic heterogeneity domains:**
To identify the fast and slow regions of dynamic heterogeneity, we used LISA clustering. 'Low-low' clusters were determined to be the slow regions of dynamic heterogeneity, and 'High-High' clusters were determined to be the fast regions of dynamic heterogeneity.
**Mean square displacement:**

$$MSD(\Delta t) = \langle | r_i(t + \Delta t) - r_i(t) |^2 \rangle$$

**Overlap function:**

$$Q(\Delta t) = \langle \frac{1}{N} \sum_{i=1}^{N} W(a - |r_i(t + \Delta t) - r_i(t)|) \rangle$$

As there are cell divisions in both experiment and simulation, we only consider those cells that are present at both $t$ and $t + \Delta t$ when we calculate the MSD and Q(t). Here, angular brackets denote average over $t$ and ensembles (in simulation). For Fig. 2f, $a = 3.25\ \mu m$ . For Fig. 3f, $a = 10\ \mu m$ . For simulation, $a = 0.1$ .

Both in MSD and overlap function, the COM drift has been subtracted during calculation.

**Actin signals:**
Actin timeseries signals were obtained from the fluorescence intensity of LifeAct-GFP images by coarse-graining the image with grid sizes comparable to cell size, i.e 30μm, using custom-written python code, from confluent monolayers. For Actin signals from single LifeAct-MDCK cells without any neighbours, per cell intensity was obtained by cell segmentation followed by measuring mean intensity within the cell outline determined by Trackmate.

**Particle Image Velocimetry:** To compute PIV, iterative PIV using FTTC registration method was computed using the MATLAB toolbox PIVlab[81]. Grid sizes were chosen based on cell size.

**Relative cell pressure analysis**: Relative cell pressures were computed using Bayesian Force inference[82] adapted using custom code for 2D monolayer data, based on the principle of balancing forces at cell vertices.

**Stress fiber analysis**: Length of stress fibers were manually measured in FIJI using the line and measure tools.

**Timeseries analysis**

**Pyboat:**

PyBoat[83] is python-based fully automatic stand-alone software that integrates multiple steps of non-stationary oscillatory time series analysis which is being used for the quantification of biochemical and physical heterogeneity spatiotemporally. Pyboat implements continuous wavelet analysis using a sliding Morlet wavelet of different frequencies and determines the power for different frequencies at each time. Then, the main oscillatory component is determined from the heatmap of frequency power at different times and frequency. It also provides optimized detrending, amplitude removal, spectral analysis, ridge detection, oscillatory parameters readout and visualization plots along with an integrated batch-processing option. Using the batch processing option, we also get the most dominant time period of all the spatial positions in the time-series. Using pyboat, we sinc-detrended the Actin time-series signal to obtain the detrended signal ,with detrending period taken to be the imaging duration to remove the trend of increasing Actin with increasing time due to increasing density.

**Statistical analysis:**

Statistical analyses were performed by Unpaired t-test with Welch's correction using GraphPad Prism 10, unless otherwise mentioned. All experiments were repeated at least three times. Lines represent the median in scatter bar plots unless otherwise mentioned. In violin plot, top line represents 75$^{th}$ percentile, middle line represents the median and the bottom line represents the 25$^{th}$ percentile.

**Data and code availability:**

Relevant data and codes are available from the authors upon reasonable request. LISA analysis codes are available online at https://github.com/sindhum98/LISA-analysis-for-epithelial-tissue

**Author contributions**

M.V conceived the project. S.M. and M.V. designed experiments. S.M. performed and analyzed all experiments except the traction force microscopy with LifeAct MDCK cells, which was performed by M.B.S. Theoretical model was contributed by S.S. and P.D. T.T. performed the LISA analysis, and T.C. contributed to oscillations analysis using pyBOAT. S.K. contributed to glassy analysis. Analysis and interpretation of data was done by S.M., P.D., Sd.M., S.S., and M.V. M.V. and S.S. developed and wrote the manuscript with help from S.M. and P.D. All authors read, discussed and commented on the manuscript.

**Acknowledgements**

We thank Sriram R. Ramaswamy, Saroj K. Nandi, Tamal Das and Srikanth Sastry for critical discussions and suggestions. M.V. is a partner group leader of the Max Planck Society (MPG), Germany, which has supported part of this work. This work is also supported by the Infosys foundation, Anusandhan National Research Foundation- previously called the Science and Engineering Research Board (project number: SERB SRG/2022/000534), and Indo German Science and Technology Centre (IGSTC WISER scheme). SS acknowledges funding from IISc, Axis Bank Center for Mathematics and Computing, and a startup grant from SERB-DST (SRG/2022/000163). We also acknowledge intramural funds at IISc Bangalore for providing support towards equipment and facilities and for salaries/fellowships of the authors. S.M. and Sd.M and acknowledge funding from Prime Minister's Research Fellowship (PMRF).

**Glassy dynamics in active epithelia emerge from an interplay of mechanochemical feedback and crowding**

Sindhu Muthukrishnan[1], Phanindra Dewan[2], Tanishq Tejaswi[1], Michelle B Sebastian[1], Tanya Chhabra[1], Soumyadeep Mondal[2], Soumitra Kolya[3], Sumantra Sarkar[2]*, Medhavi Vishwakarma[1]*

1. Department of Bioengineering, Indian Institute of Science, Bangalore, India
2. Department of Physics, Indian Institute of Science, Bangalore, India
3. Tata Institute of Fundamental Research, Hyderabad, India

Correspondence: medhavi@iisc.ac.in and sumantra@iisc.ac.in

**Table of contents:**

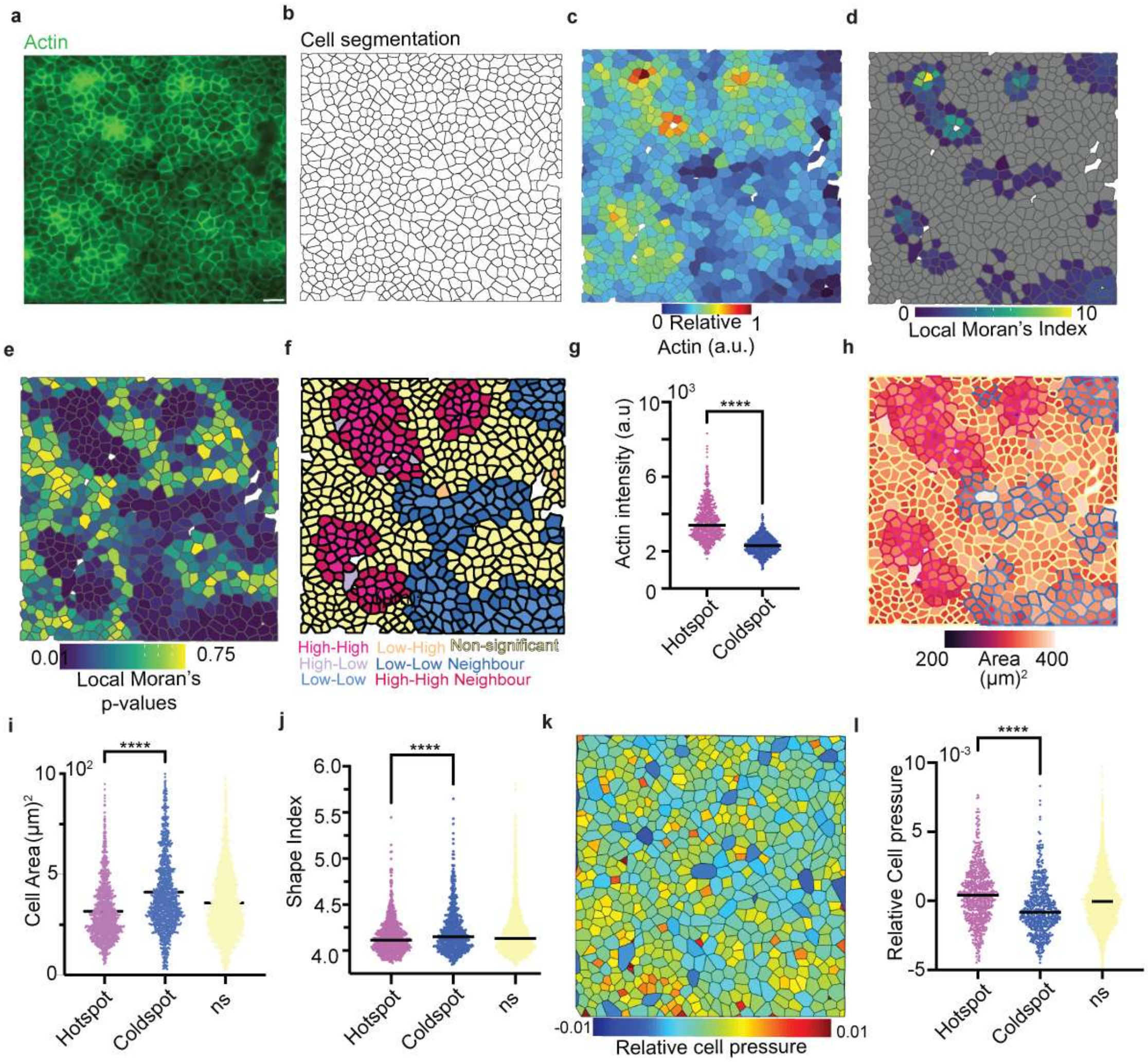

***Supplementary Fig 1- Spatial characterization of Actin distribution and Mechanochemical feedback:*** *a) Raw microscopy image of cells stained for Actin with Phalloidin, b) Cell boundaries obtained from cell segmentaion, c) Cells pseudocolored for normalized F-Actin levels, d) Heatmap of Local Moran's Index of cells, e) Heatmap of p-values of cells of Local Moran's Index obtained from , f) LISA cluster map of F-Actin, g) Actin intensity of cells classified as High-High (Hotspotspink) and Low-Low (Coldspots-blue), h) Cells color-coded for cell area and outlines colored for LISA cluster category, i) Scatter dot plot of single cell area in High-High, Low-Low and Nonsignificant clusters, j) Scatter dot plot of Shape Index in High-High, Low-Low and Non-significant clusters, k) Heatmap of relative cell pressures obtained by Bayesian Force Inference, l) Scatter dot plot of relative cell pressures in High-High, Low-Low and Non-significant LISA clusters. Scale bars= 50μm. Lines represent the median.*

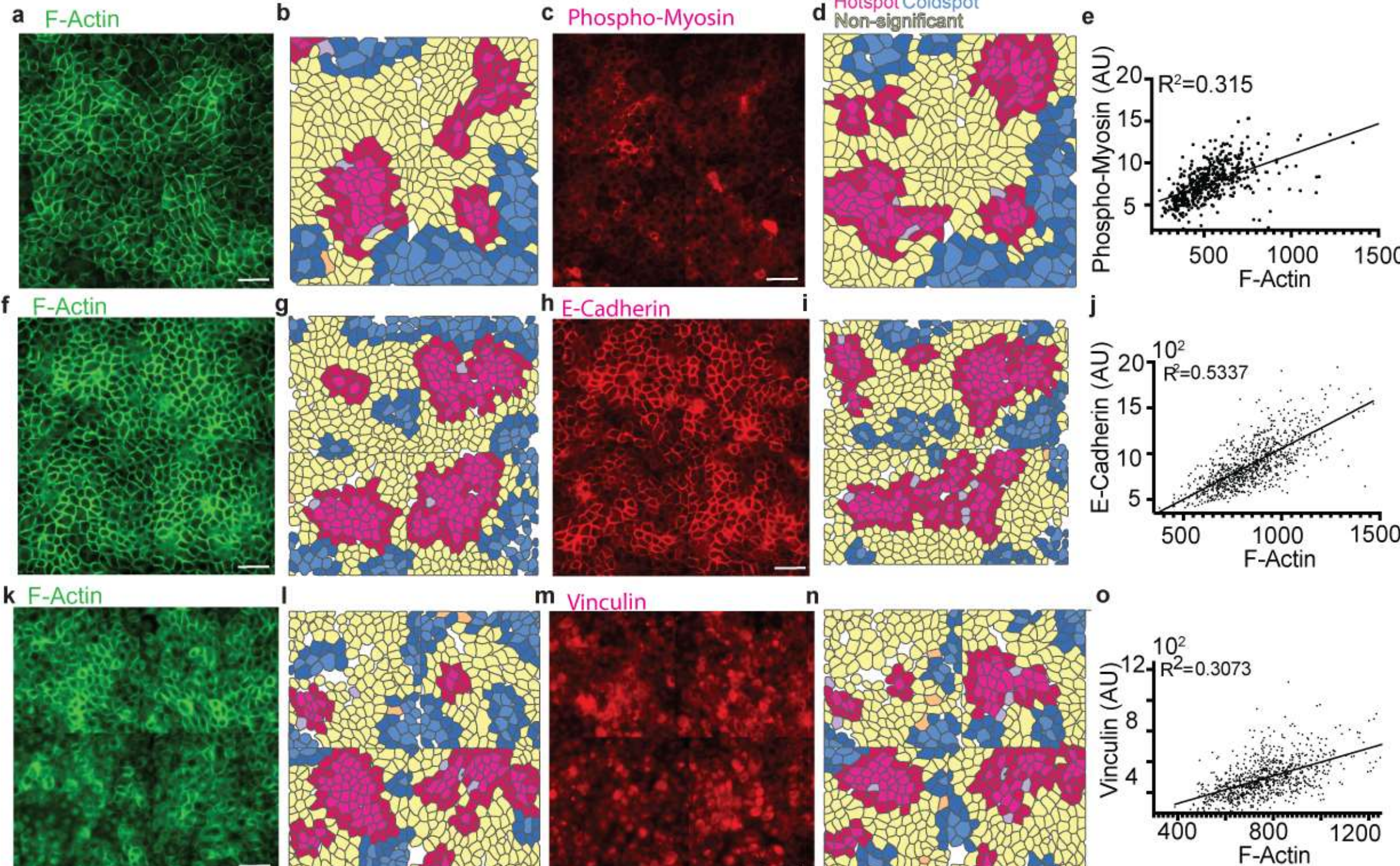

***Supplementary Figure 2: Spatial correlation in cytoskeletal proteins correspond to the spatial clustering in Actin:*** *a) Immunostaining image of F-Actin, b) LISA cluster map of F-Actin, c) Immunostaining image of P-Myosin corresponding to a), d) LISA* cluster map of P-Myosin, e) Scatterplot of F-Actin vs P-Myosin, *f) Immunostaining image of F-Actin corresponding to h, g) LISA cluster map of F-Actin corresponding to f, h) Immunostaining image of E-Cadherin corresponding to f), i) LISA* cluster map of E-Cadherin, j) Scatterplot of F-Actin vs E-Cadherin, k*) Immunostaining image of F-Actin corresponding to m, l) LISA cluster map of F-Actin corresponding to k, m) Immunostaining image of E-Cadherin corresponding to k), n) LISA* cluster map of Vinculin, o) Scatterplot of F-Actin vs Vinculin.

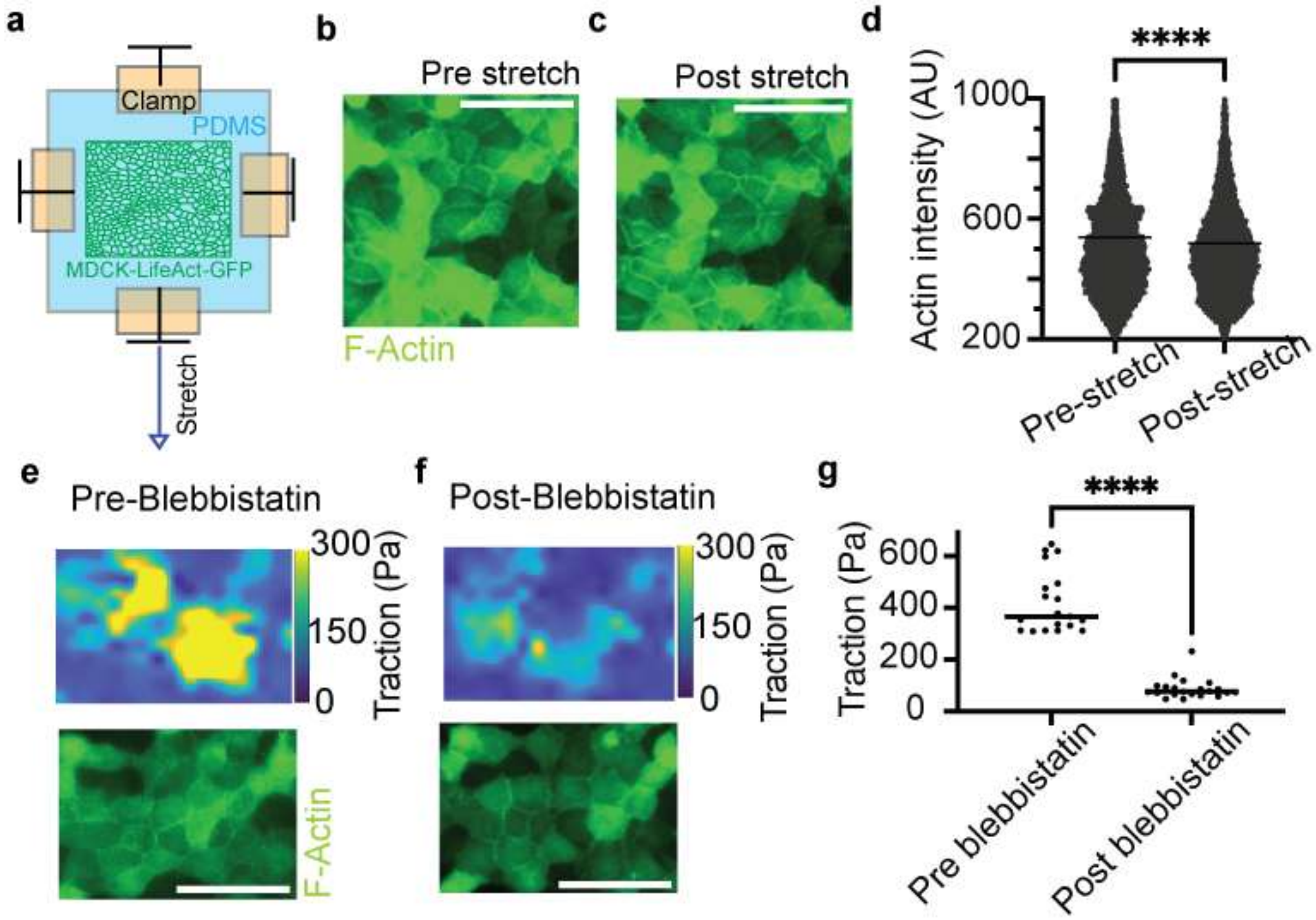

***Supplementary Figure 3: Dynamic Mechanochemical feedback****: a) Schematic sketch of the cell stretching experiment, b) Representative LifeAct MDCK cells before(b) and after (c) stretching the substrate, d) Comparison of Actin intensity in cells pre and post stretch, e) Traction force (top) and Actin from a representative region before (e) and after (f) Blebbistatin addition, g) Comparison plot of traction pre and post Blebbistatin addition.*

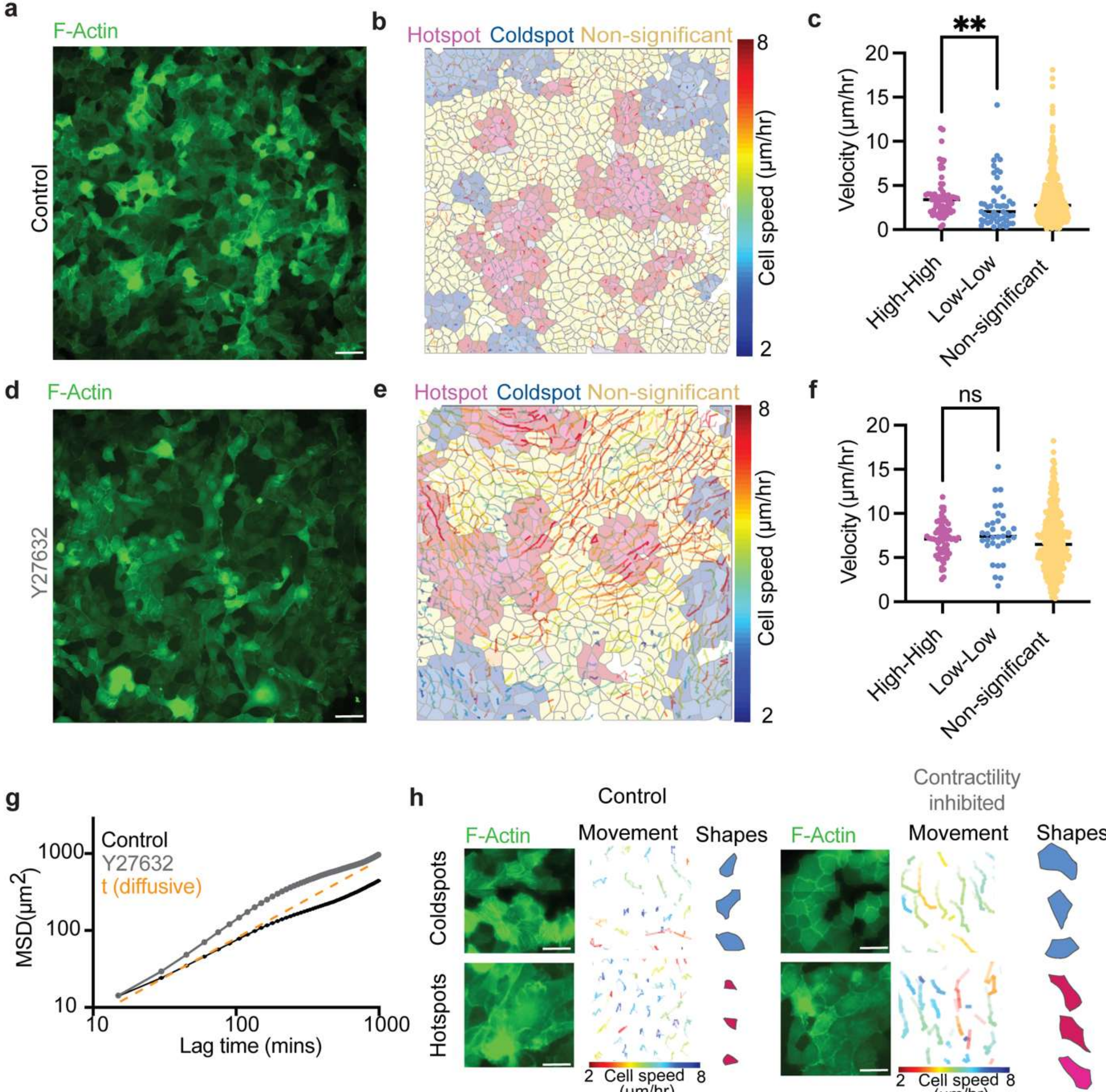

***Supplementary Figure 4: Disrupting contractility leads to loss of mechanochemical feedback and fluidises the tissue:*** *a) LifeAct MDCK cells from a representative control sample at the first timepoint, b) LISA cluster map corresponding to a and cell tracks overlaid on top, c) Comparison of cellular velocity at the different Actin LISA clusters, d) LifeAct MDCK cells from a representative Y27632 treated sample at the first timepoint, e) LISA cluster map corresponding to f and cell tracks overlaid on top, f) Comparison of cellular velocity at the different Actin LISA clusters, g) Mean Square Displacement plots for control and contractility inhibited monolayers, h)* Left/top- *High-resolution images of F-Actin at a representative coldspot from control sample, cell tracks and cell shapes;* Left/ bottom- *High-resolution images of F-Actin at a representative hotspot, cell tracks and cell shapes ;* Right/ top- *High-resolution images of F-Actin at a representative coldspot from Y27632 sample, cell tracks and cell shapes;* Right/ bottom- *High-resolution images of F-Actin at a representative coldspot from control sample, cell tracks and cell shapes.*

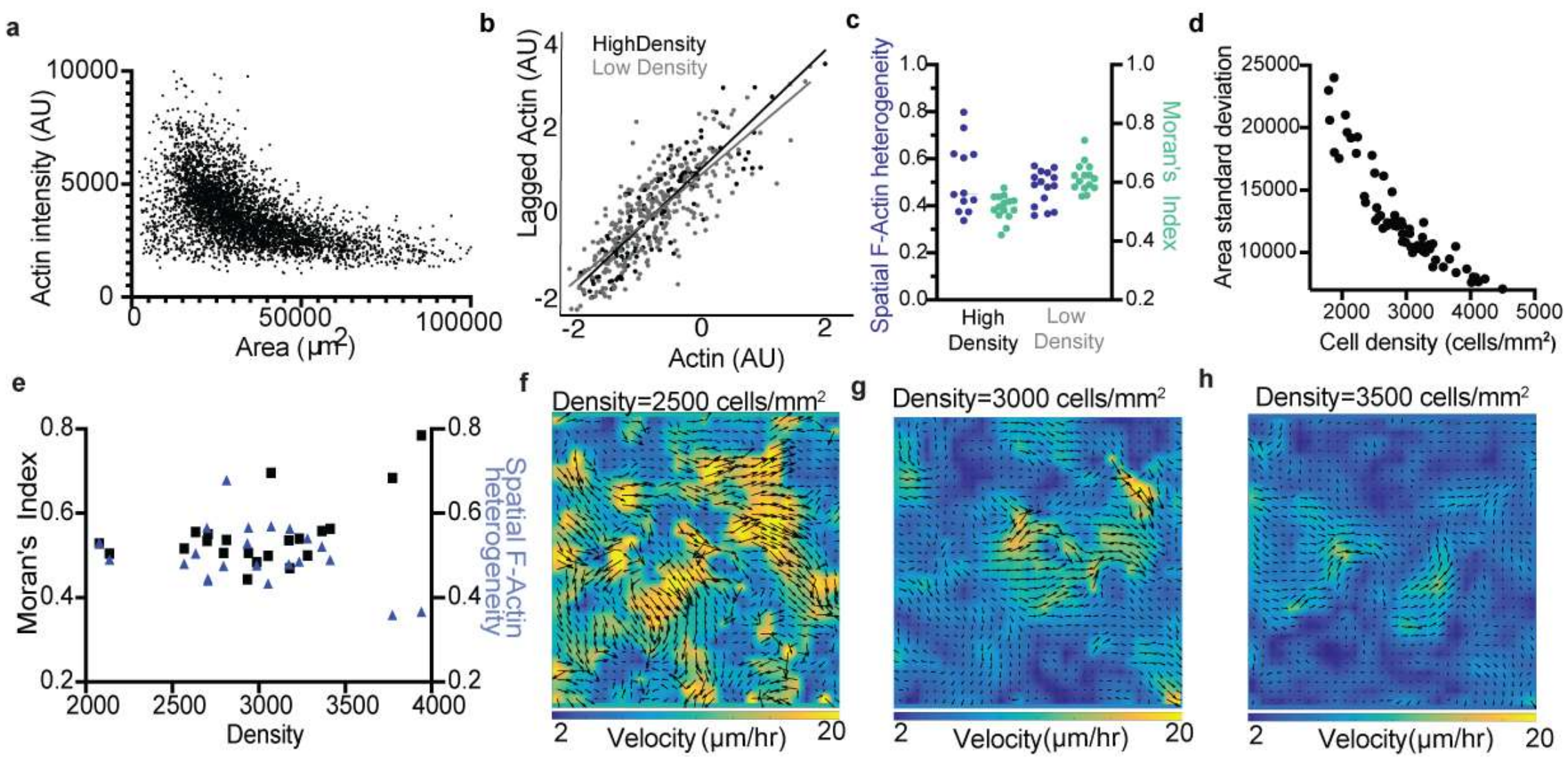

***Supplementary Fig 5: Cytoskeletal changes and Moran analysis of loosely and densely packed monolayers:*** *f) Plot of Actin vs Area at low and high density, g) Moran's plot of Actin vs lagged Actin at low and high density, h) Spatial F-Actin heterogeneity (percentage of Non-significant cells) and Global Moran's Index at high and low density, i) Standard deviation of cell area vs density, j) Spatial F-Actin heterogeneity and Moran's Index vs density, k-m) Velocity heatmap from Particle Image Velocimetry map at low (k), medium (l) and high density (m). Scale bars= 50µm. Lines represent the median.*

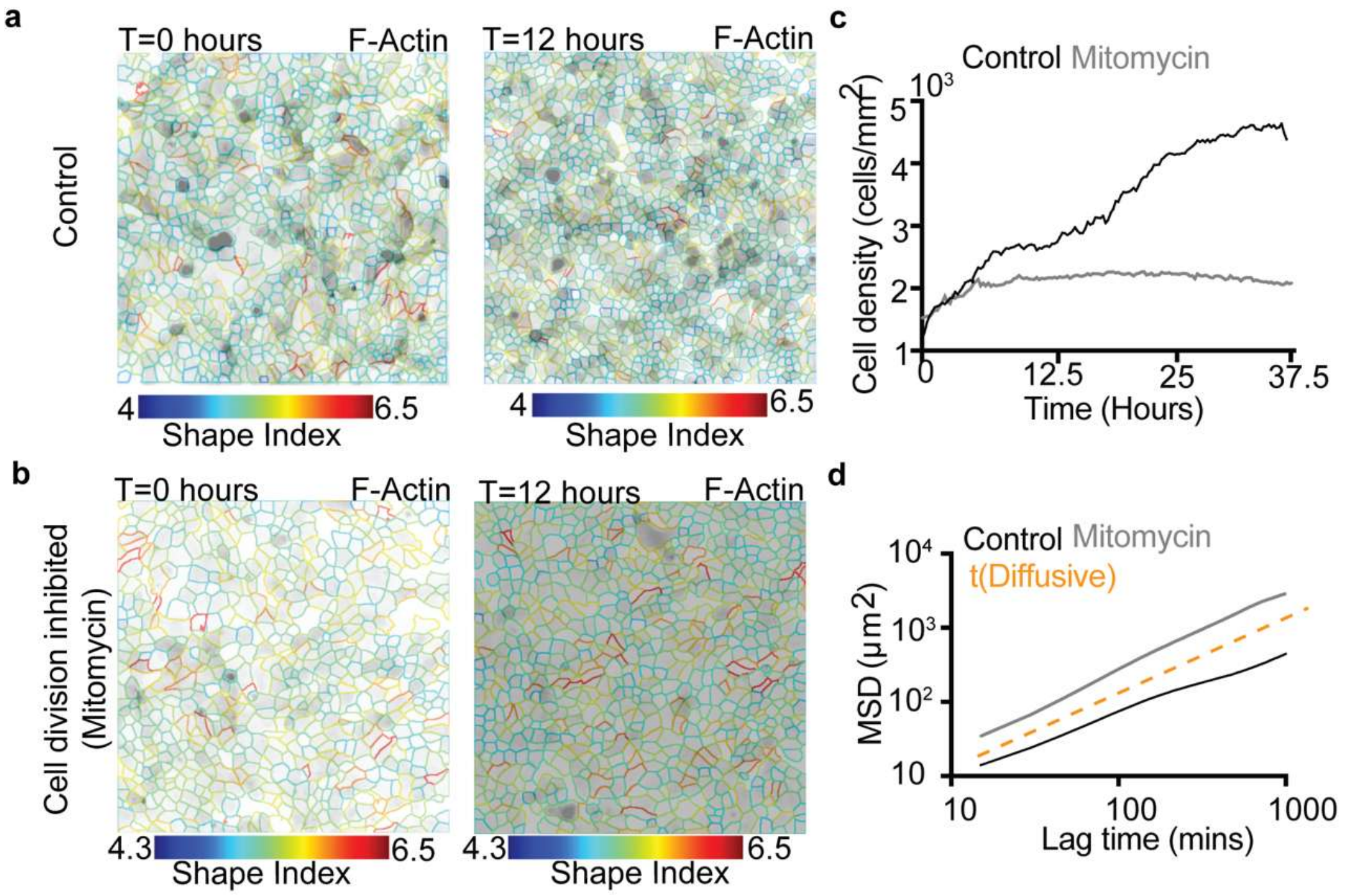

***Supplementary Figure 6: Fluidisation in the absence of crowding:*** *a) Timelapse images at T=0 hours (left) and T=12 hours (right) from a control sample, with cell outlines pseudocolored for Shape index, b) Timelapse images at T=0 hours (left) and T=12 hours (right) from a Mitomycin-C treated sample with cell outlines pseudocolored for Shape index, c) Plot of cell density over time for control and Mitomycin-C treated samples, d) Mean Square Displacement plots for control and Mitomycin-C treated samples*

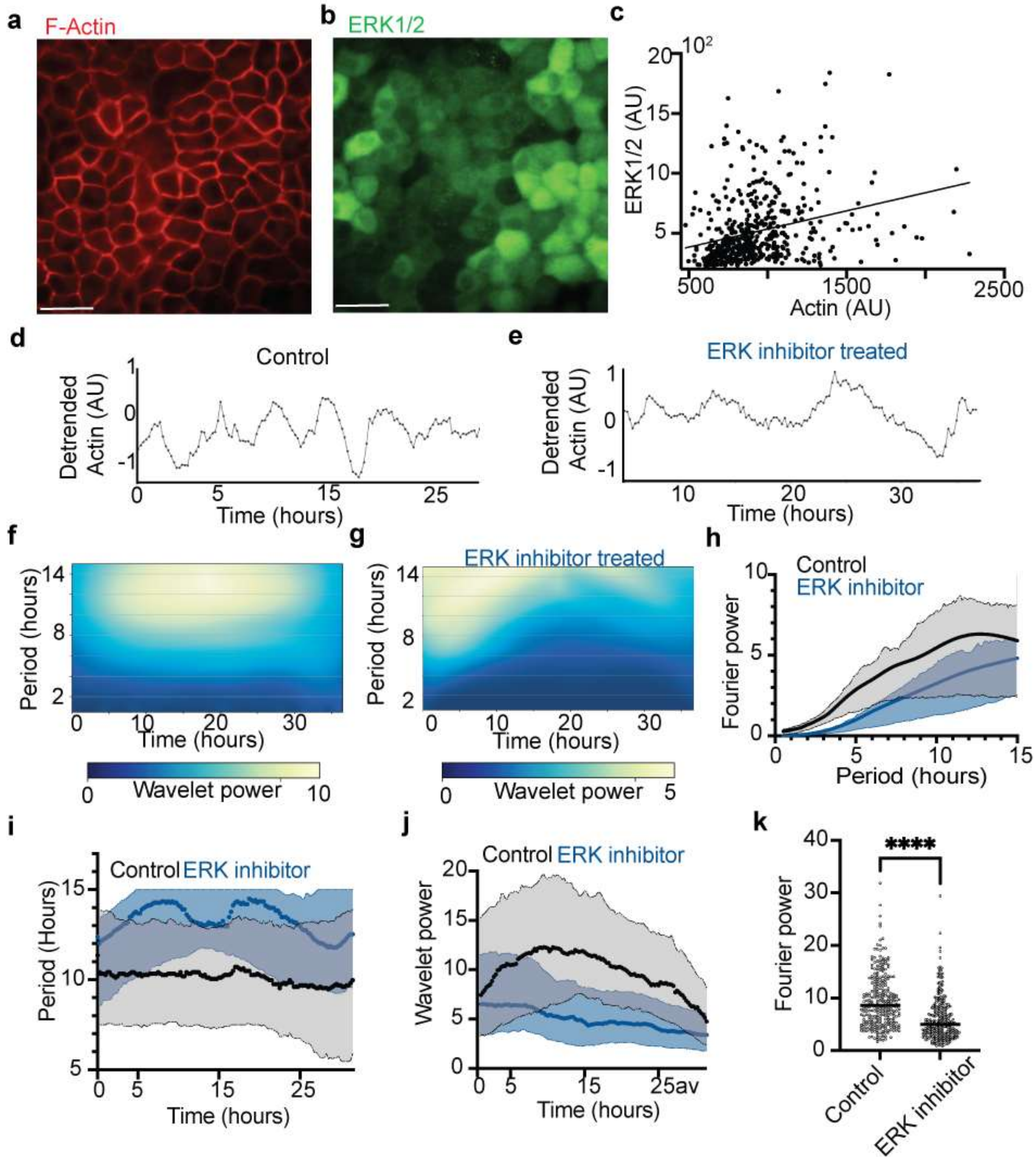

***Supplementary Fig 7: ERK inhibition leads to loss of Actin oscillations:*** *a) Representative image of monolayer stained for F-Actin (a) and ERK1/2 (b), c) Per-cell F-Actin plotted against per cell P-ERK1/2, d) Representative local Actin signal from a control monolayer, e) Representative local Actin signal from ERK inhibitor treated monolayer, f) Wavelet heatmap of control Actin signal, g) ) Wavelet heatmap of ERK-inhibitor treated Actin signal, h) Trend of power over time, i) Period determined by wavelet analysis as a function of time in the control (black) and ERK inhibitor treated (blue) Actin signals, j) Wavelet power as a function of time in the control (black) and ERK inhibitor treated (blue) Actin signals, k) Comparison of Fourier powers in the control and ERK inhibitor treated Actin signals. In h-j, lines represent the median and the shaded regions correspond to the quarters.*

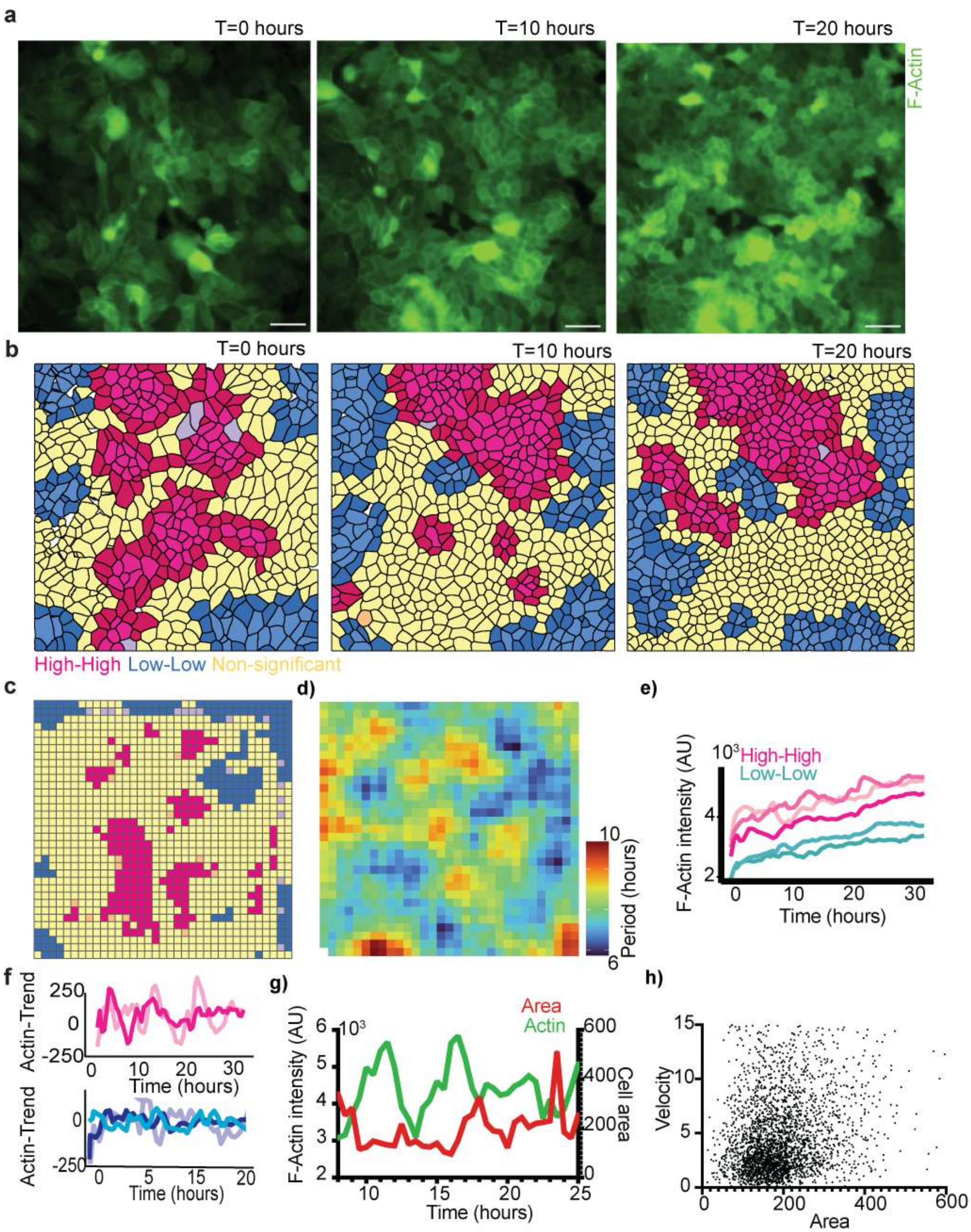

*Supplementary figure 8:* ***Hotspots and coldspots are stable over time and show local Actin oscillations with distinct periods that are spatially continuous with non-significant regions, related to area, velocity and local density fluctuations*****:** *a) Representative timelapse images from homeostatic MDCK-LifeAct-GFP monolayers at 0 hours (left), 10 hours (middle) and 20 hours (right), b) LISA cluster maps of Actin at 0 hours (left), 10 hours (middle) and 20 hours (right), c) Representative LISA cluster map from the gridding analysis (methods section) used for Actin signal analysis, d) Heatmap of timeperiod at different locations in the monolayer demonstrating continuity in timeperiods, e) Actin signals without detrending from hotspots (pink) and coldspots (blue) as a function of time, f) Detrended Actin signals from hotspots (top) and coldspots (bottom) as a function of time, g)Single-cell Area (red- right y axis)and F-Actin (Green, left y-axis) obtained from cell segmentation and tracking plotted as a function of time, h) Single cell velocity plotted against single cell area, obtained from cell segmentation and tracking.*

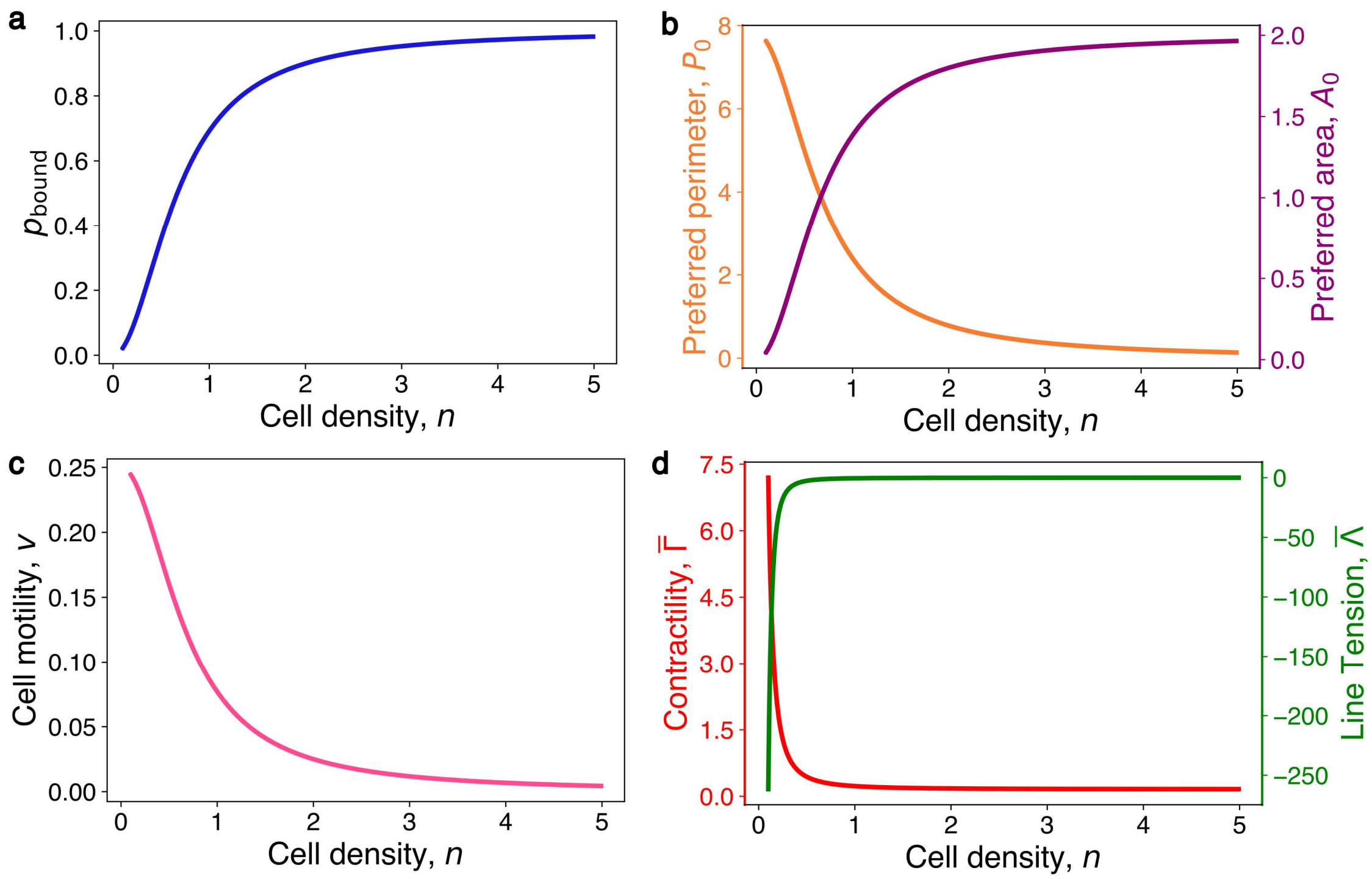

***Supplementary Fig 9: Mechanochemical feedback couples vertex model parameters to density:*** *a) Probability of bound cortical actin ($p_{bound}$) vs. density. b) The vertex model parameters $P_0$ and $A_0$ vs. density. c) Motility vs. density. d) Normalized contractility and normalized line tension vs, density in the model.*

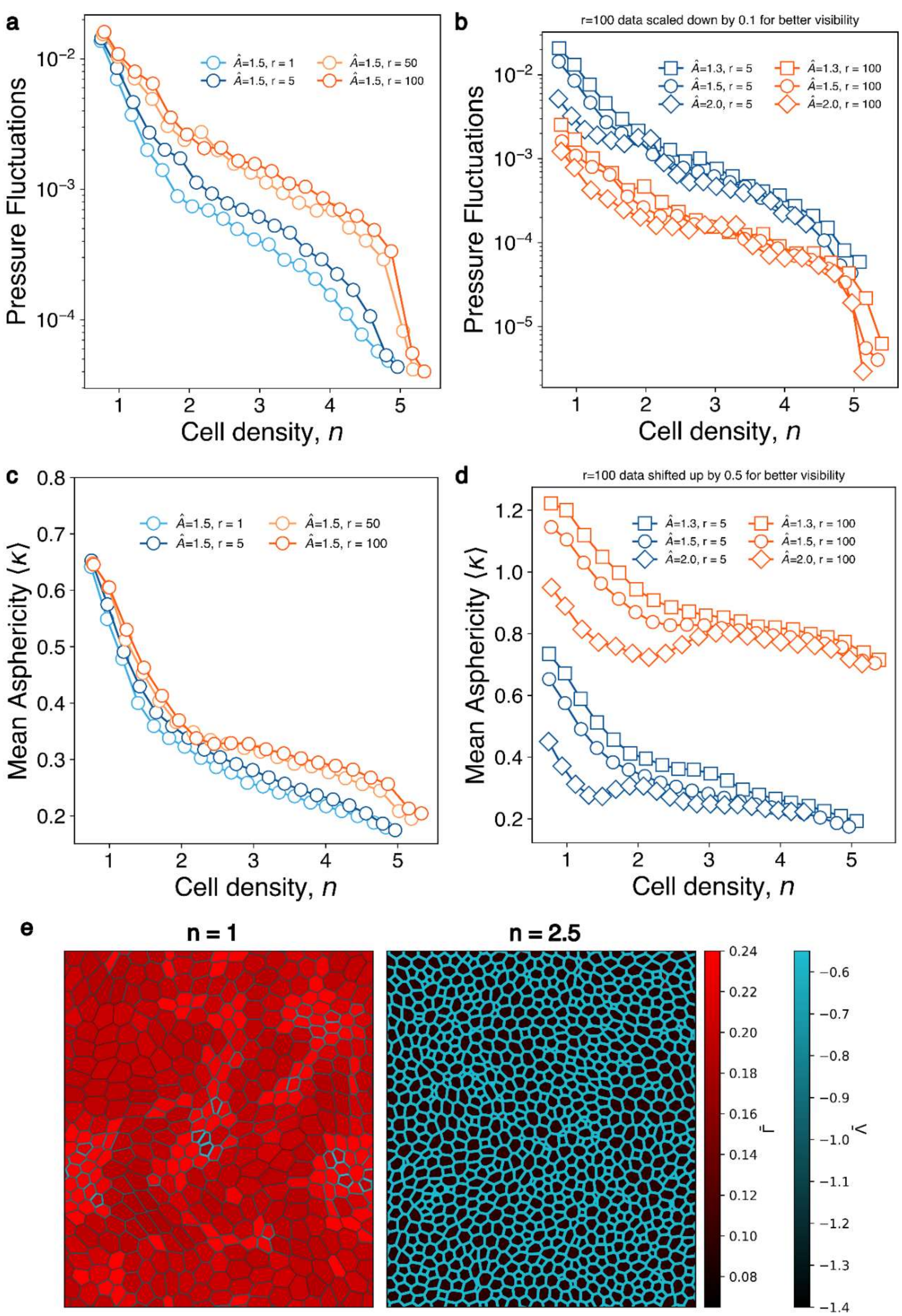

***Supplementary figure 10: Glass transition in the vertex model:*** *(a) Pressure fluctuations vs. density for* $\hat{A} = 1.5$ *shows rate dependence. (b) Pressure fluctuations vs. density for different* $\hat{A}$ *but one rate value. (c) Mean asphericity vs. density for* $\hat{A} = 1.5$ *shows clear rate dependence. (d) Mean asphericity vs. density for different* $\hat{A}$ *but one rate value. e) Contractility and line tension plots at different densities from a vertex model simulation with cell divisions and mechanochemical feedback.*

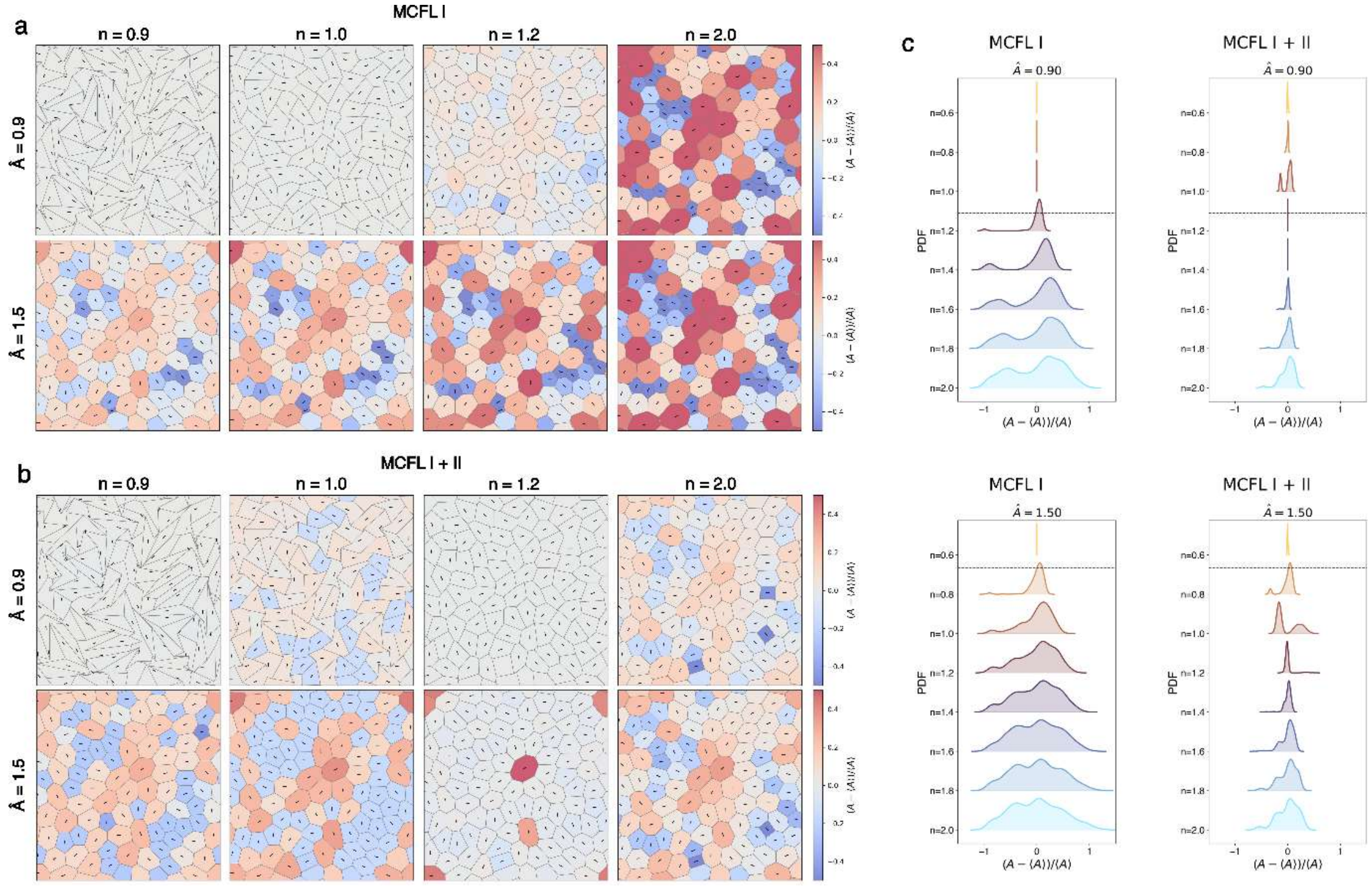

***Supplementary Fig 11: Constant density steady state configurations of the vertex model with mechanochemical feedback: a) Configurations with MCFL-I: At low  , the configurations are floppy at , which progressively becomes regular at higher . For , the configurations are regular for most densities. The cells are coloured according to area strain  b) Configurations with MCFL-I+II: The trend with increasing  is similar to (a). c) Area strain distributions for configurations with MCFL-I and MCFL-I+II.***

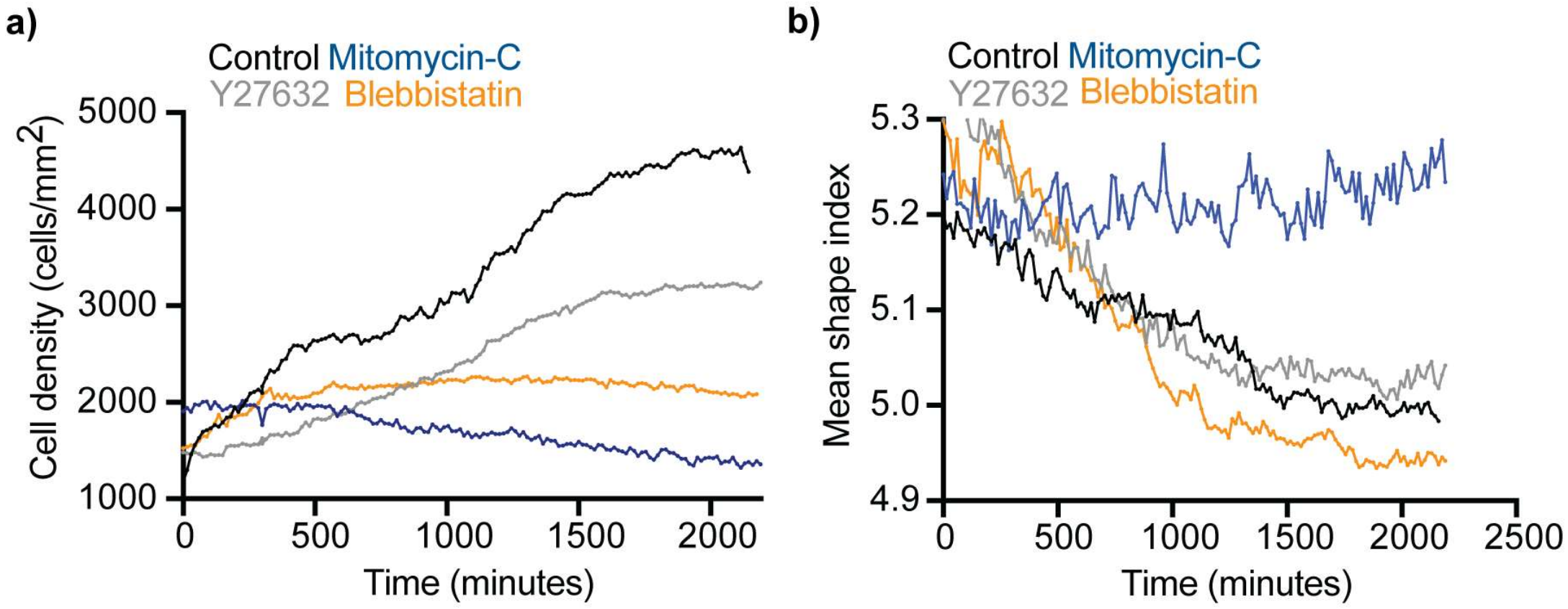

**Supplementary figure 12**: **Effect of drugs on division rate:** a) Plot of cell density over time in control, Mitomycin-C, Blebbistatin and Y27632 treated monolayers as a function of time, b) Change in division rate is also reflected in the mean shape index, as observed in the plot of Mean shape index over time in control, Mitomycin-C, Blebbistatin and Y27632 treated monolayers

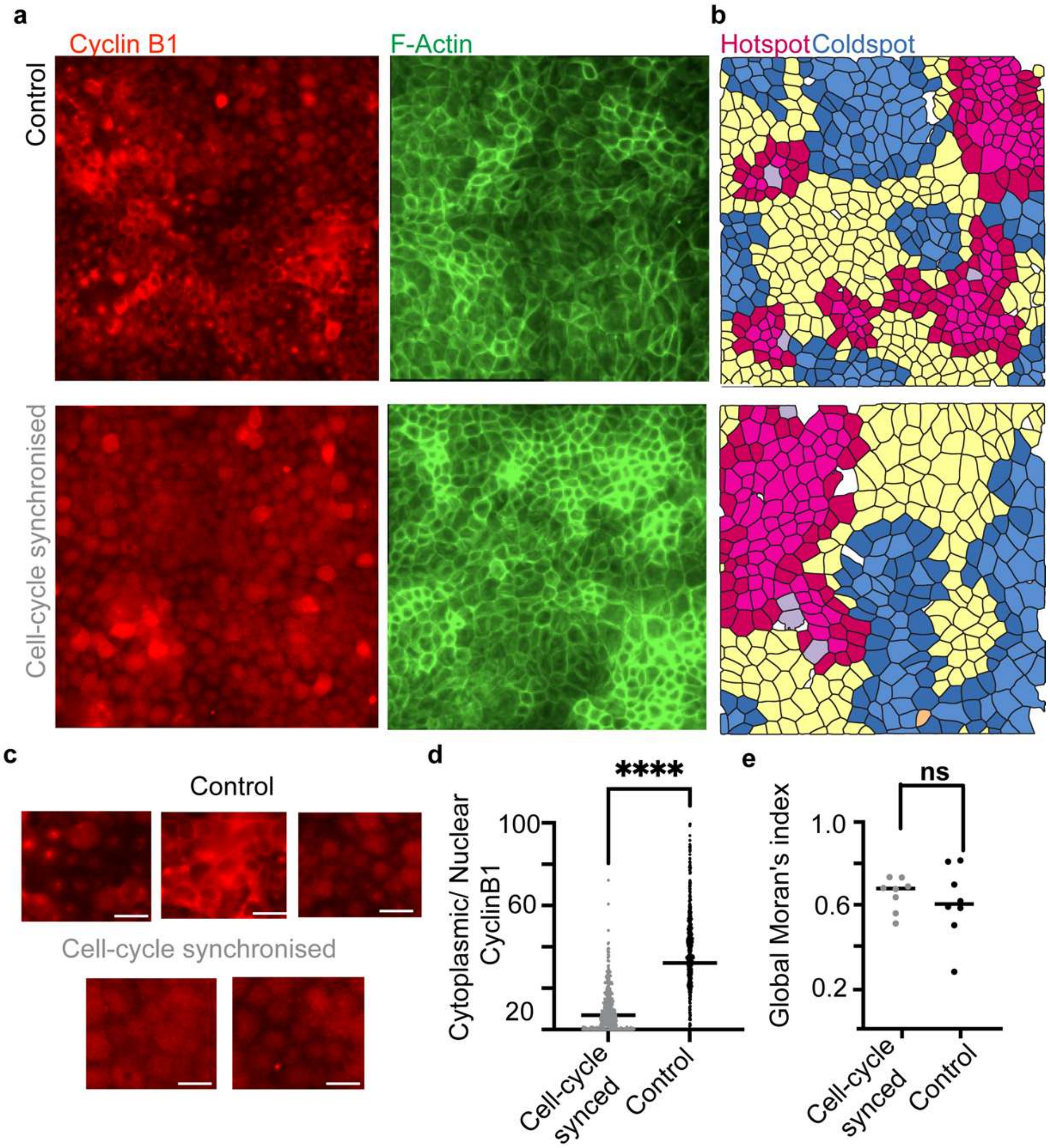

**Supplementary figure 13**: **Spatial Actin heterogeneity persists after cell cycle synchronization with Thymidine double block:** a) Confluent dishes of MDCK cells stained Cyclin B1 (Left-red) and F-Actin (Right green in cell-cycle synchronized by Thymidine double block (Bottom) and control (Top) samples, b) F-Actin LISA clusters maps corresponding to the images in panel A. c) Zoomed in images of Cyclin B1 corresponding to images in panel A demonstrating nuclear, cytoplasmic, perinuclear distribution of Cyclinb1at different locations in a control monolayer (top), compared to a more uniform distribution in nuclear or cytoplasmic compartments in the cell-cycle synchronized samples. d) Ratio of Cyclin B1 intensity at the nucleus and cytoplasm shows lesser variance in the cell-cycle synchronised samples suggesting uniformity in their cell-cycle state and e) No significant difference in the Global Moran's Index of Control and Cell cycle synchronized samples suggesting that inhomogeneous cell cycle state does not contribute to the observed heterogeneity in Actin levels. Lines represent the median. Scale bars=50μm.

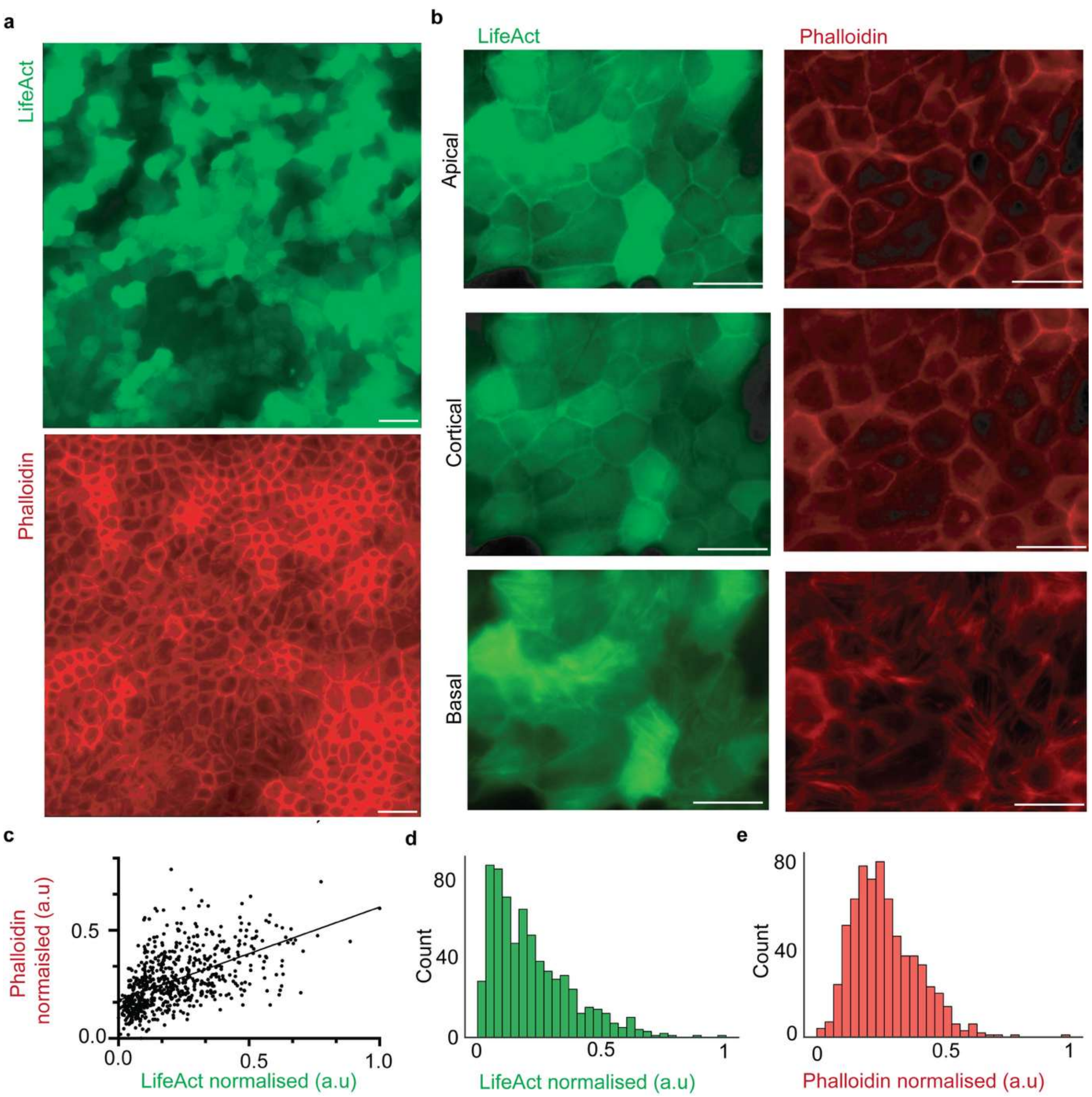

**Supplementary figure 14: Comparison of Actin staining with LifeAct and Phalloidin shows similar distributions:**

Confluent dishes of MDCK cells stained both for LifeAct-GFP (top) and Phalloidin 594 (bottom) imaged in 3D Z-stacks b) Zoomed images of LifeAct (Left-green) and Phalloidin (Right-red) at the top plane (Apical- top), middle plane (Cortical- middle), and bottom plane (Basal plane- bottom), c) Scatter plot of Min-Max Normalised LifeAct and Phalloidin intensities; $R^2$= 0.2925, d) Histogram of LifeAct and e) Phalloidin intensities. Scale bars=50μm.

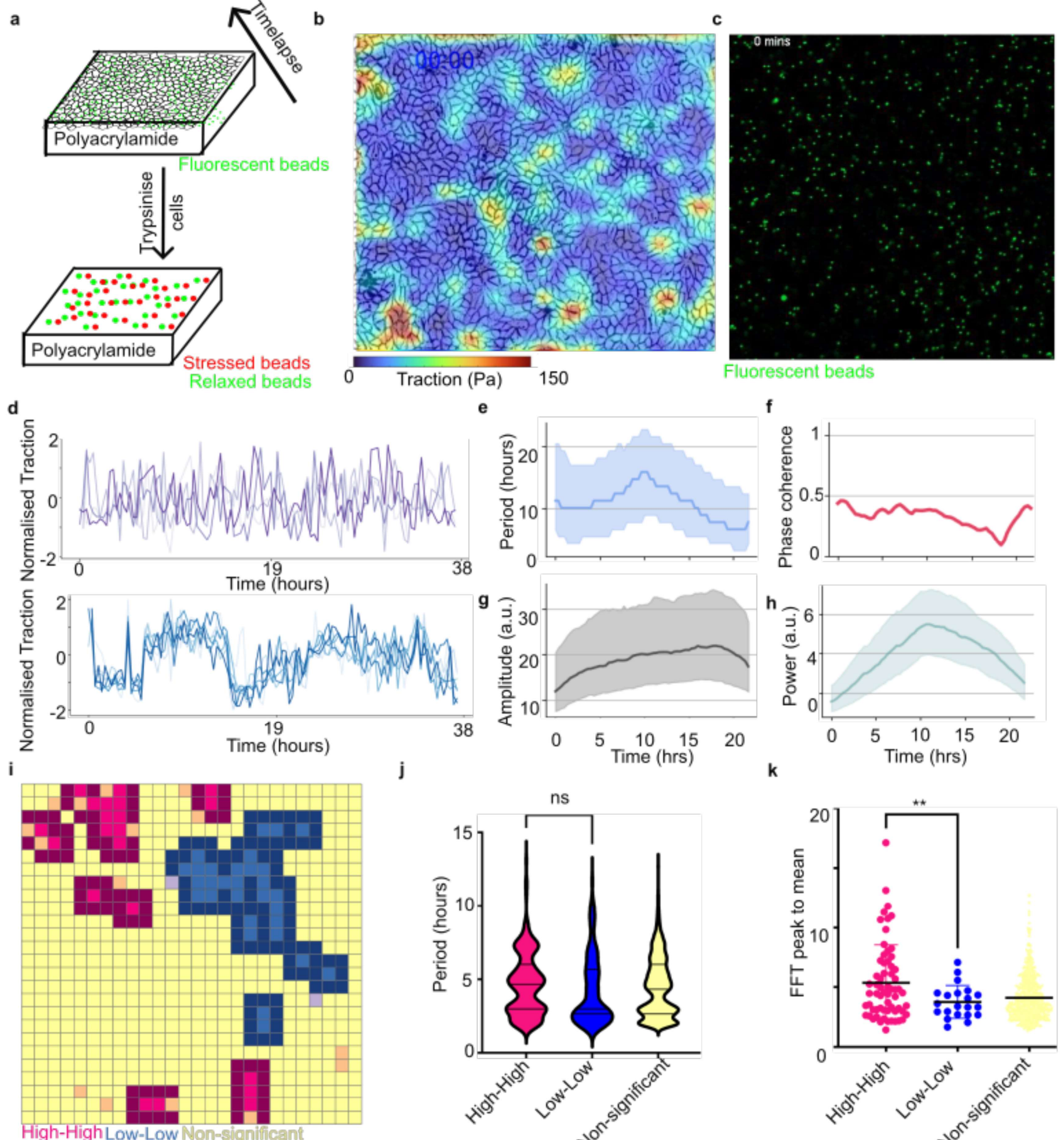

**Supplementary Figure 15: Spatiotemporal dynamics of traction force shows hours-scale oscillations**: a) Schematic of timelapse traction force Microscopy experiment, b) Traction force heatmap overlaid on cell image, c) Image of beads used in traction force experiment, d) Bottom- Representative locally oscillatory traction signals and top- Representative locally non- oscillatory traction signals, e) Period of oscillation obtained from Wavelet transform using PyBoat, f) Phase coherence over time obtained from Wavelet transform using PyBoat, g) Amplitude of traction signal over time obtained from Wavelet transform using PyBoat, h) Wavelet power over time obtained from Wavelet transform using PyBoat, i) LISA plot of traction force at T=20 hours, j) Plot comparing period of hotspots, coldspots and non-significant regions, k) Plot comparing FFT (Fast Fourier Transform) peak/mean ratio, a metric of oscillation strength at hotspots, coldspots and non-significant regions. Lines represent the median.

**Supplementary Video S1**: Timelapse movie of homeostatic LifeAct MDCK cells (left) with cell segmentation and tracking (right) used for MSD and Q(t) analysis, imaged every 2.5 minutes. Cell outlines are pseudocoloured for cell area, corresponding to the colourbar in the video. Cell tracks are coloured based on the average track speed, with the cell speed linearly increasing from 2 to 10 µm per hour from blue to red colour. Dark spaces correspond to cells that were not identified during segmentation. Scale bar=50µm.

**Supplementary Video S2**: Timelapse movie of homeostatic control LifeAct MDCK monolayer (left) and monolayer treated with Mitomycin C (right) to inhibit cell division (right).

**Supplementary video S3:** Timelapse movie of homeostatic LifeAct MDCK cells pseudocoloured for per-cell Actin intensity that were used to analyse the timeperiod of actin clusters. Dark spaces correspond to cells that were not identified during segmentation. Scale bar=50µm.

**Supplementary video S4:** Timelapse LISA segmentation of Actin intensity- LISA analysis performed for homeostatic LifeAct MDCK cells imaged every 30 minutes. The colours corresponding to LISA classification scheme are as follows: Pink- High-High, Yellow- Non-significant and Blue- Low- Low. Dark spaces correspond to cells that were not identified during segmentation. Scale bar=20µm.

**Supplementary video S5:** Timelapse movie of homeostatic single cell LifeAct-MDCK cells without cell-cell contacts imaged every two minutes, that were used for the analysis of Actin oscillation period in single cells.

**Supplementary video S6:** Cell tracking of low (left) and high (right) density monolayers compared side-by-side. The final timepoint shows the tracks for the cells over all timepoints. Cells are pseudocoloured for cell area, corresponding to the colourbar in the video. Cell tracks are coloured based on the average track speed, with the cell speed linearly increasing from 2 to 10 µm per hour from blue to red colour. Scale bar=50µm.

**Supplementary Video S7**: Cropped regions showing Actin oscillations in neighbouring patches. Cell outlines are pseudocoloured for cell area, corresponding to the colour bar below

200 Cell Area ($\mu m^2$) 1000

**Supplementary Video S8:** Experimental videos depicting oscillations of different periods at the actin hotspots and coldspots. Left- Evolution of LISA clusters over time, Middle- Zoomed in view of cells from the coldspots (topleft) and hotspots (bottom left) and corresponding segmented and tracked cells pseudocoloured for Actin intensity (right). Right- Representative detrended Actin signals from coldspots (bottom) and hotspots (top). Scale bar=50µm.

**Supplementary Video S9:** Video showing vertex model simulation with cell divisions: i) without mechanochemical feedback, ii) with only MCFL-I, iii) with MCFL-I+II.

**Supplementary Video S10:** Vertex model simulation with cell divisions and MCFL-I+II. The colormap shows the local cell density or 1/ (cell area) and the arrows indicate the PIV or the average velocity around that grid point.

# Supplementary Information: Theory

Sindhu Muthukrishnan,[1] Phanindra Dewan,[2] Tanishq Tejaswi,[1] Michelle B Sebastian,[1] Tanya Chhabra,[1] Soumyadeep Mondal,[2] Soumitra Kolya,[3] Sumantra Sarkar,[2,*] and Medhavi Vishwakarma[1,†]

[1]*Department of Bioengineering, Indian Institute of Science, Bengaluru, Karnataka, India, PIN 560012*
[2]*Department of Physics, Indian Institute of Science, Bengaluru, Karnataka, India, PIN 560012*
[3]*Tata Institute of Fundamental Research, Hyderabad, India, PIN 500046*

## I. RATIONALE BEHIND THE MODEL AND THE CONSTITUTIVE RELATIONS

### A. The canonical vertex model

The canonical vertex model is described by the following hamiltonian [1, 2]:

$$H = \sum_{\alpha} \frac{1}{2} K (A_\alpha - A_0)^2 + \sum_{\alpha} \frac{1}{2} \Gamma P_\alpha^2 + \sum_{i,j} \Lambda_{i,j} l_{i,j}, \tag{S1}$$

where $A_\alpha$, $P_\alpha$ are the area and perimeter of the cell $\alpha$. $K, \Gamma$, and $A_0$ are the area elasticity coefficient, perimeter elasticity coefficient, and preferred area of the cells, respectively, which we take to be identical for all the cells. $l_{ij}$ is the length of the interface or edge between vertices $i$ and $j$, and $\Lambda_{ij}$ is the line tension along this interface. If we assume that the line tension is the same across all the interfaces, such that $\Lambda_{ij} = \Lambda$, the vertex model can be written in another well-known canonical form:

$$H = \sum_{\alpha} \frac{1}{2} K (A_\alpha - A_0)^2 + \sum_{\alpha} \frac{1}{2} \Gamma (P_\alpha - P_0)^2, \tag{S2}$$

where $P_0 = -\Lambda/\Gamma$. The parameters of the vertex model can be normalized to define contractility, $\bar{\Gamma}$ and the normalized tension, $\bar{\Lambda}$ as follows:

$$\begin{aligned} \bar{\Gamma} = & \left(\frac{\Gamma}{K}\right) \frac{1}{A_0} \\ \bar{\Lambda} = \frac{\Lambda}{K A_0^{3/2}} = & -\left(\frac{\Gamma}{K}\right) \frac{P_0}{A_0^{3/2}} \end{aligned} \tag{S3}$$

*a. Ground state phase diagram:* The ground state of the vertex model has been analyzed from which a phase diagram can be constructed (Fig. T1). We use this phase diagram as a starting point for our analysis.

### B. Biological origin of contractility and tension

The actin cytoskeleton of the cell controls the contractility of the cell and the line tension of cell-cell contacts in epithelial tissues. However, they originate from different morphologies of the actin cytoskeleton. Contractility controls the length of the perimeter, and it arises from the formation of contractile actomyosin rings and stress fibers. In contrast, the line tension arises from the accumulation of cortical actomyosin along the cell-cell junctions. Therefore, we can assume that high line tension implies higher accumulation of the cortical actin, whereas high contractility implies higher accumulation of the stress fibers and the contractile rings. Translating these observations into the parameters of the vertex model (Eq. S3), we can construct the following picture (Fig. T2). This picture implies that for fixed $\Gamma/K$, increasing $A_0$ decreases the stress fibers and increases the junctional actins. Similarly, an increase in line tension is associated with decreasing $P_0$ or increasing $A_0$ or both.

As a tissue matures through cell division and cell-cell interactions, we expect it to explore some trajectories in this phase diagram. In our experiments, the tissue solidifies through cell division. Time-lapse imaging of actin, myosin, and E-cadherin during this solidification process reveals that the cells contain many elongated stress-fibers and low accumulation of junctional actin at low densities. The opposite trend is seen at high density, in which the cells have highly mature cell-cell junctions with high accumulation of actin, myosin, and E-cadherin at the junctions. In contrast, the actin is highly depleted in the bulk, and stress fibers are hardly seen. Similar observations were also reported earlier. Hence, during tissue solidification through cell division, the trajectory should start at the upper left part of the phase diagram and move to the lower right part, as shown in the figure below (Fig. T3).

* sumantra@iisc.ac.in
† medhavi@iisc.ac.in

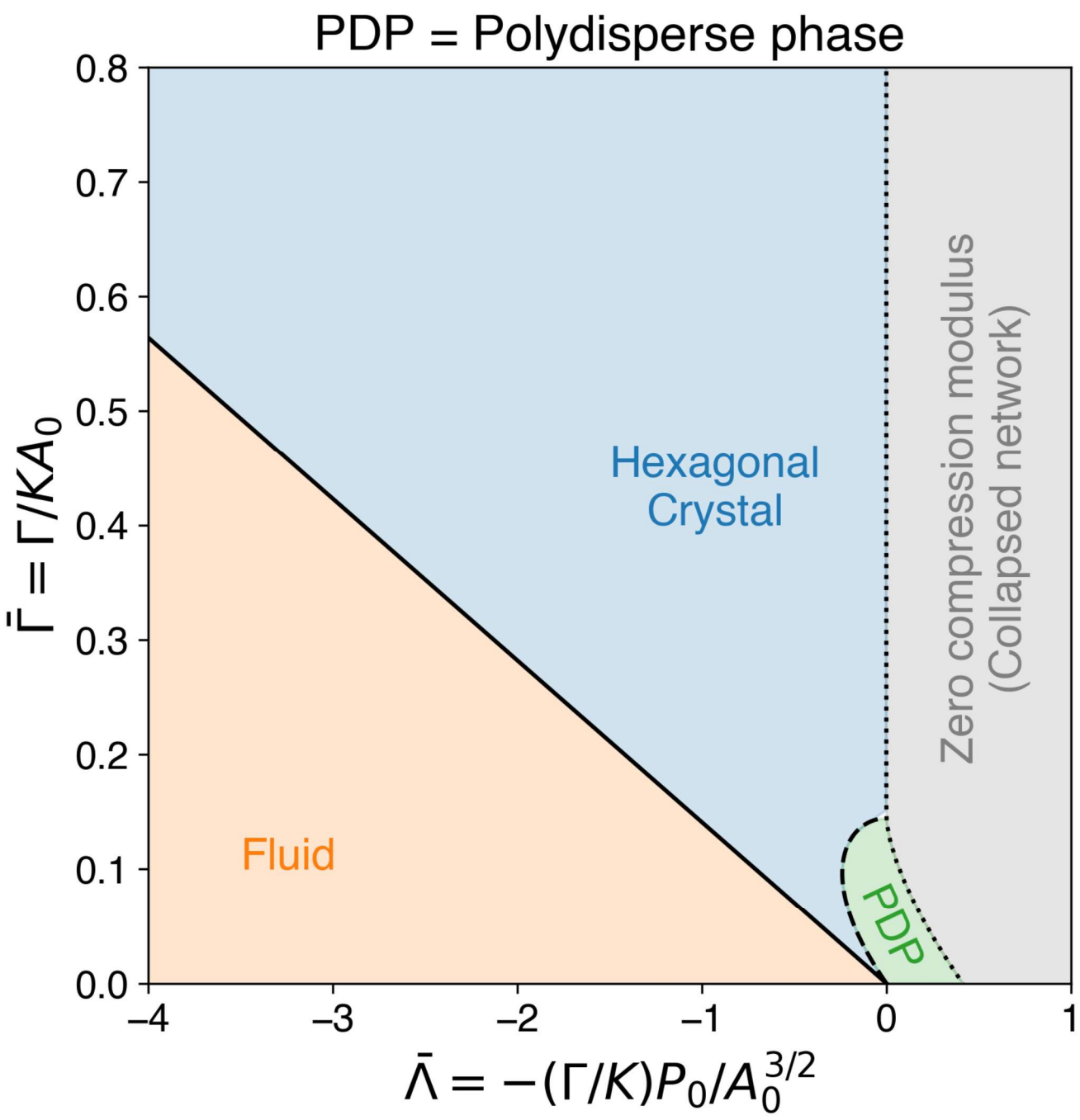

FIG. T1. The ground state phase diagram of the canonical vertex model. The solid and the dashed black lines denote regions where the shear modulus of the hexagonal crystal vanishes. The orange fluid state is composed of soft networks, whereas the green polydisperse phase is composed of polygons of various shapes and sizes. In the canonical vertex model, the PDP can contain 4-8 or 3-12 crystalline lattices.

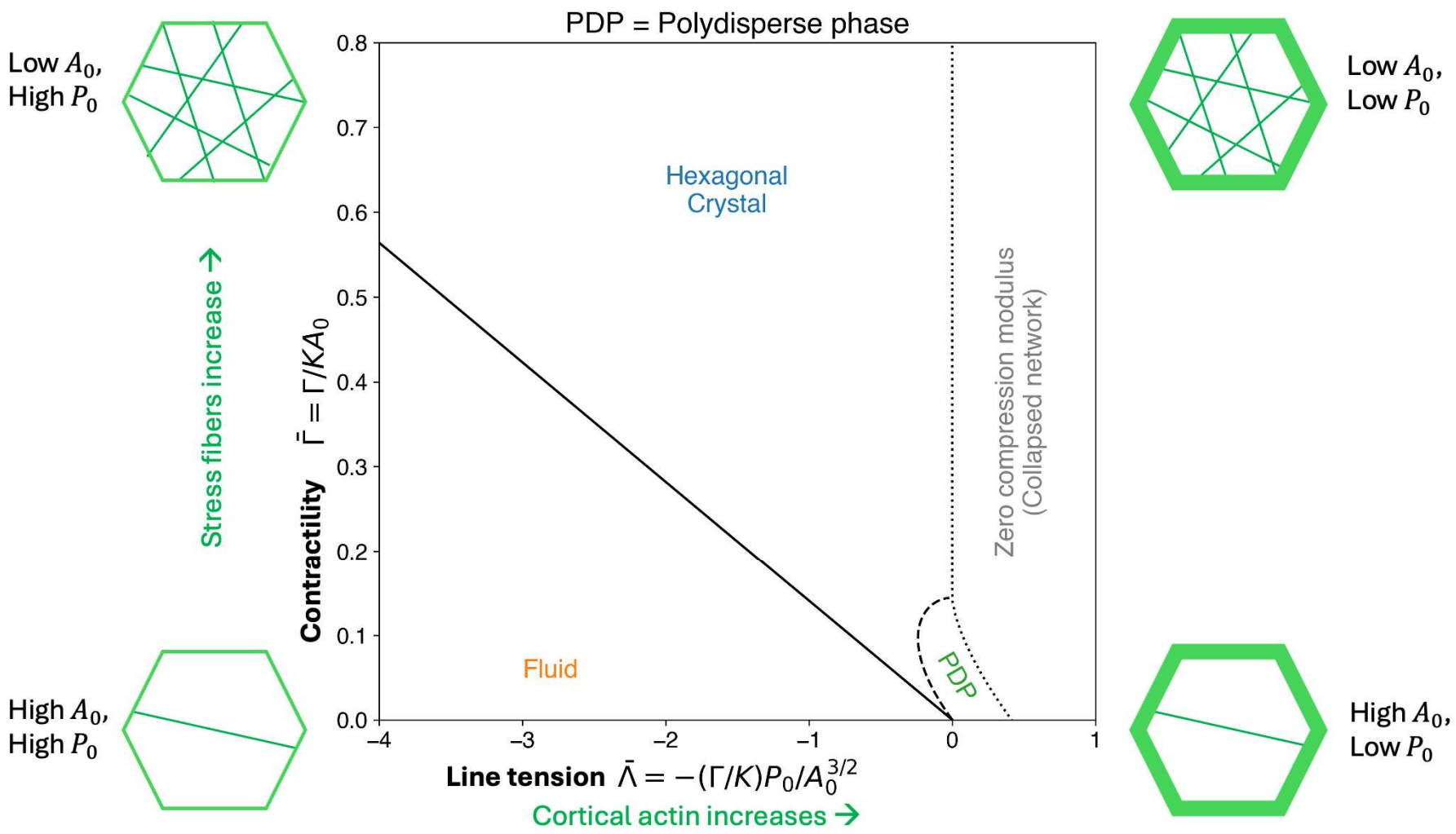

FIG. T2. The biological origin of the contractility and tension.

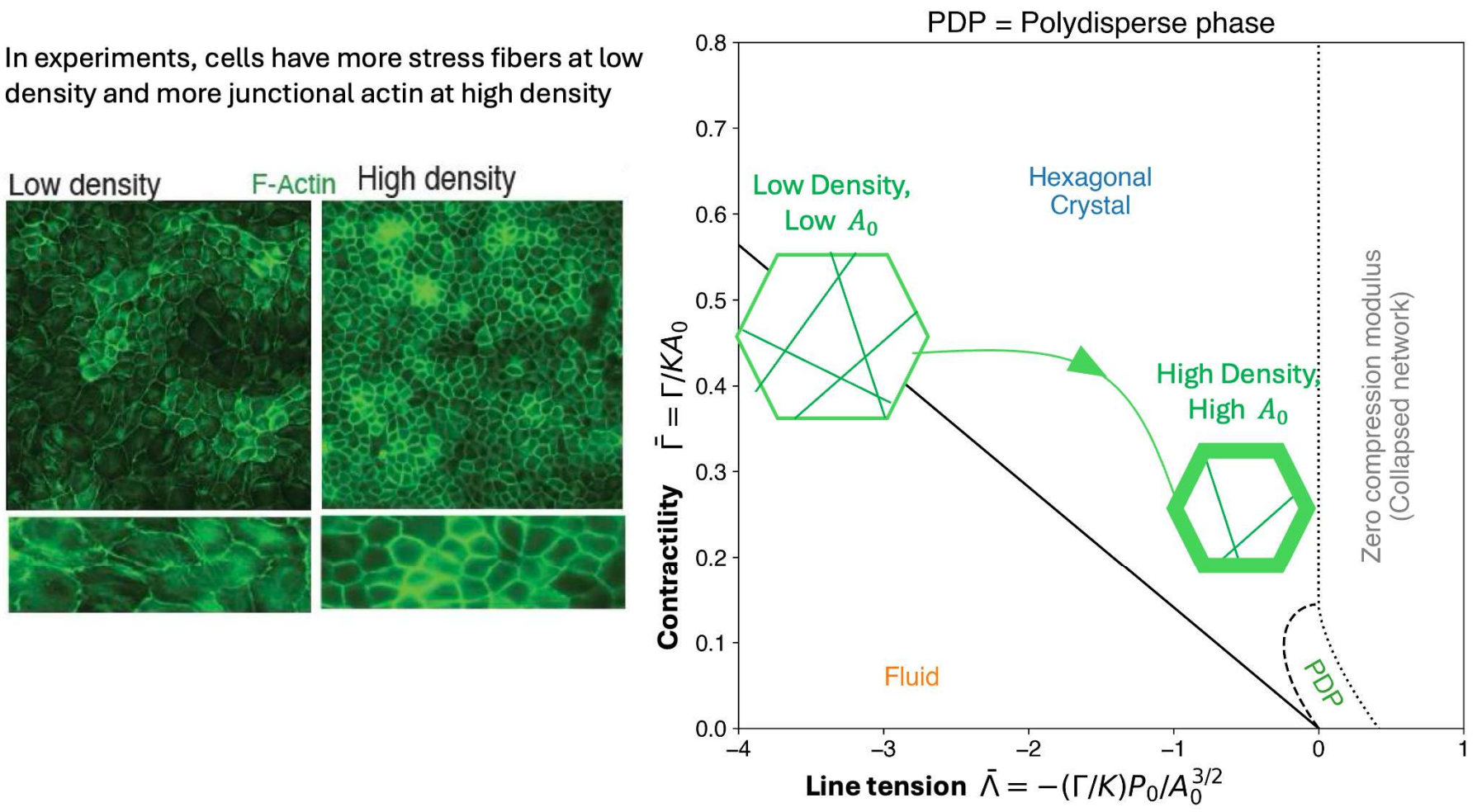

FIG. T3. Trajectory of our experiment on the phase diagram. A theoretical description of tissue solidification through cell division will require a similar change in the contractility and line tension.

### C. Constitutive laws for tissue solidification through cell division

Based on the analysis above, it is clear that any model of crowding-induced tissue solidification, such as through cell division, would require constitutive relations that ensure high contractility & low tension at low density and low contractility & high tension at high density. One way to achieve this phenomenology is to consider the load-dependent binding of myosin to actin in the cell-cell junction. As cell density increases, cells touch each other and form adhesive interfaces. This affects cell mechanics in two ways. First, the density-dependent binding of E-cadherin catalyzes the production of Rho-GAP [3], which inhibits Rho-dependent actomyosin contractility in the cell cytoplasm, inhibiting the formation and the maintenance of the stress fibers [4]. Second, E-cadherin, recruits Rho at the cell-cell junction, increasing the actomyosin contractility through a density-dependent positive feedback [5]. The abrogation of the stress fibers in the cytoplasm pays for the enhanced junctional/cortical actomyosin. We can capture the fraction of cortical actomyosin, $p_{bound}$, as a function of cell density from this consideration.

#### 1. *Load-dependent binding of myosin*

The unbinding rate of myosin from actin depends on the local strain. The higher the contraction of the actomyosin network, the lower the unbinding rate. In fact, to a good approximation, the unbinding rate, $k_u$, decays exponentially with the strain, $\epsilon$. Specifically, for an isotropic deformation of the cell in the 2D vertex model, the strain is a scalar:

$$\epsilon = \frac{A - \hat{A}}{\hat{A}}, \tag{S4}$$

where $\hat{A}$ is a reference area, whose form will be determined later. The unbinding rate can be written as [6, 7]:

$$k_u = k_u^0 \exp(\chi\epsilon) \tag{S5}$$

$$= k_u^0 \exp\Big(\chi(A/\hat{A} - 1)\Big) \tag{S6}$$

$$\equiv k_u^0 \exp(\chi(x-1)) \tag{S7}$$

$$x = A/\hat{A} \tag{S8}$$

Hence, for $\chi > 0$, which is what is observed experimentally [6, 7], the unbinding rate decreases with increasing contraction, $x < 1$, leading to increased actomyosin binding, which increases the contraction further. This mechanochemical feedback loop, which we call MCFL-I, changes the bound fraction of actomyosin in a density-dependent manner. To see this, consider the following reaction:

$$M + F \underset{k_b}{\overset{k_u}{\rightleftharpoons}} MF \tag{S9}$$

Here $M$ and $F$ denotes myosin and F-actin, respectively, and $MF$ is their bound state. Assuming that the reactions occur at a much faster rate than appreciable changes in the cell area, the bound fraction of myosin can be calculated assuming chemical equilibrium.

$$p_{bound} = \frac{[MF]}{[M] + [MF]} \tag{S10}$$

In chemical equilibrium,

$$k_b[M][F] = k_u[MF] \tag{S11}$$

$$\therefore p_{bound} = \frac{1}{1 + \frac{k_u}{k_b[F]}} \tag{S12}$$

$$= \frac{1}{1 + \frac{k_u^0}{k_b}\frac{\exp[\chi(x-1)]}{[F]}} \tag{S13}$$

$$= \frac{1}{1 + K_D^0\frac{\exp[\chi(x-1)]}{[F]}} \tag{S14}$$

The equation simplifies significantly when $\chi = 1$, and $x \approx 1$. In this limit,

$$p_{bound} \approx \frac{1}{1 + \frac{K_D^0}{N_F/A}x} \tag{S15}$$

$$\tag{S16}$$

where $N_F = [F]A$ is the number of f-actin in the cell. Our experiments show that the total amount of actin remains constant. Hence, $K_D^0/N_F$ is a constant. In fact, we define:

$$\hat{A} = N_F/K_D^0. \tag{S17}$$

Clearly, $\hat{A}$ is inversely proportional to the dissociation constant of the actomyosin complex. Hence, it can be controlled by controlling the binding affinity of the myosin. In general, the lower the dissociation constant, the higher the $\hat{A}$, and hence the higher the binding affinity of myosin to actin. With this definition, we can write the bound fraction as:

$$p_{bound} = \frac{1}{1 + \left(\frac{A}{\hat{A}}\right)^2} = \frac{1}{1 + x^2} \tag{S18}$$

Eq. S18 works remarkably well for a range of area, $A$ (Fig. T4). Furthermore, this formula is much easier to work with in analytical calculations than the more general equation:

$$p_{bound} = \frac{1}{1 + x\exp\{[\chi(x-1)]\}}. \tag{S19}$$

Hence, we use this equation for all our simulations and analytical calculations.

### 2. *Constitutive relations*

The line tension, $\bar{\Lambda}$ should be proportional to the bound fraction. Because $\bar{\Lambda} \propto -P_0$, $P_0$ should decrease with $p_{bound}$, the fraction of junctional actin and increase with $1 - p_{bound}$, the fraction of stress fibers. $p_{bound}$ increases with decreasing area, that is, increasing density. Hence, the fraction of stress fibers decreases with increasing density. This reduction in stress fiber arises from the conservation of the total actin in the cell and allocation of more F-actin into the cell-cell junctions as the density increases. This observation implies that contractility, $\bar{\Gamma}$, should decrease with increasing $p_{bound}$. Finally, because $\bar{\Gamma} \propto 1/A_0$, $A_0$ should increase with $p_{bound}$. Taken together, we propose the following constitutive relations:

$$\begin{aligned} A_0(A) &= \quad 2a_0\frac{1}{1 + \left(\frac{A}{\hat{A}}\right)^2} \\ P_0(A) &= \quad 2\hat{q}_0\sqrt{a_0}\frac{\left(\frac{A}{\hat{A}}\right)^2}{1 + \left(\frac{A}{\hat{A}}\right)^2}. \end{aligned} \tag{S20}$$

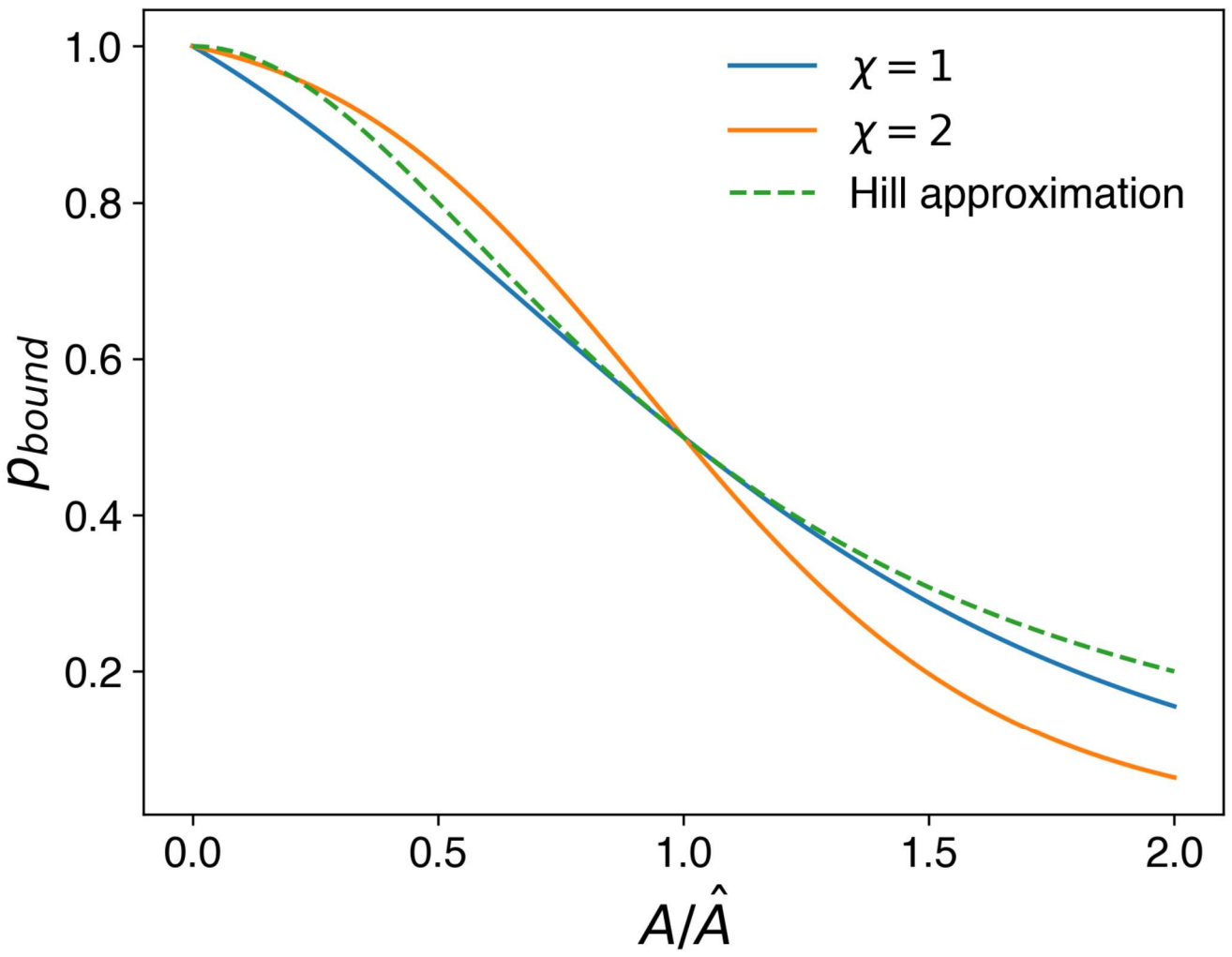

FIG. T4. Comparison of $p_{bound}$ computed from the true expression (Eq.S19) vs the hill approximation (Eq.S18).

Here,

$$a_0 = A_0(\hat{A}) = 1 \tag{S21}$$

$$\hat{q_0} = P_0(\hat{A})/\sqrt{A_0(\hat{A})} \tag{S22}$$

$$\text{Cell Density, } n = 1/A \tag{S23}$$

These two constitutive relations are sufficient to reproduce the "static" properties of the tissue, such as its zero-frequency elastic moduli. To understand tissue dynamics, such as glass transition through cell division, we need to quantify the observed dependence of cell motility on the intensity of stress fibers. Specifically, it has been observed that the cell motility is directly proportional to the stress fibers present in the cell [8, 9]. Because in our model $(1 - p_{bound})$ is the fraction of stress fibers, we postulate the following constitutive relation relating cell motility to the fraction of stress fibers.

$$v(A) = 2v_0 \frac{\left(\frac{A}{\hat{A}}\right)^2}{1 + \left(\frac{A}{\hat{A}}\right)^2} \tag{S24}$$

This relation ensures that at high densities the motility is zero. Although we have used this specific relationship for the motility, any decreasing function of cell density works.

These constitutive relations, shown in Fig. T5, reproduce the experimentally observed behavior with fluid phase at low densities and solid phase at high densities, as shown in Fig. T6. In the next section, we will use them to construct the ground state phase diagram as a function of cell density. However, it is important to consider alternate constitutive relations, which we do next.

### D. Alternate constitutive relations

#### 1. *Constant $A_0$ and $P_0$, but density-dependent $\Gamma/K$*

If $A_0$ and $P_0$ are constants, then a simple linear relationship exists between $\bar{\Gamma}$ and $\bar{\Lambda}$:

$$\bar{\Lambda} = -\left(\frac{\Gamma}{K}\right) \frac{P_0}{A_0^{3/2}} = -\frac{P_0}{\sqrt{A_0}} \frac{\Gamma}{K A_0} \tag{S25}$$

$$\therefore \bar{\Lambda} = -\hat{q}_0 \bar{\Gamma}. \tag{S26}$$

The line separating the fluid region from the hexagonal crystal in Fig. T1 satisfies the equation:

$$\bar{\Lambda} = -4\sqrt{\pi}\bar{\Gamma}. \tag{S27}$$

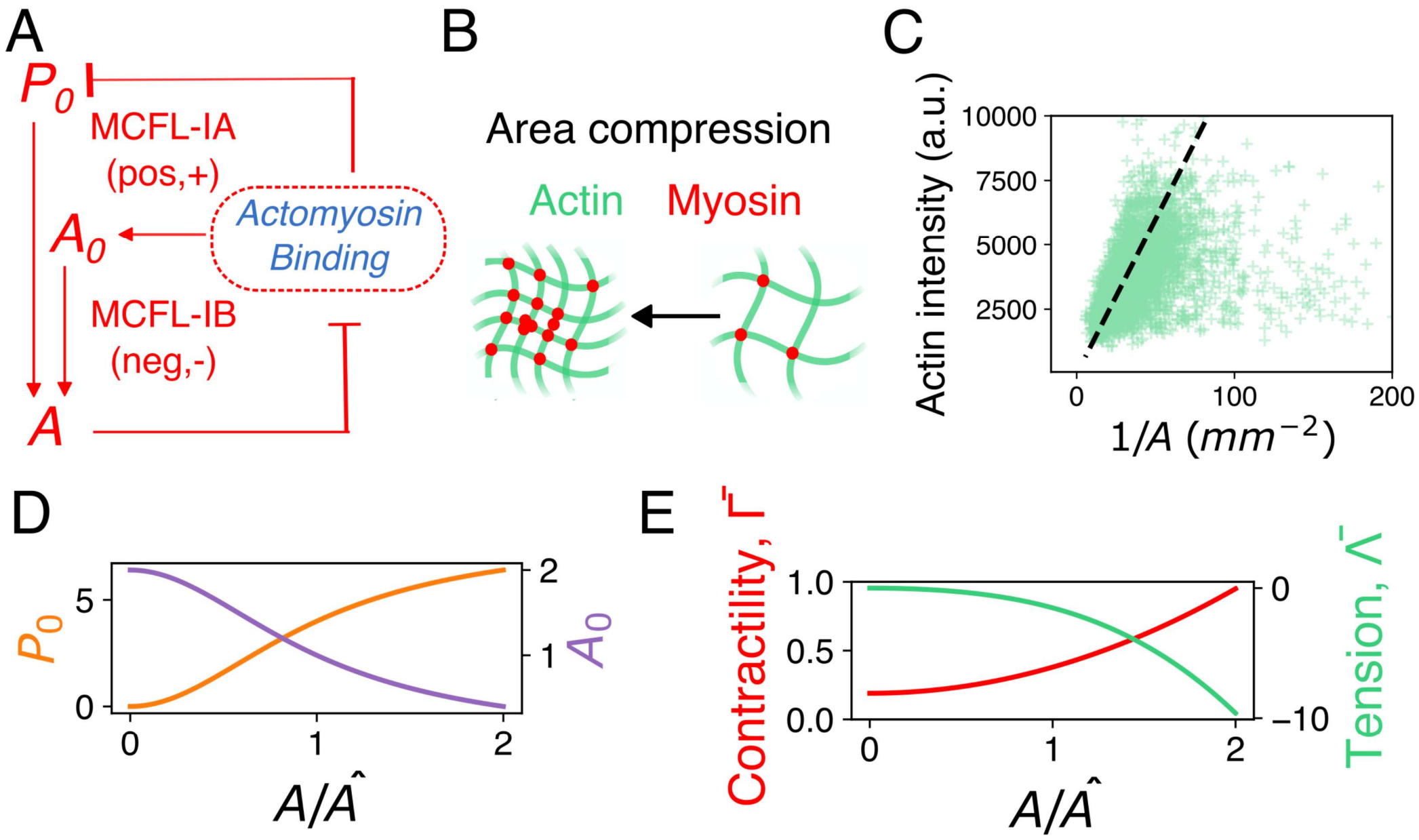

FIG. T5. (A)MCFL-I consists of two interacting positive and negative feedback loops. MCFL-IA leads to bistability, which creates coexisting cells with larger and smaller areas. MCFL-IB stabilizes the bistable areas from increasing or decreasing indefinitely, creating a stable distribution of cells even at high densities. (B) It arises from the load-dependent binding of myosin to actin. (C) Actin intensity is inversely proportional to the area of the cell, implying that the total actin is constant. (D) $A_0$ and $P_0$ as a function of $A/\hat{A}$ shows the density-dependence of the constitutive relations. (E) Contractility and tension as a function of $A/\hat{A}$ computed from these constitutive relations.

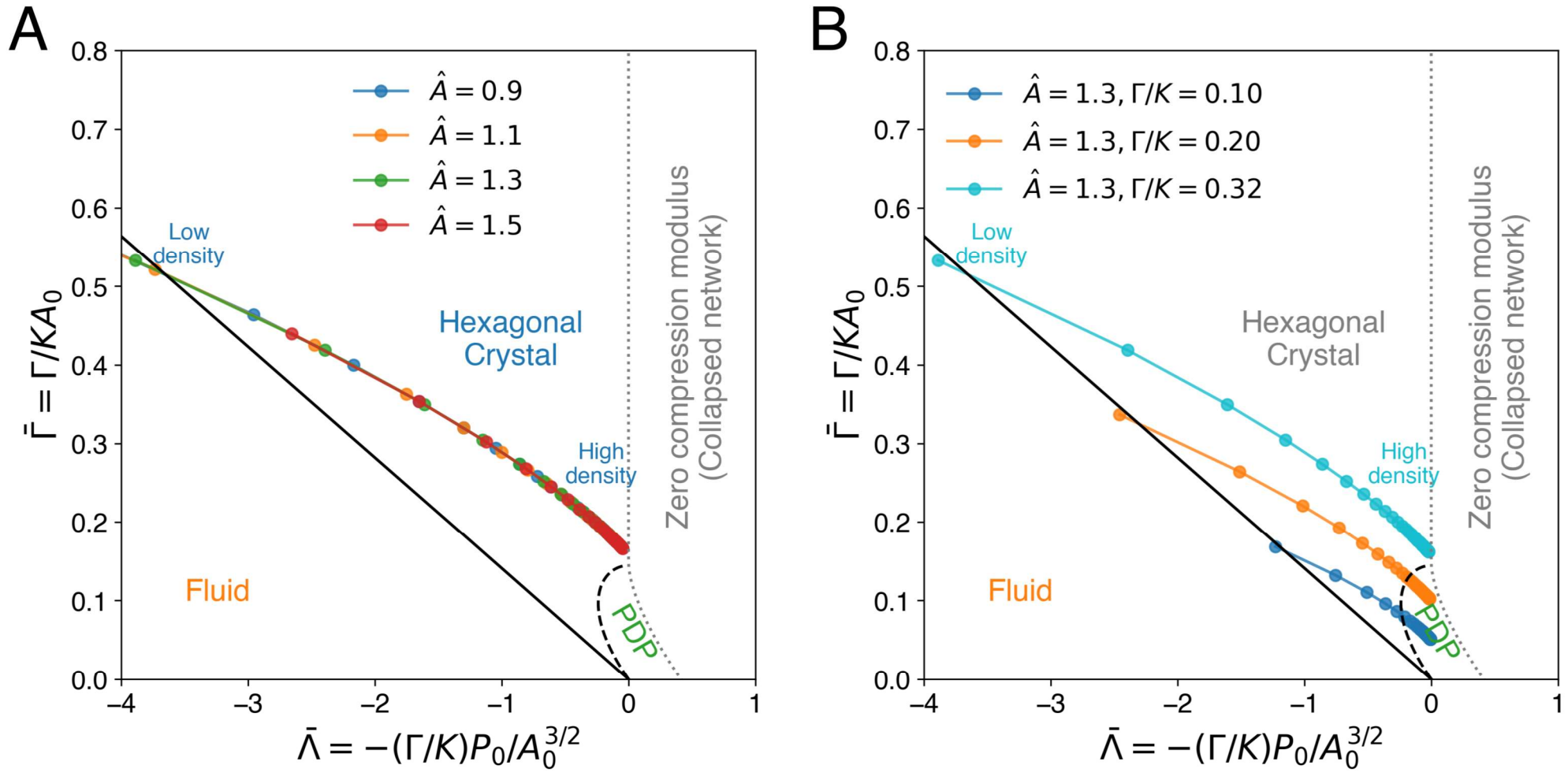

FIG. T6. (A)The constitutive relations (Eq. S20) reproduce density dependence consistent with experimental trajectories. Here, trajectories for a few $\hat{A}$ values are shown. (B) Changing $\Gamma/K$ shifts these trajectories along the $\bar{\Gamma}$ axis. For $\Gamma/K < 1/2\sqrt{3}$, the trajectories can access the PDP phase at high density. The rigidity of this state is yet to be determined.

Hence, as long as the shape parameter $\hat{q}_0 < 4\sqrt{\pi} \approx 7.1$, the trajectory always remains in the crystalline region. Conversely, if $\hat{q}_0 > 4\sqrt{\pi}$, then the trajectory strictly remains in the fluid region of the phase diagram. For sufficiently small $\hat{q}_0$, the trajectories in the crystalline phase can reach the PDP phase at small $\bar{\Gamma}$ values. Clearly, this protocol does not match experimental observation, where a single trajectory remains in the fluid phase at a low density and reaches the crystalline phase at high density. Doing so requires both $\bar{\Gamma}$ and $\hat{q}_0$ to change with density, in which case, they are equivalent to the constitutive relations Eq. S20.

### 2. $A_0 \propto 1/n$, $P_0$ and $\Gamma/K$ constant

As cell density, $n$, increases, the area of the cells, $A$, decreases. The average area scales as $1/n$. One possible way to achieve this is to make the preferred area, $A_0 = 1/n$, while keeping all other parameters constant. Hence, in this protocol, the contractility, $\bar{\Gamma}$, increases with $n$, and the line tension decreases with $n$, leading to fluidization at high density. Furthermore, the trajectory is exactly opposite to what is observed in experiments, including ours. Hence, this constitutive relation is unphysical and should not be used to model a proliferating tissue.

## E. Effect of Blebbistatin

With the assumption that the only way blebbistatin ($B$) can interact with myosin is via competitive inhibition, we can write the kinetic equations as:

$$M + F \underset{k_b}{\overset{k_u}{\rightleftharpoons}} MF \tag{S28}$$

$$M + B \underset{k_1}{\overset{k_2}{\rightleftharpoons}} MB \tag{S29}$$

At equilibrium, we get

$$k_b[F][M] = k_u[FM] \tag{S30}$$

$$k_1[B][M] = k_2[BM] \tag{S31}$$

We can calculate the fraction of unbound myosin as

$$\frac{[M]}{[FM]} = \frac{k_u}{k_b[F]} = \frac{k_D^F}{[F]} \tag{S32}$$

and,

$$\frac{[BM]}{[FM]} = \frac{k_u}{k_2}\frac{k_1[B]}{k_b[F]} = \frac{k_u}{k_b[F]}\frac{k_1[B]}{k_2} = \frac{k_D^F}{[F]}\frac{[B]}{k_D^B} \tag{S33}$$

And finally we can calculate the fraction of myosin bound to actin as

$$\frac{[FM]}{\text{Total M}} = \frac{[FM]}{[M+[FM]+[BM]]} \tag{S34}$$

$$= \frac{1}{1+\frac{[M]}{[FM]}+\frac{[BM]}{[FM]}} \tag{S35}$$

$$= \frac{1}{1+\frac{k_D^F}{[F]}+\frac{k_D^F}{[F]}\frac{[B]}{k_D^B}} \tag{S36}$$

$$= \frac{1}{1+\frac{k_D^F}{[F]}\left(1+\frac{[B]}{k_D^B}\right)} \tag{S37}$$

$$= \frac{1}{1+\left(\frac{A}{\hat{A}}\right)^2\left(1+\frac{[B]}{k_D^B}\right)} \tag{S38}$$

$$= \frac{1}{1+(A/\hat{A}')^2} \tag{S39}$$

where, $\hat{A}' = \frac{\hat{A}}{\sqrt{1+[B]/k_D^B}}$, is the modified $\hat{A}$. From this calculation we can see that the presence of blebbistatin can reduce $\hat{A}$.

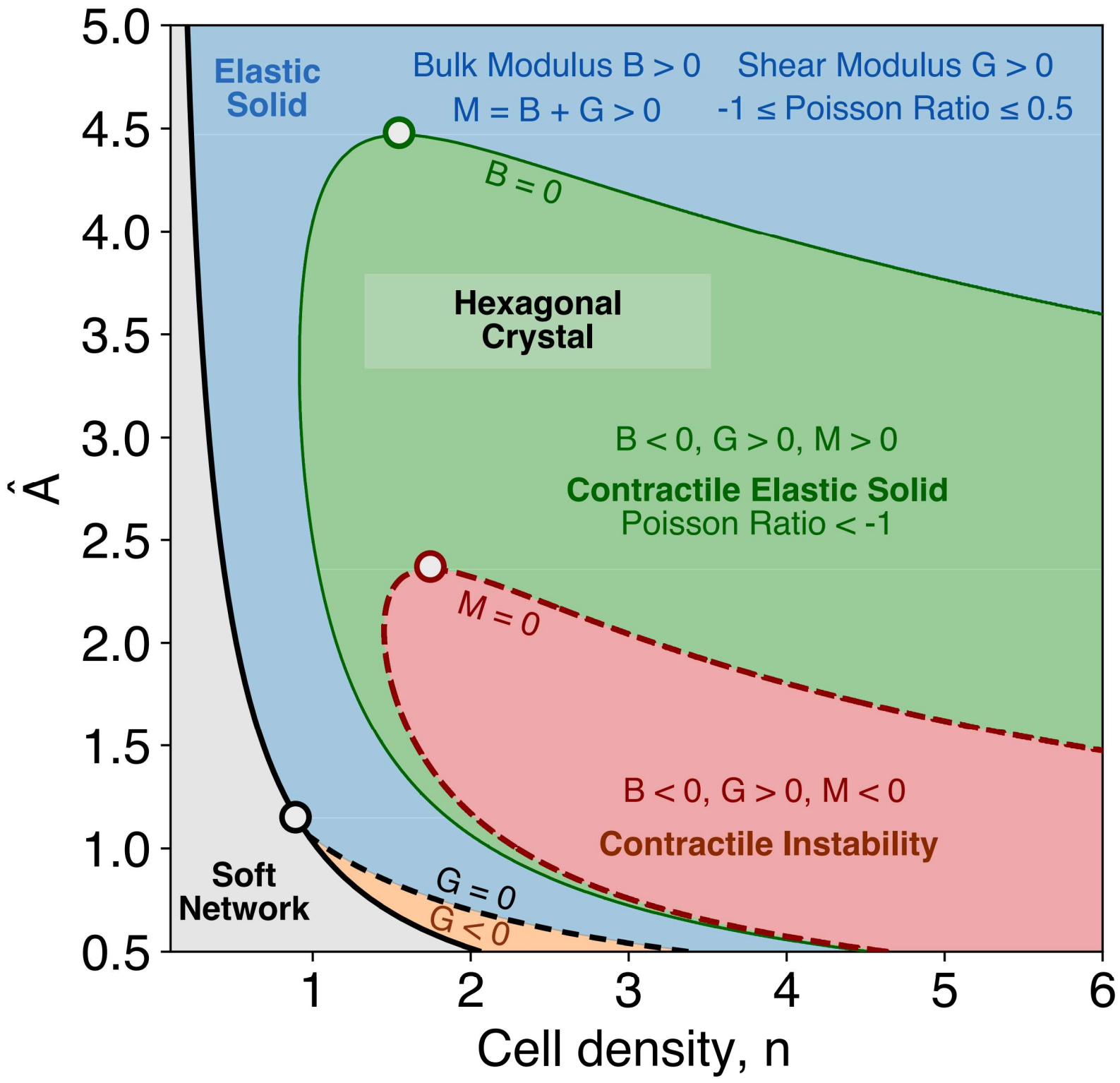

FIG. T7. The ground state phase diagram of the vertex model with MCFL-I in the absence of any motility. There are five different regions, marked by whether $q_0 = P_0/\sqrt{A_0} = 3.722$ (solid black line), which is also the jamming density $n_J$, and the elastic moduli of the crystalline ground state. In the gray region, $q_0 = P_0/\sqrt{A_0} > 3.722$, and the ground state is a floppy soft network. Above a density (the non-gray regions), $n_{lo}(\hat{A})$, marked by the solid black line, the ground state is a hexagonal crystal. In the light blue region, $B, G, M > 0$, and the ground state is a conventional crystalline elastic solid. In the orange region, the bulk modulus, $B > 0$, but the pure shear modulus, $G < 0$, while the P-wave modulus, $M = B + G > 0$. We expect the ground state to be fluid and contain highly anisotropic polygons. In the green region, $B < 0$, but $G, M > 0$. In the red region, $G > 0$, but $B, M < 0$. In this region, the P-wave speed is imaginary, and hence, the tissue must undergo a contractile instability of the homogeneous crystalline state and generate regions of high and low densities, which maybe identified with actin hotspots and coldspots. The white markers with black, red, and green borders mark the maximum $\hat{A}$ values at which $G = 0$, $M = 0$, and $B = 0$, respectively.

## II. GROUND STATE PHASE DIAGRAM OF THE VERTEX MODEL WITH MCFL-I

### A. Vertex Model

We use the following vertex model with mechanochemical feedback.

$$E = \frac{1}{2}\left[K(A - A_0)^2 + \Gamma(P - P_0)^2\right] \tag{S40}$$

$$A_0(A) = 2a_0 p_b(A) \tag{S41}$$

$$P_0(A) = 2\hat{q}_0\sqrt{a_0}[1 - p_b(A)] \tag{S42}$$

$$p_b(A) = \frac{1}{1 + \left(\frac{A}{\hat{A}}\right)^2} \tag{S43}$$

### B. Elastic moduli of the hexagonal ground state

Following the procedure described in [10], we can calculate the pure shear and the bulk moduli of the hexagonal crystal. Applying this approach to our model, we get:

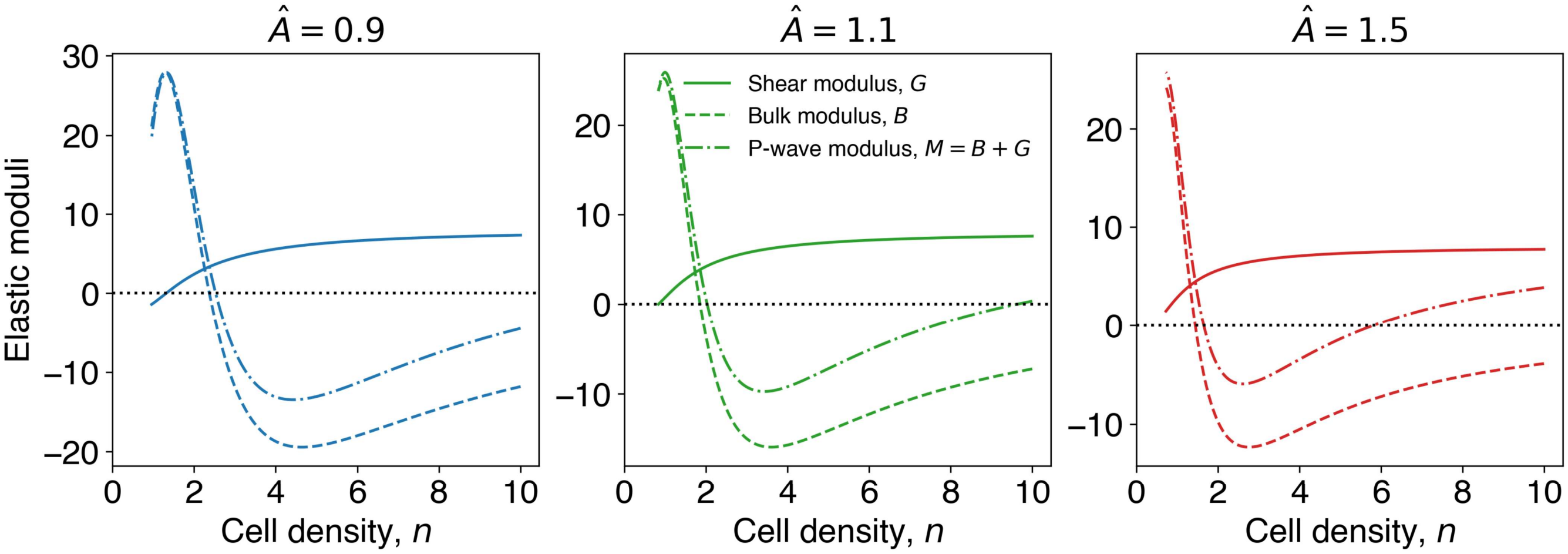

FIG. T8. The elastic moduli of the hexagonal ground state for different $\hat{A}$. At the lowest permissible density, $n_{min}$, the shear moduli can have negative, zero, or positive values depending on $\hat{A}$. These three cases correspond to an elastic fluid, fluid, and a solid phase.

*a. Shear Modulus*

$$G = \sqrt{3}\Gamma\left(12 - \frac{2}{\sqrt{\frac{2A}{3\sqrt{3}}}}P_0(A)\right) \tag{S44}$$

This expression is slightly different from what was obtained in [2, 10], because of the slightly different form of the energy. There, the following expression is obtained:

$$G_F = \sqrt{3}\Gamma\left(12 - \frac{P_0}{\sqrt{\frac{2A}{3\sqrt{3}}}}\right) \tag{S45}$$

The key difference between $G$ and $G_F$ is that the former can be negative, but the latter is always positive. In fact, in the pioneering work of Bi et.al. [11], $G$ was used to mark the boundary between the solid and the liquid phase.

*b. Bulk Modulus* The expression for the bulk modulus, $B$, is cumbersome due to the $A$-dependence of $A_0$ and $P_0$. The expression also does not provide any intuition. Hence, here we provide a sympy code to calculate the expression, and show the results. The key surprise is that the bulk modulus can become negative, while the shear modulus remains positive. In a range of density, the P-wave modulus, $M = B + G$, becomes negative. In this range, no longitudinal (P) waves can propagate, but transverse (S) waves can, which may explain why the amplitude of the actin oscillations dampen with density.

SymPy code for the bulk modulus

```
### Hexagonal lattice Bulk Moduli
import sympy as sp
import numpy as np

sp.init_printing()
x, A, Ahat, P, Phat, a0, qhat0, K, Gamma,m, a, Lambda, A0, P0, n = sp.symbols('x A Ahat P
↪  Phat a0 qhat0 K Gamma m a Lambda A0 P0 n',positive=True)

### Side length of a hexagon
a = sp.sqrt(2*A/(3*sp.sqrt(3)))
### Box dimensions
# x  = 1+epsilon is the scaling factor. For bulk modulus calculation, both Lx and Ly are
↪  scaled by x.
Lx = x*sp.sqrt(3)*a
Ly = x*3*a
l1 = Ly/3
l2 = sp.sqrt((Lx/2)**2 + (Ly/6)**2)

### Area of a hexagon
Ab = Lx*Ly/2
### Perimeter of a hexagon
Pb = 2*l1 + 4*l2
Sb = (Ab/Ahat)**2
pb = 1/(1+Sb)
A0 = 2*a0*pb
P0 = 2*qhat0*sp.sqrt(a0)*(1-pb)

### There are two hexagons in the box
Ea = K*(Ab - A0)**2

Ep = Gamma*(Pb - P0)**2 ## Canonical vertex
E = (Ea + Ep)

EpF =  Gamma*Pb**2 + Lambda*Pb ### Farhadifar
EF = Ea + EpF

### Bulk modulus from the Vertex Model + MCFL-I
B2 = sp.diff(E,x,2).subs({x:1})/(2*A)

#### Farhadifar bulk modulus
BF = sp.diff(EF,x,2).subs({x:1})/(2*A)

fB2 = sp.lambdify([A, Ahat, P, Phat, a0, q0, K, Gamma],B2,'numpy')

latex_B2 = sp.latex(B2)
sp.simplify(B2,deep=True)
```

### C. Effect of topological transitions and active forces

The ground state analysis presented here excludes topological transitions, such as cell intercalation through the T1-transitions and the node-switching algorithm used to prevent cell overlaps. Furthermore, we also do not consider the effect of active forces, such as from cell motility and MCFL-II, in these calculations. These transitions and forces can perturb the integrity of the crystalline phase, and we expect them to have nontrivial effects on the stability of the phases. A detailed exploration and characterization of these perturbations is beyond the scope of the current manuscript.

Simulations with just MCFL-I and motility indicate that the motility smoothens out the discontinuous transition. It also indicates that the floppy region below $n_J$ is susceptible to shear deformations.

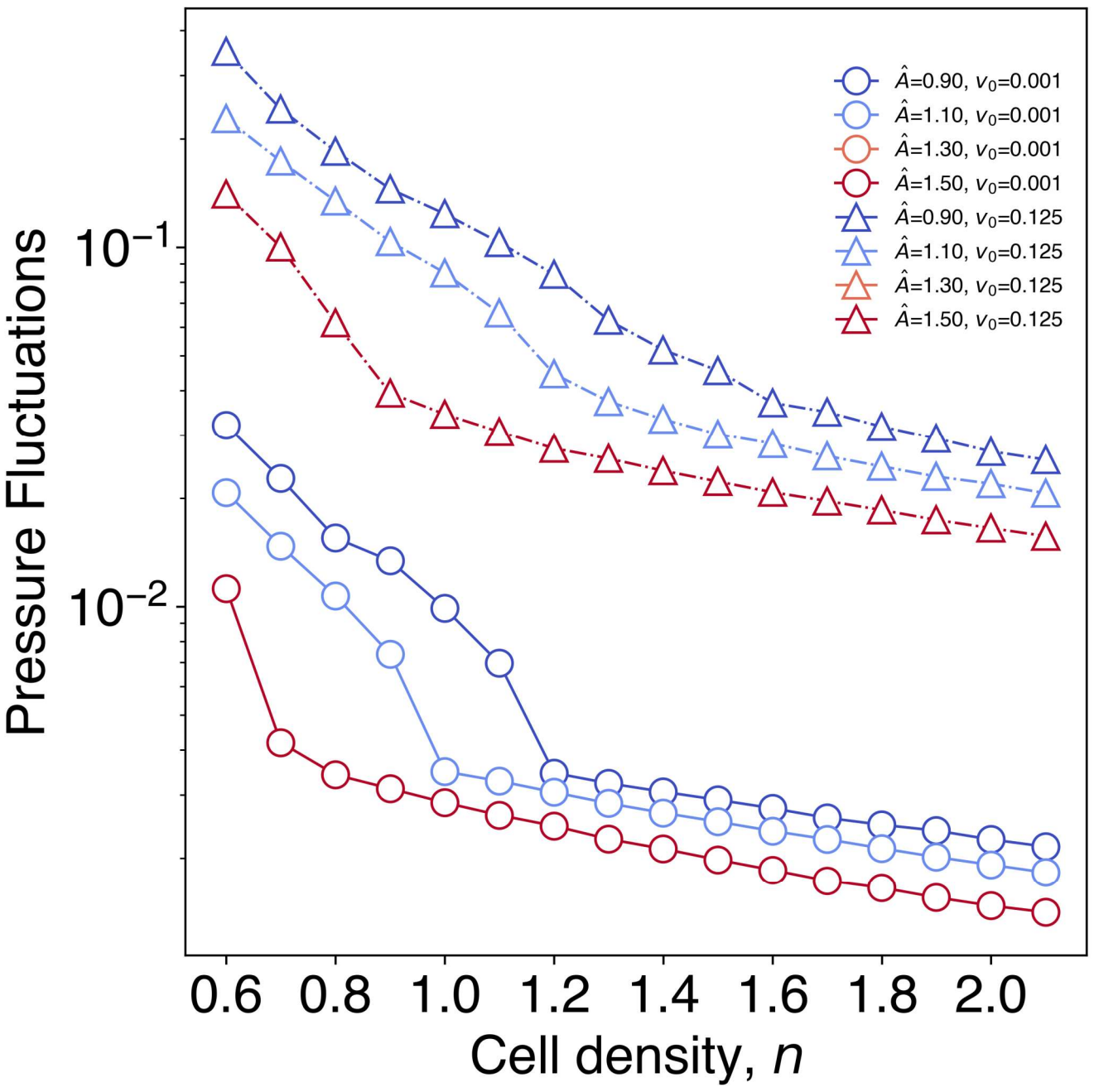

FIG. T9. Pressure fluctuation vs. $n$ for different $\hat{A}$ and motility.

## III. NUMERICAL SIMULATION OF THE MODEL

To reproduce the diverse behaviors observed experimentally in MDCK cell monolayers, we implement an active vertex model with mechanochemical feedback. An essential feature of the model is that cells can divide within the vertex framework. The mechanochemical feedback consists of two mechanochemical feedback loops (MCFLs). Each component of the model is described below.

### A. Active Vertex Model

The vertex model represents tissues as a tessellation of polygons, with each polygon representing a cell. The energy of the vertex model for cells labelled by $\alpha$ is written as:

$$E_{VM}(t) = \sum_{\alpha} \left[ \frac{K_\alpha}{2}(A_\alpha - A^0_\alpha(t))^2 + \frac{\Gamma_\alpha}{2}(P_\alpha - P^0_\alpha(t))^2 \right] \tag{S46}$$

Here, the area energy comes from the volume incompressibility of cells whereas the perimeter energy term comes from the actomyosin contractility of cells. The constant parameters $K_\alpha$ is the area modulus which gives the area elasticity and $\Gamma_\alpha$ is the perimeter modulus which is the elasticity due to the actomyosin cortex. The cells can be either solid-like or fluid-like according to the shape index defined as $q = P/\sqrt{A}$.

The vertices in the vertex model undergo overdamped dynamics. So, the equation of motion of the $i$th cell is given by:

$$\zeta \frac{\partial \vec{r}^{(i)}}{\partial t} = \vec{F}^{(i)} + \vec{F}^{(i)}_{act} + \vec{F}^{(i)}_{motility} \tag{S47}$$

Here, $\vec{r}^{(i)}$ is position of the $i$th vertex, and $\vec{F}^{(i)}$ is the equilibrium force on each vertex given by the gradient of the vertex model energy: $\vec{F}^{(i)} = -\nabla_{\vec{r}^{(i)}} E_{VM}$. Additionally, there can be other active forces $\vec{F}^{(i)}_{act}$ and motility force $\vec{F}^{(i)}_{motility}$. We have $\vec{F}_{act} = 0$ in our simulations. $\vec{F}^{(i)}_{motility}$ is a force which propels individual cells by $\vec{F}^{(i)}_{motility} = v_0 \vec{p}_\alpha$, where $\vec{p}_\alpha$ is the polarity of a given cell given by $\vec{p}_\alpha = (\cos\theta_\alpha, \sin\theta_\alpha)$, where this angle $\theta_\alpha$ performs rotational diffusion:

$$\partial_t \theta_\alpha = \sqrt{2D_r}\,\eta_\alpha(t) \tag{S48}$$

where $\eta_\alpha(t)$ is a Gaussian white noise with zero mean and correlation $\langle \eta_\alpha(t)\eta_{\alpha'}(t')\rangle = \delta(t-t')\delta_{\alpha\alpha'}$.

## B. Mechanochemical Feedback

### 1. MCFL-I

The first mechanochemical feedback loop arises from the load-dependent binding of myosin to actin. When a cell is compressed, i.e., when the cell area $A$ decreases, then the unbinding rate of myosin decreases, which changes the preferred area $A_0$ and the preferred perimeter $P_0$ of the cell in the vertex model. In our model, the preferred area and perimeter undergo changes in an area-dependent manner. As mentioned in the main text, actin reorganizes and concentrates at cell-cell junctions as the tissue matures. This means that junctional tension should increase, and the contractility should decrease as a tissue matures. Such an effect must be accounted for by one positive and one negative MCFL. Since junctional tension is given by $-P_0/A_0^{3/2}$ and the contractility is given by $A_0^{-1}$ [1], the feedback should be such that $P_0$ decreases with areal compression and $A_0$ should increase. Such feedback can be accounted for in the following way:

$$\begin{aligned} \tau_A \dot{A}_0 &= -[A_0 - \hat{A}_0(A)] \\ \tau_P \dot{P}_0 &= -[P_0 - \hat{P}_0(A)] \end{aligned} \tag{S49}$$

where we see that the vertex model parameters $A_0$ and $P_0$ are time-dependent and they relax to area-dependent values $\hat{A}_0(A)$ and $\hat{P}_0(A)$. The $\hat{A}_0$ and $\hat{P}_0$ depend on the binding probability of myosin to actin. If the myosin-binding probability is given by $p_{bound} = 1/(1+(A/\hat{A})^2)$. Here, $\hat{A}$ is related to the dissociation constant $K_d$ of myosin. Then, we know the constitutive relations from before:

$$\begin{aligned} \hat{A}_0 &= 2a_0 p_{bound} \\ \hat{P}_0 &= 2\hat{q}_0\sqrt{a_0}(1-p_{bound}) \end{aligned} \tag{S50}$$

### 2. MCFL-II

To model the actin oscillations observed in experimental MDCK monolayers, we model ERK as a Hopf oscillator, which, in this case, is a Brusselator. Other Hopf oscillators also exhibit similar collective behaviours, as we have checked using the FitzHugh-Nagumo model [12]. The chemistry of the Hopf oscillator is coupled to the mechanics via areal compressions, just like in MCFL-I. The 1D version of the model has been studied before to model for ERK oscillations [12, 13]. The above Eq. S49 will now be modified as:

$$\begin{aligned} \dot{M} &= a - (b+1)M + cM^2E \\ \dot{E} &= bM - cM^2E - DE \\ \tau_D \dot{D} &= -(D - D_0) - \beta D(A-1) \\ \tau_A \dot{A}_0 &= -[A_0 - \hat{A}_0(A)] - \alpha(E - E_0) \\ \tau_P \dot{P}_0 &= -[P_0 - \hat{P}_0(A)] - \frac{\alpha}{2\sqrt{A_0}}(E - E_0) \end{aligned} \tag{S51}$$

where $E$ is ERK, $M$ is MEK, and $D$ is a degrader of ERK, which couples the mechanics to the chemistry via the term $\beta D(A-1)$. The chemistry is coupled back to the mechanics via the terms $\alpha(E-E_0)$ and $\frac{\alpha}{2\sqrt{A_0}}(E-E_0)$.

#### 3. Cell Divisions

To introduce cell division in our model, we assume each cell has a cyclin level, $c$. The cyclin increases with time according to the following growth law:

$$\frac{\partial c}{\partial t} = 1 \tag{S52}$$

A cell is picked with a probability rate $r$, and only if $c > c_{th}$ and $A > A_{th}$, where $c_{th}$ is a threshold cyclin value and $A_{th}$ is a threshold area, the picked cell is divided. The selected cell is divided by a line perpendicular to its long axis.

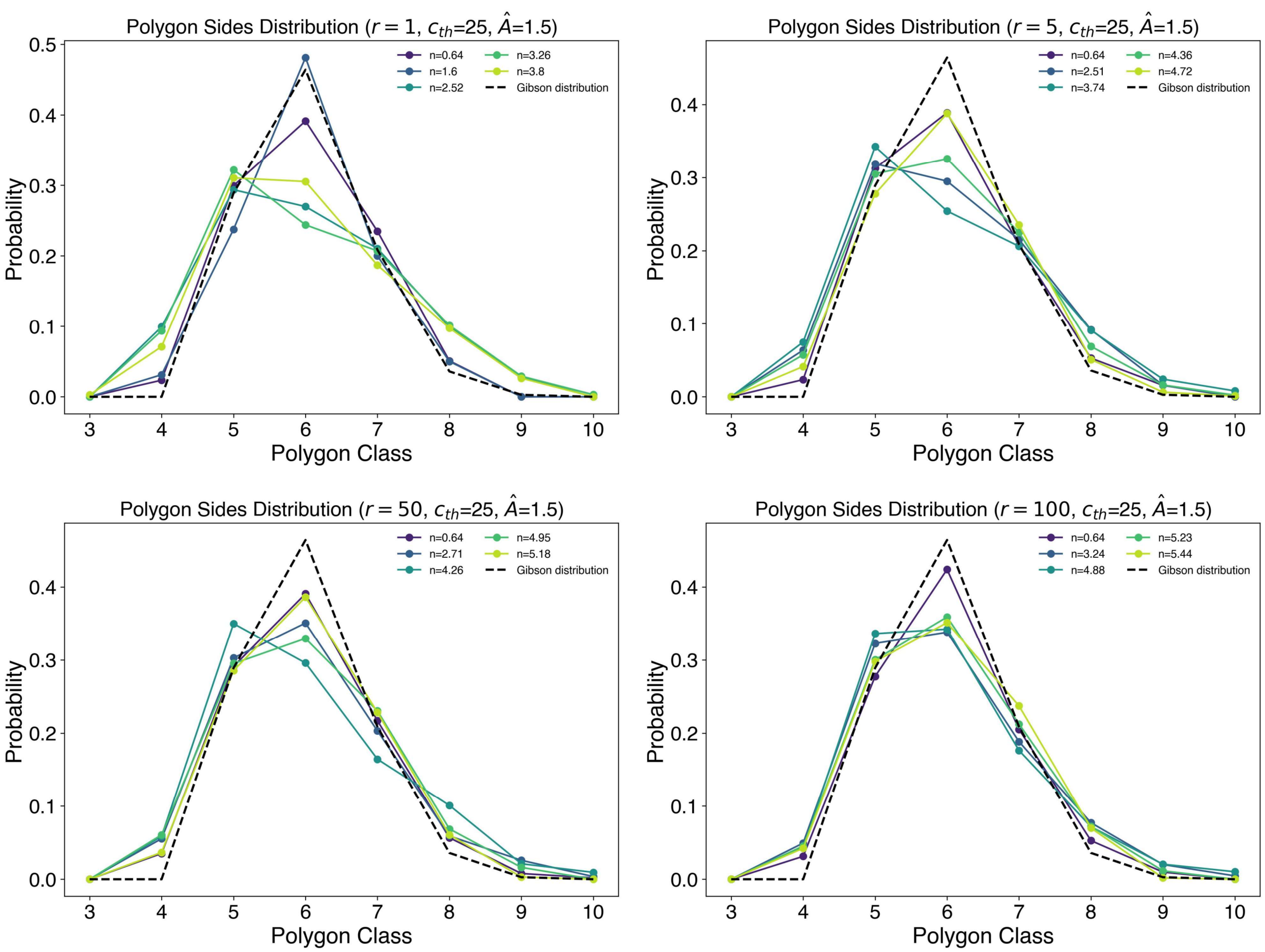

FIG. T10. The polygon distribution at different densities for different $r$ values, compared to the distribution given by Gibson et al.

## IV. ANALYSIS OF THE EXPERIMENTAL AND SIMULATION DATA

### A. Ageing

To show ageing in simulation, we have calculated the overlap function $Q(\Delta t)$ and self-intermediate scattering function $F_s(k, \Delta t)$, for cell trajectories with different time origins $t_{origin}$. The $Q(\Delta t)$ plot is shown in Fig. T11 and the $F_s(k, \Delta t)$ plot is shown in Fig. T13.

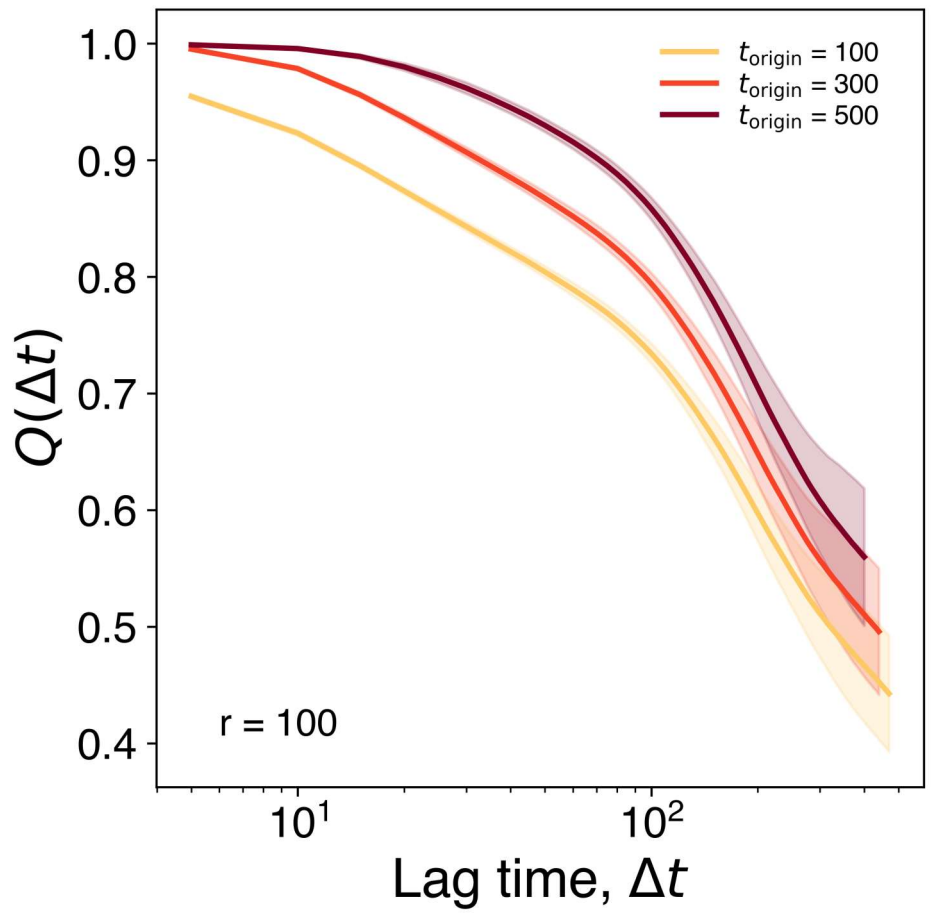

FIG. T11. $Q(\Delta t)$ for different $t_{\text{origin}}$, for $r = 100 \times 10^{-5}$ and $\hat{A} = 1.50$.

To calculate the self-intermediate scattering function, we need to pick a wavevector, $k$. We use the $k$ at which the static structure function $S(k)$ peaks [Fig. T12].

### B. Force-moment tensor of a cell

Forces in the vertex model act on the vertices of a cell, $c$. If the position vector of vertex $i$ is $\vec{r}_i$ and the force acting on it is $\vec{f}_i$, then the force moment tensor, $\Sigma$, is defined as:

$$\Sigma = \sum_{i \in V(c)} \vec{r}_i \otimes \vec{f}_i \tag{S53}$$

$$\therefore \Sigma_{mn} = \sum_{i \in V(c)} r_{i,m} f_{i,n}. \tag{S54}$$

Here, $V(c)$ is the set of vertices belonging to the cell $c$, and $m, n$ are the components of the vectors. The force-moment tensor serves as the analog of the stress tensor for a single cell. The trace (sum of the eigenvalues) of $\Sigma$ measures the compressive stress at the cell, whereas the difference of the eigenvalues measures the shear stress. In thermal equilibrium, the standard deviation of $Tr.\Sigma$ is inversely proportional to the bulk viscosity, and the standard deviation of the shear stress is inversely proportional to the shear viscosity. In out-of-equilibrium systems, such as the vertex model in this paper, fluctuations and viscosities are inversely related, but this relationship does not arise from thermal equilibrium.

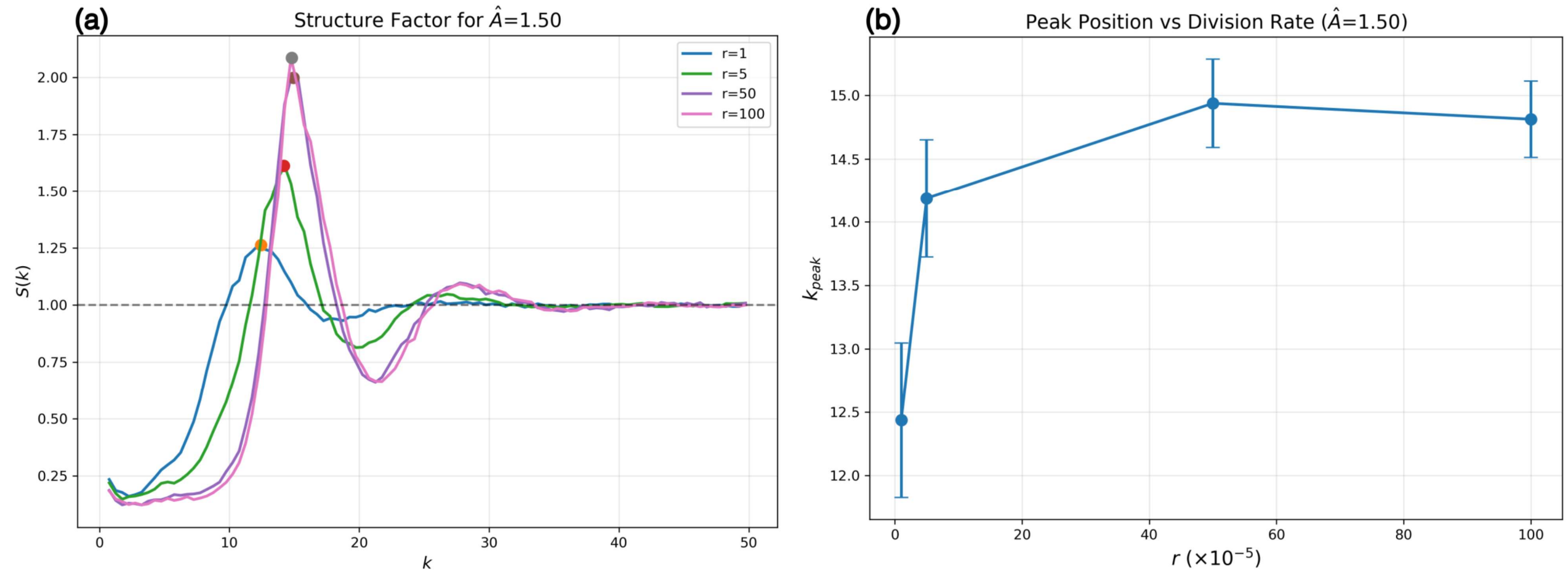

FIG. T12. Time-averaged structured factor for $\hat{A} = 1.50$.(b) $k$ values corresponding to the peaks of $S(k)$.

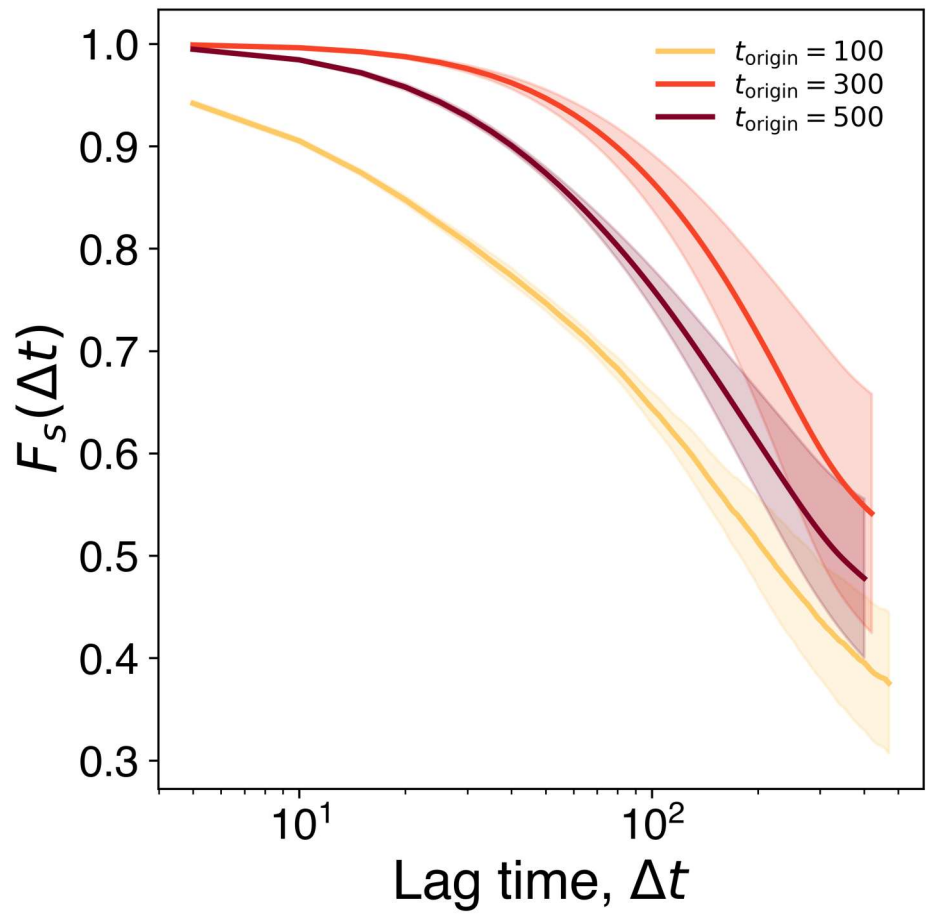

FIG. T13. $F_s(k, \Delta t)$ for different $t_{origin}$, for $k =$, $r = 100 \times 10^{-5}$ and $\hat{A} = 1.50$.

### C. Asphericity of cell shapes

We quantify cell shape anisotropy using asphericity. To compute asphericity, we define the shape-moment tensor:

$$M = \sum_{i \in V(c)} \vec{r}_i \otimes \vec{r}_i \tag{S55}$$

$$\therefore M_{mn} = \sum_{i \in V(c)} r_{i,m} r_{i,n}. \tag{S56}$$

Next, we compute the eigenvalues of this matrix. In 2D, these are $\lambda_1$ and $\lambda_2$. The asphericity, $\kappa$, is defined as:

$$\kappa = \left| \frac{\lambda_1 - \lambda_2}{\lambda_1 + \lambda_2} \right|. \tag{S57}$$

For a nearly isotropic shape, such as a circle or a regular hexagon, $\kappa \to 0$, whereas for anisotropic shapes $\kappa > 0$ and approaches 1 for shapes with high aspect ratios. Geometrically, it is equivalent to fitting an ellipse to the underlying shape, whose semimajor and semiminor axis lengths are $\lambda_1$ and $\lambda_2$. We use this equivalence to calculate the asphericity of the MDCK cells in the experiments. Each segmented cell is fitted to an ellipse in TrackMate, from which we compute the asphericities.

The asphericity here is related to the aspect ratio (AR) as follows:

$$\kappa = \left| \frac{\lambda_1 - \lambda_2}{\lambda_1 + \lambda_2} \right| = \left| \frac{\lambda_1/\lambda_2 - 1}{\lambda_1/\lambda_2 + 1} \right| = \left| \frac{AR - 1}{AR + 1} \right|. \tag{S58}$$

where the aspect ratio is given by $AR = \lambda_1/\lambda_2$. The asphericity $\kappa$ is related to the aspect ratio, which has been shown to have a universal distribution for confluent epithelial monolayers [14].

### D. Dynamic Heterogeneity Size Calculation

First, we grid the system into a $25 \times 25$ grid and average the cell velocities in each box. This gives a velocity map of the system. After this, we find the grid points with the top 20% velocity. Following [15], we classify the dynamic heterogeneity region as the largest contiguous region (here, a collection of boxes) as the dynamic heterogeneity region and the size of that region as the dynamic heterogeneity size.

### E. Wavelet Analysis of the signal

We analyzed the spatiotemporal dynamics of both experimental and theoretical datasets using a unified framework. In the experimental system, actin-labeled images were acquired continuously over a 30-hour period with a temporal resolution of 30 minutes, yielding a sequence of spatially resolved intensity maps. In parallel, from simulations of our mechanochemical feedback model, we obtained the temporal evolution of the projected inverse area over the same time window. To make the two datasets directly comparable, the field of view in both cases was divided into a uniform $20 \times 20$ grid. Within each grid element, local time series data were extracted by averaging pixel intensities from actin images in the experiment, and inverse area images from the simulations.

A central consideration in this comparison is the correspondence between actin intensity in experiment and inverse area in simulation. Experimentally, actin accumulation reflects structural compaction: higher actin levels are associated with smaller areas. Accordingly, we treated inverse area from simulations of our model as a proxy for actin concentration in the experimental data. This mapping enabled a consistent interpretation across modalities, thereby allowing theoretical predictions to be grounded in experimentally measurable quantities.

To isolate oscillatory components, the raw time series from both experiment and theory were first detrended using the PyBOAT software package. PyBOAT (A Biological Oscillations Analysis Toolkit) uses smoothing splines to remove low-frequency baseline trends, ensuring that slow signal variations do not mask the underlying oscillatory dynamics. After detrending, PyBOAT applies a continuous wavelet transform using the Morlet wavelet, generating a time–frequency representation of the signal. This approach captures transient and non-stationary oscillations that standard Fourier analysis would miss. From the resulting wavelet spectra, we extracted the dominant period and quantified the oscillation power for each grid element and then averaged the power values to obtain a representative measure of oscillatory strength, which was compared across simulated and experimental datasets. In the main text, to compare with the experiment, we choose $r = 5 \times 10^{-5}$, $c_{th} = 25$, and $\hat{A} = 1.50$ from our simulation in wavelet analysis.

## V. CHOICE OF PARAMETERS

### A. Choice of $\alpha$ and $\beta$

MCFL-II generates oscillations in the model, which has the mechanochemical couplings $\alpha$ and $\beta$. From simulations, we have found that the time period distributions of oscillations do not change with $\alpha$ and $\beta$ but change with $\tau_D$, $\tau_A$ and $\tau_P$. We have chosen values $\alpha = 2.25$ and $\beta = 2.88$.

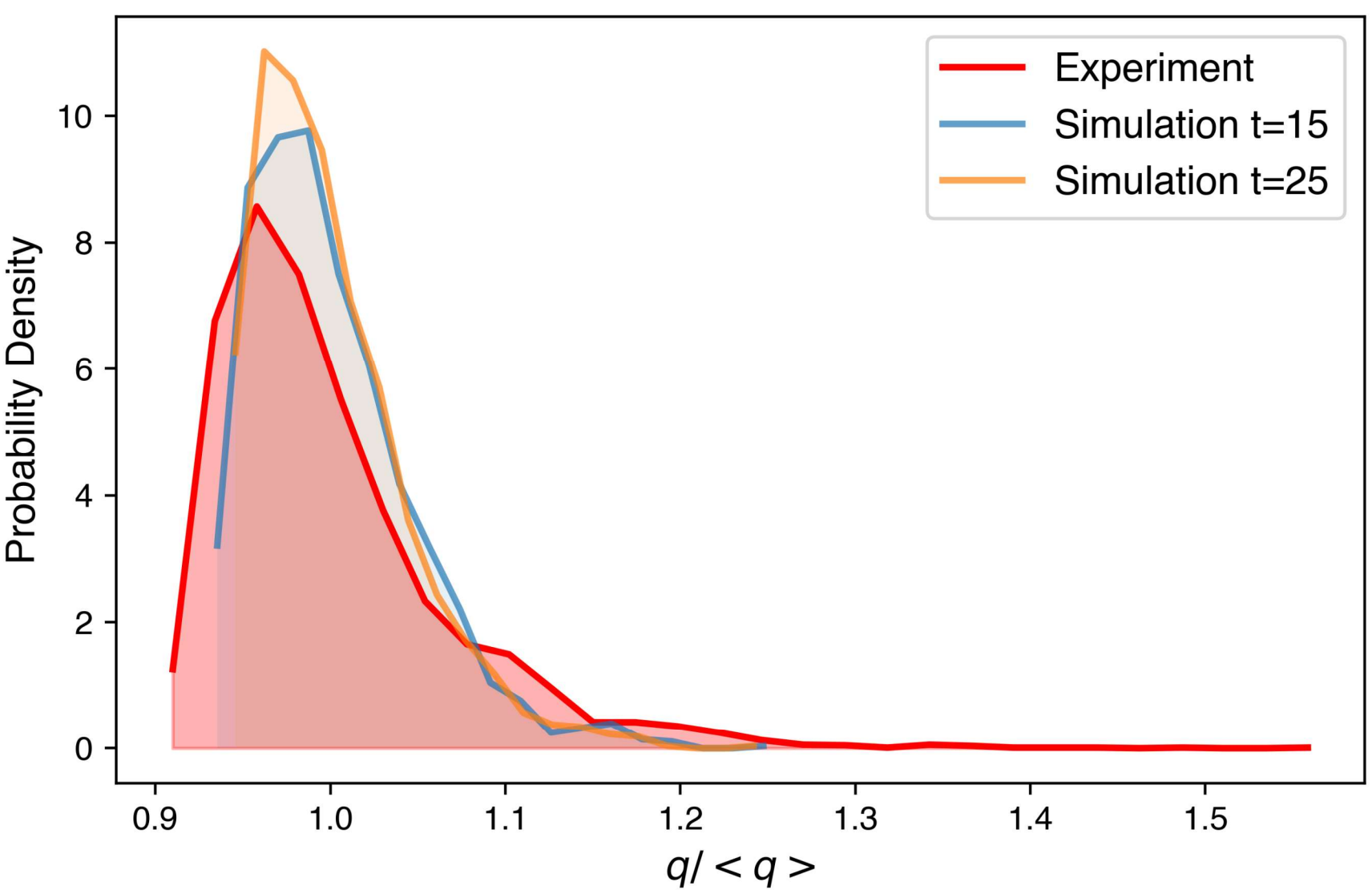

FIG. T14. Comparison of shape index distributions from experiment and simulation for $\alpha = 2.25$ and $\beta = 2.88$.

### B. Choice of $\tau_D$, $\tau_A$, $\tau_P$

The time period of oscillations generated by MCFL-II depends on $\tau_D$, $\tau_A$ and $\tau_P$. We have chosen the values by comparing the distributions with the experimentally observed distributions.

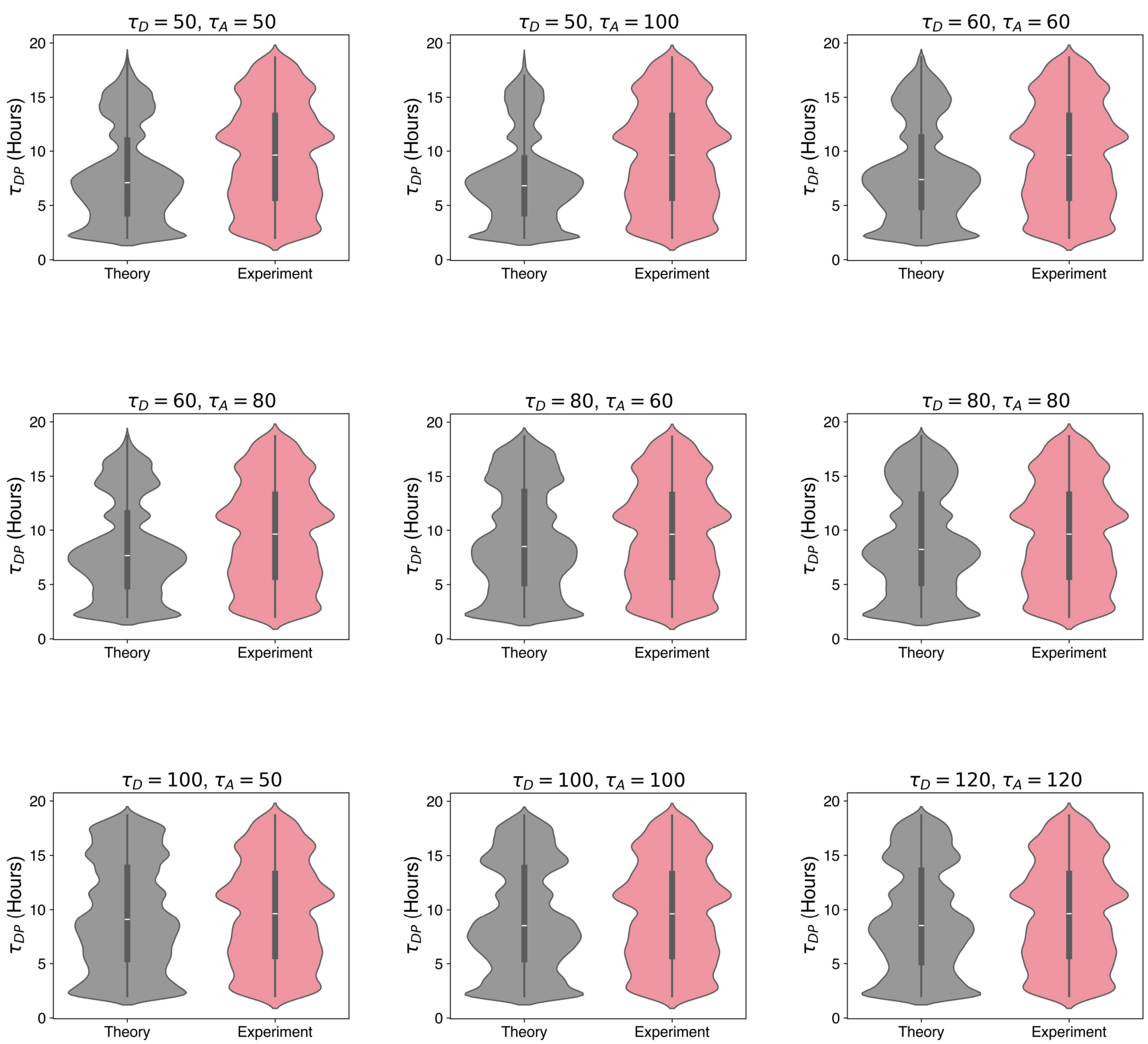

FIG. T15. Violin plots showing the dominant time period distributions compared with those obtained from experimental data. Here, $\tau_A = \tau_P$, $\hat{A} = 1.50$, $r = 50 \times 10^{-5}$.

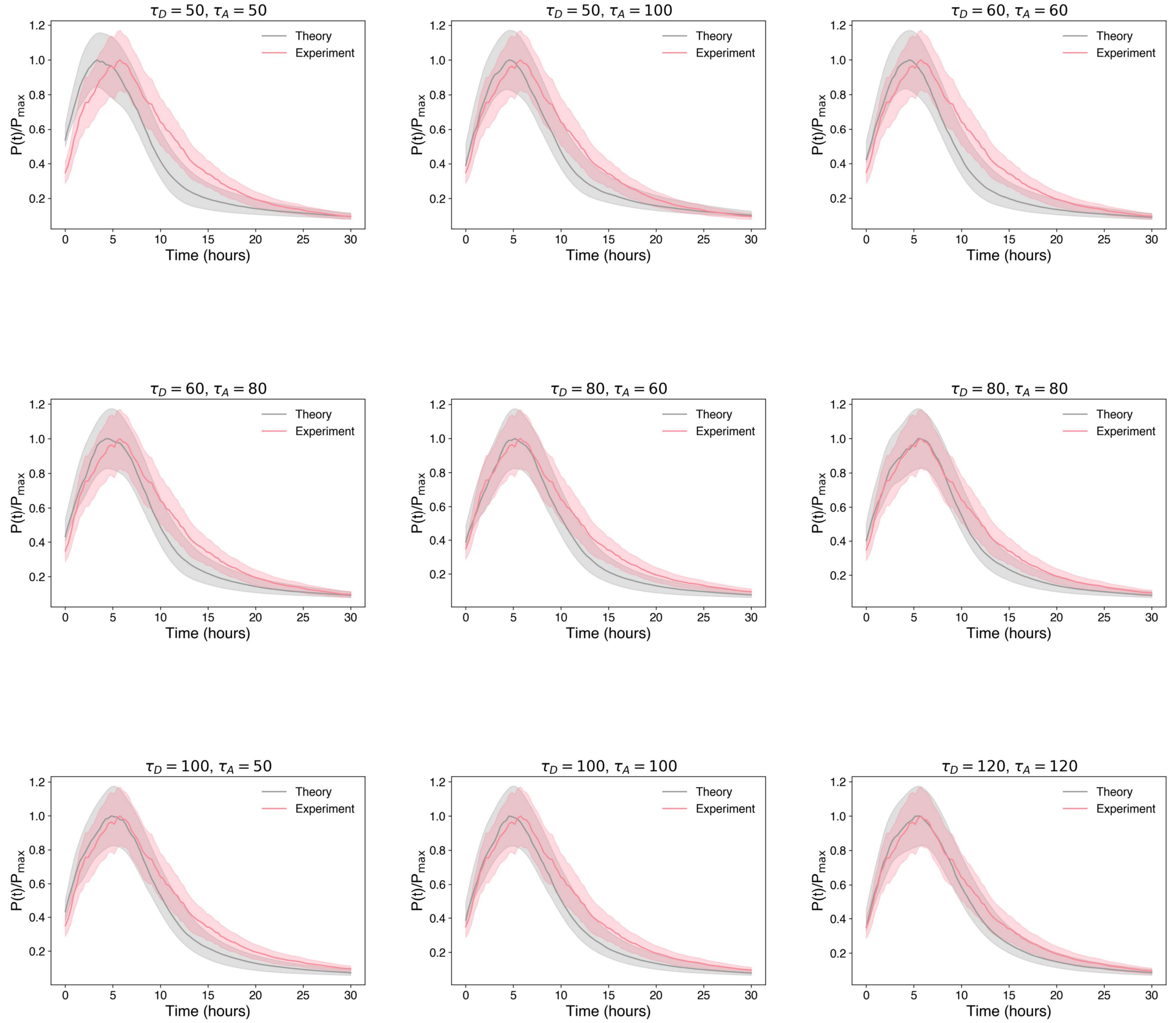

FIG. T16. Power time series plots comparison between experiment and simulation. Here, $\tau_A = \tau_P$, $\hat{A} = 1.50$, $r = 50 \times 10^{-5}$.

By observing that the distributions match best for $\tau_D = \tau_A = \tau_P = 120$, we use those values for the simulations.

### C. Choice of $r$, $c_{th}$ and $A_{th}$

We have chosen the values of $r$ and $c_{th}$ by comparing the simulation data with the experimentally observed growth rate.

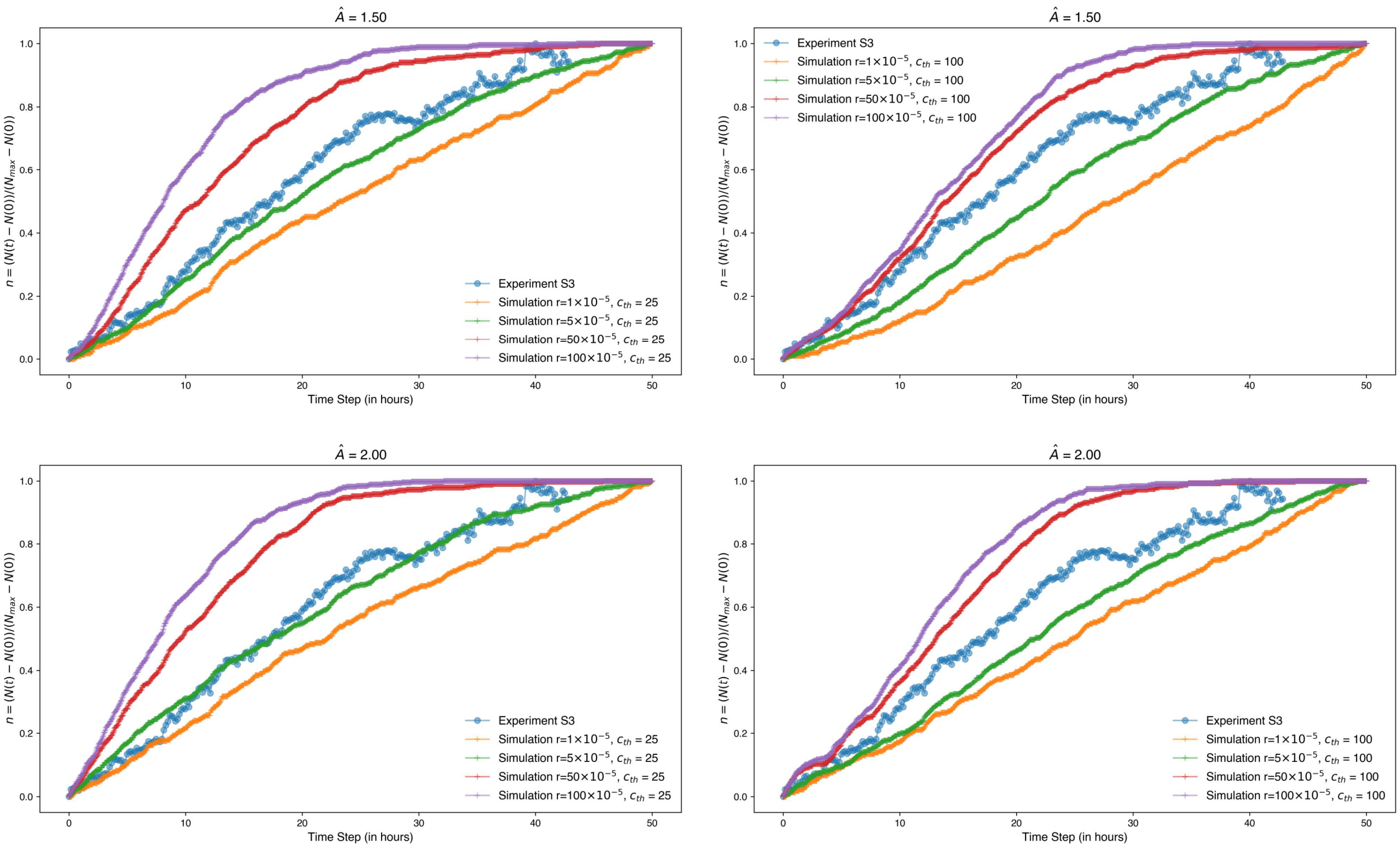

FIG. T17. Cell number growth for different division rates, $r$, for $c_{th} = 25, 100$, and $\hat{A} = 1.5, 2$.

The $N(t)$ curve in simulations plateaus as the cell areas continuously decrease over time [Fig. T17], as seen in the decrease of $dN/dt$ with time [Fig. T18]. We have obtained $\frac{dN}{dt}(t)$ by fitting a function $N(t) = A(1 - e^{-kt}) + N_0$ to the $N(t)$ data and then finding $\frac{dN}{dt} = Ake^{-kt}$. The time at which the $N(t)$ starts to plateau depends on the value of $A_{th}$. The value for simulations is set to $A_{th} = 0.3$.

Moreover, we see that the time period distribution fits better for the $r \approx 5 \times 10^{-5}$, which match the experimental growth rate [Fig. T19].

The growth rate $dN/dt$ in simulations varies with the parameters $r$ and $c_{th}$ as given in Fig. T20.

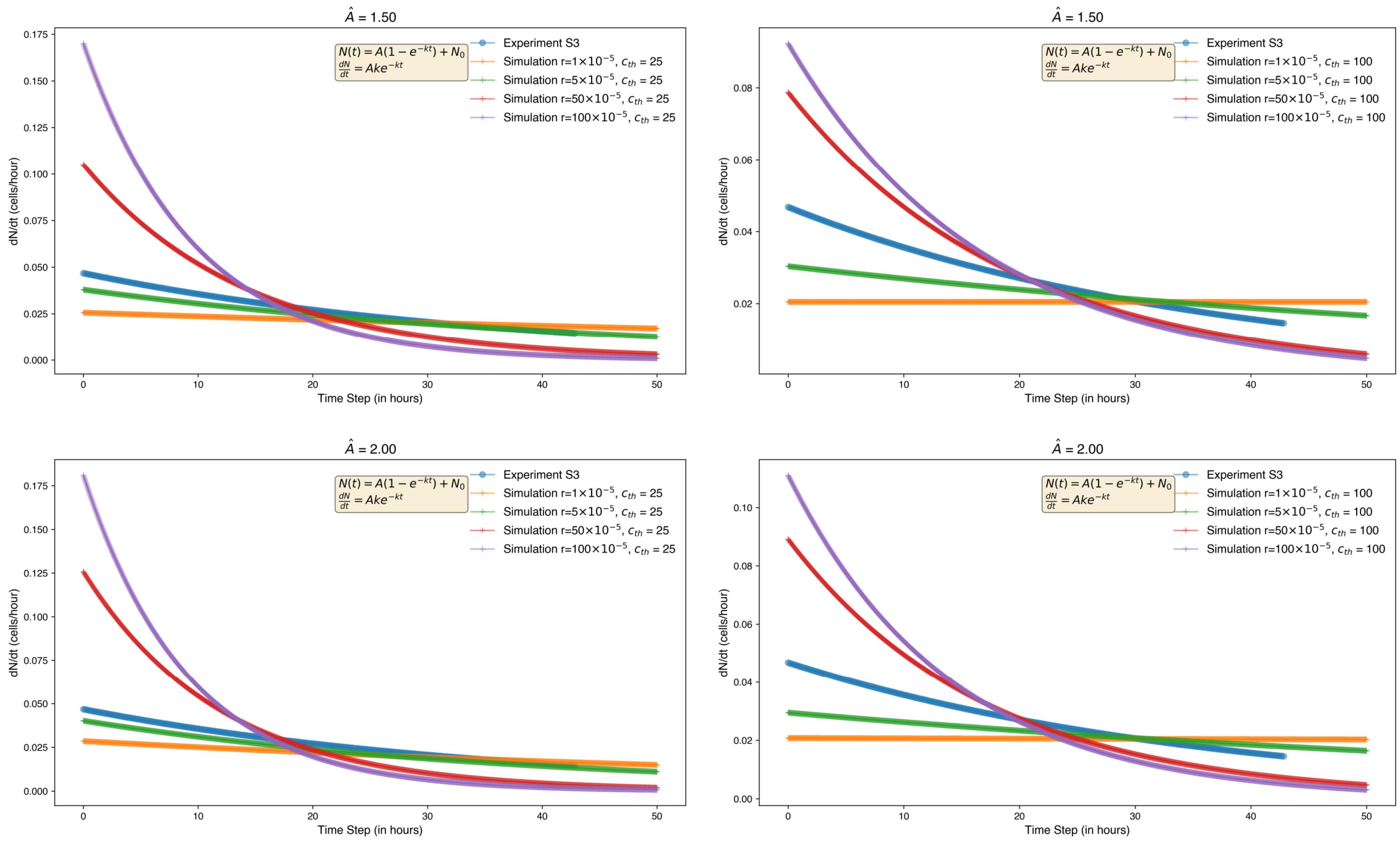

FIG. T18. Cell number growth rate for different division rates, $r$, for $c_{th} = 25, 100$, and $\hat{A} = 1.5, 2$.

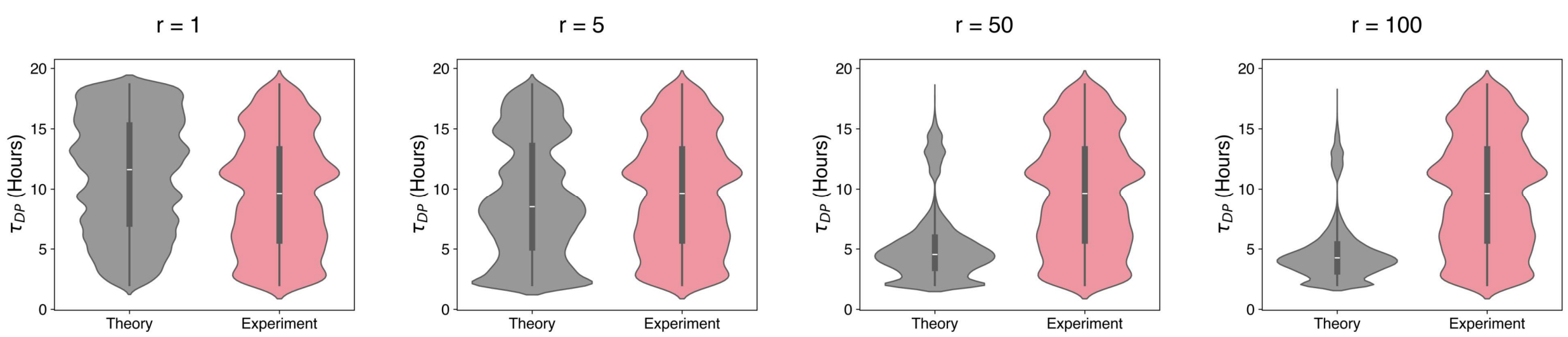

FIG. T19. Comparison between the dominant time period distributions from simulation and experiment for different $r$ (in $\times 10^{-5}$), $\hat{A} = 1.5$, and $c_{th} = 25$

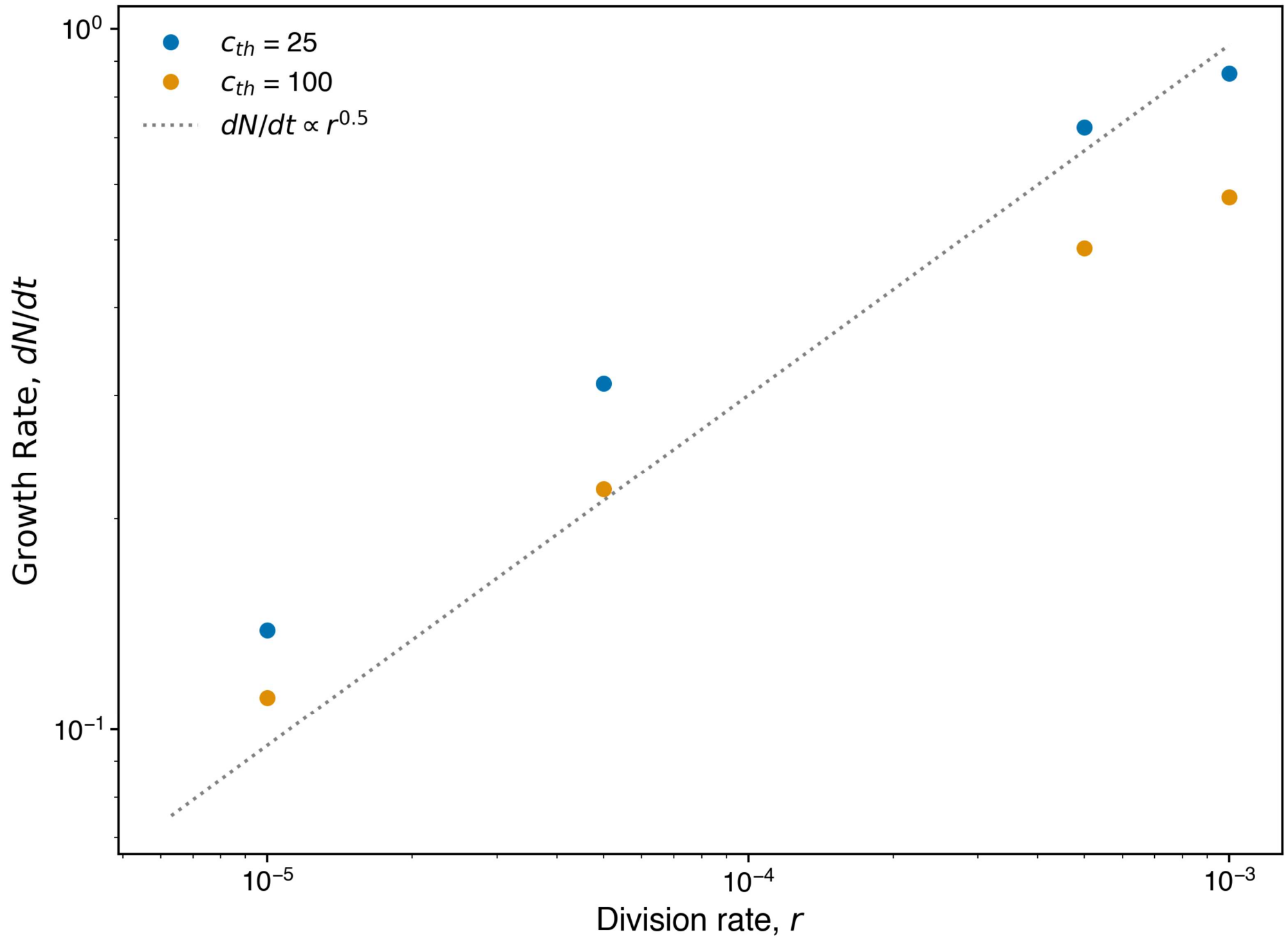

FIG. T20. Growth rates for different $r$ and $c_{th}$.

The growth rate here is calculated by fitting a straight line to the initial increase in the $N(t)$ data from simulations.

### D. Parameters Used

| Parameter | Value |
|---|---|
| $K$ | 1.2 [16] |
| $\Gamma$ | 0.38 [16] |
| $\zeta$ | 1 |
| $\hat{q}_0$ | 3.9 |
| $v_0$ | 0.125 |
| $D_r$ | 0.5 |
| $\tau_D = \tau_A = \tau_P$ | 120 |
| $\Delta t$ | 0.01 |

- **Simulation and Experimental timescales:** 1 simulation time step = 2 minutes in experiment.
- The values of other parameters are mentioned in the caption of the main text figures.
- **Topological transitions:** The simulation has the following topological transitions:
  - T1 transitions
  - T2 transitions: cell divisions
  - A node switch operation has also been implemented to prevent overlaps [17].

---